\documentclass[review]{elsarticle}

\usepackage{graphicx}
\usepackage{amssymb}

\usepackage[margin=2.5cm]{geometry}
\usepackage{graphicx}
\usepackage{dcolumn}
\usepackage{bm}
\usepackage{braket}
\usepackage{graphicx}
\usepackage{epstopdf}
\usepackage{mwe}    
\usepackage{subfig}
\usepackage[T1]{fontenc}
\usepackage{caption}
\usepackage{amssymb}
\usepackage{amsmath}
\usepackage{booktabs}
\usepackage{tabularx}
\usepackage{multirow}
\usepackage{floatrow}
\usepackage{xcolor}

\newcommand{\tensor}[1]{\bm{\mathsf{#1}}} 
\newcommand{\ms}{\scriptscriptstyle}

\newcommand{\subscript}[1]{_{\ms #1}}

\newcommand{\sg}{\hat{\sigma}}

\newcommand{\Kps}[1]{k^{'}\subscript{#1}}

\newcommand{\widefbox}[1]{\fbox{\hspace{2em}#1\hspace{2em}}}

\journal{Journal Name}

\begin{document}

\begin{frontmatter}

\title{Preconditioned Central Moment Lattice Boltzmann Method on a Rectangular Lattice Grid for Accelerated Computations of Inhomogeneous Flows}


\author{Eman  Yahia}
\ead{eman.yahia@ucdenver.edu}
\author{Kannan N. Premnath}
\ead{kannan.premnath@ucdenver.edu}

\address{\normalsize Department of Mechanical Engineering\\ College of Engineering, Design and Computing\\ University of Colorado Denver\\ 1200 Larimer street, Denver, Colorado 80217 , U.S.A.}



\begin{abstract}
Convergence acceleration of flow simulations to their steady states at lower Mach numbers can be achieved via preconditioning the lattice Boltzmann (LB) schemes that alleviate the associated numerical stiffness, which have so far been constructed on square lattices. We present a new central moment LB method on rectangular lattice grids for efficient computations of inhomogeneous and anisotropic flows by solving the preconditioned Navier-Stokes (PNS) equations. Moment equilibria corrections are derived via a Chapman-Enskog analysis for eliminating the truncation errors due to grid-anisotropy arising from the use of the rectangular lattice and the non-Galilean invariant cubic velocity errors resulting from an aliasing effect on the standard D2Q9 lattice for consistently recovering the PNS equations. Such corrections depend on the diagonal components of the velocity gradients, which are locally obtained from the second order non-equilibrium moments and parameterized by an associated grid aspect ratio $r$ and a preconditioning parameter $\gamma$, and the speed of sound in the collision model is naturally adapted to $r$ via a physically consistent strategy. We develop our approach by using a robust non-orthogonal moment basis and the central moment equilibria are based on a matching principle, leading to simpler expressions for the corrections for using the rectangular grids and for representing the viscosities as functions of the relaxation parameters, $r$ and $\gamma$, and its implementation is modular allowing a ready extension of the existing LB schemes based on the square lattice. Numerical simulations of inhomogeneous and anisotropic shear-driven bounded flows using the preconditioned rectangular central moment LB method demonstrate the accuracy and significant reductions in the numbers of steps to reach the steady states for various sets of characteristic parameters.

\end{abstract}




\end{frontmatter}


\section{Introduction} \label{sec:1}
With the rapid development in the computational fluid dynamics (CFD) techniques and their engineering and scientific applications, the need for accurate and efficient simulations of fluid flows has become vitally important. The lattice Boltzmann (LB) method has proved especially advantageous in simulating a variety of fluid flows, including complex flows such as multiphase and multicomponent systems, porous media flows, and turbulence~\cite{mcnamara1988use,benzi1992lattice,lallemand2021lattice}. Its firm basis resulting from certain a discretization of the Boltzmann equation and its ability to accommodate considerations beyond hydrodynamics, such as the higher order kinetic moments, contributed to many refinements of this approach. The LB method is naturally parallelizable, flexible in adopting models from kinetic theory, and it facilitates boundary condition implementations on Cartesian grids with relative ease, which have paved the way for its growing number of applications. Briefly, the method involves tracking the spatial and temporal evolution of the distribution of the particle populations due to collisions and advection along the characteristic discrete velocity directions referred to as a lattice. The collision step is often represented by a model involving the relaxation to certain equilibria, either directly involving the distribution functions~\cite{qian1992lattice} or their raw moment~\cite{d2002multiple}, central moment~\cite{geier2006cascaded} or cumulant~\cite{geier2015cumulant} representations, performed with using a single relaxation time~\cite{qian1992lattice} or multiple relaxation times~\cite{d2002multiple}, or by a model that is compliant with certain notions of entropy~\cite{karlin1999perfect}. The asymptotic continuum limit of such collide-and-stream steps on a lattice satisfying the necessary symmetry and isotropy considerations correspond to the dynamics of the fluid flow represented by the Navier-Stokes (NS) equations.

The LB method typically uses uniform Cartesian grids, resulting, for example, from the choice of a square lattice in two-dimensions (2D). Real world problems are often dominated by inhomogeneous and anisotropic flows, including wall-bounded shear flows. For example, in turbulent boundary layers or flow through channels or ducts, the eddy sizes are markedly different in different coordinate directions, and which progressively increase in the direction normal to the wall. Similar situations arise in simulating flows within enclosures with high geometric skewness characterized by large disparities in the length scales or the aspect ratios. Thus, the use of square/cubic grids are to solve such problems of practical interest are associated with high computational costs both in terms of time and memory, which can be orders of magnitude more than efficient algorithms based on nonuniform, stretched grids. Hence, it becomes highly important to develop more efficient approaches that use grids which naturally accommodate the spatial variations in the flow features. As a result, much attention has been paid to extend the LB schemes involving non-uniform grids. While unstructured grids or schemes involving interpolations for considering stretched grids can be used in this regard (see e.g., the monograph~\cite{kruger2017lattice} for a survey of such methods), it can result in complicated implementations or introduce additional numerical dissipation~\cite{lallemand2000theory}. On the other hand, one of the hallmarks of the LB methods is that the perfect lock-step advection or streaming used in their standard formulations incurs minimal overall dissipation while maintaining the simplicity of its implementation. For more efficiently simulating anisotropic flows while preserving this important feature, it becomes natural to utilize rectangular lattices rather than square lattices in 2D.

Thus, significant focus has been shown towards developing LB methods using rectangular lattice grids during the last decade following the initial investigation in this direction by Koelman~\cite{koelman1991simple}. For example, Hegele \emph{et al}~\cite{hegele2013rectangular}, Peng \emph{et al}~\cite{peng2016lattice}, Wang \emph{et al}~\cite{wang2019simulating} presented rectangular LB algorithms based on a single relaxation time (SRT) model via an extended lattice set, corrections to equilibrium distribution functions, and counteracting source terms, respectively, to recover the NS equations. Moreover, rectangular LB schemes based on raw moments using multiple relaxation times (MRT) were developed by Bouzidi \emph{et al}~\cite{bouzidi2001lattice} and Zhou \emph{et al}~\cite{zhou2012mrt}, which were analyzed and an improved LB scheme on a rectangular grid with the necessary correction terms that is consistent with the NS equations was constructed by Peng \emph{et al}~\cite{peng2016hydrodynamically}. However, many of these methods involved cumbersome implementations, complicated expressions for the corrections, and numerical stability issues when the grid aspect ratio of the rectangular lattice (defined later) is significantly far off from unity (i.e., characterizing strong grid stretching in one of the directions relative to the other) or for simulating flows with relatively low viscosities or high Reynolds numbers. On the other hand, recognizing that the use of central moments, which naturally preserves the Galilean invariance of those moments independently supported by the lattice, can significantly improve the stability and accuracy when compared to the use of raw moments~\cite{geier2006cascaded,asinari2008generalized,premnath2009incorporating,premnath2011three,ning2016numerical,de2017non,de2017nonorthogonal,fei2017consistent,fei2018three,Hajabdollahi201897,HAJABDOLLAHI2018838,chavez2018improving,hajabdollahi2019cascaded,fei2020mesoscopic,hajabdollahi2021central,hajabdollahi2021central,adam2019numerical,adam2021cascaded},
we recently constructed a rectangular central moment LB method (RC-LBM)~\cite{yahia2021central}, which was then further extended to three-dimensions with an improved implementation strategy~\cite{yahia2021three}. While the original central moment LB scheme was constructed using an orthogonal moment basis~\cite{geier2006cascaded}, Geier \emph{et al.}~\cite{geier2015cumulant} in 2015 provided a detailed discussion on the role of the moment basis in their development of a cumulant LB method and also constructed a variety of collision models, including those based on raw moments, central moments and cumulants using non-orthogonal moment basis and presented them in the various appendices of~ \cite{geier2015cumulant}. Moreover, the numerical stability advantages of using such non-orthogonal moment basis relative to the orthogonal moment basis were demonstrated via a linear stability analysis in~\cite{dubois2015stability}.  Besides, earlier studies on cascaded LB schemes performed mathematical analysis and demonstrated consistency to the Navier-Stokes equations using such simpler basis~\cite{asinari2008generalized, premnath2009incorporating}. The use of non-orthogonal central moments in algorithmic implementations in LB schemes was later adopted in Refs.~\cite{de2017non,de2017nonorthogonal}. Hence, in contrast to the prior rectangular LB schemes, the RC-LBM used a natural non-orthogonal moment basis, a matching principle to construct the equilibria involving higher order velocity terms resulting in a simpler and significantly more robust implementation. Also, in Ref.~\cite{yahia2021three}, we also explicitly demonstrated the computational advantages of using a rectangular lattice in lieu of a square lattice in solving inhomogeneous and anisotropic flows. Moreover, the central moment LB method on a cuboid lattice presented in Ref.~\cite{yahia2021three} is modular in construction thereby allowing ready extension of the existing algorithms on a cubic lattice to cuboid lattices, and provides a unified formulation with corrections that are applicable for a wide variety of all the standard collision models.

Since the LB methods, which are time-marching and weakly compressible flow solvers, represent the fluid motion in the incompressible limit asymptotically, the smaller the Mach number is used simulations, the better is their accuracy, which also enhances their numerical stability. However, a general issue in CFD, including those for the LB methods, for computing flows via reducing the Mach number to relatively small values is the associated increase in the stiffness that causes slower convergence to the steady state. This is due to the large condition number of evolution equations resulting from wide contrasts in the flow speeds and the acoustic speeds in such cases. One approach to reduce the number of steps taken for convergence is to precondition the system of flow equations, wherein such disparities between the characteristic speeds are reduced at the cost of sacrificing the temporal accuracy. As shown by Turkel (see e.g.,~\cite{turkel1987preconditioned,turkel1999preconditioning}), this has been accomplished in the context of classical CFD methods by solving the so-called preconditioned NS equations which involve an adjustable preconditioning parameter. Guo \emph{et al}~\cite{guo2004preconditioned} introduced the first preconditioned LB scheme using an SRT model, which was then extended to a MRT model involving raw moments and forcing terms by Premnath \emph{et al}~\cite{premnath2009steady}. Izquierdo and Fueyo~\cite{izquierdo2009optimal} demonstrated optimal preconditioning of a MRT-LB scheme, while Meng \emph{et al}~\cite{meng2018preconditioned} introduced a preconditioned MRT-LB algorithm for simulations of steady two-phase flows in porous media. More recently, Hajabdollahi and Premnath~\cite{hajabdollahi2019improving} developed a cascaded central moment LB scheme for simulation of preconditioned NS equations, which was then further improved by eliminating the non-Galilean invariant cubic velocity errors that are dependent on the preconditioning parameter and demonstrating significant reductions in the number of steps for convergence of a variety of flows in Ref.~\cite{Hajabdollahi201897}. Moreover, we note that, recently, a preconditioned SRT-LB approach based on a finite-volume discretization on unstructured grids was developed and studied by Walsh and Boyle~\cite{walsh2020preconditioned}. However, generally, prior investigations constructed preconditioned LB algorithms on square lattices, and rectangular LB schemes for the solution of preconditioned NS equations while maintaining the collide-and-stream steps with perfect lock-step advection have not yet been discussed in the literature. Development and analysis of such preconditioned LB schemes on rectangular lattices could enable convergence acceleration of inhomogeneous and anisotropic flows thereby further improving the computational efficiency achieved with the use of rectangular lattice grids, which is the focus of this paper.

In this work, we aim to construct a new preconditioned rectangular central moment lattice Boltzmann method (referred to as the PRC-LBM in what follows). In this regard, we employ a simpler non-orthogonal moment basis and the central moment equilibria are constructed by matching with those of the continuous Maxwell distribution with appropriate modifications in order to consistently recover the preconditioned NS equations. By performing a Chapman-Enskog analysis, we will derive corrections to the equilibria that eliminate the truncation errors due to grid anisotropy and the non-Galilean invariant cubic velocity terms arising due to aliasing effects on the standard D2Q9 lattice appearing in the emergent equations under the asymptotic limit when compared to the preconditioned NS equations. The resulting corrections will be shown to depend on the preconditioning parameter, grid aspect ratio, and the normal components of the velocity gradient tensor, where the latter will be expressed in terms of second-order non-equilibrium moments which will allow their computations locally without using finite difference approximations. It may be noted that in our previous 2D rectangular LBM~\cite{yahia2021central}, the transformation matrices for mappings between the distribution functions and raw moments, which depend on the grid aspect ratio, are constructed to separate the trace of the second order moments from its other components that allows independent specification of the bulk viscosity and shear viscosity, respectively. By contrast, in this paper, following our recent work in Ref.~\cite{yahia2021three}, the PRC-LBM will segregate the bulk viscosity from the shear viscosity only within the step involving relaxation of moments to preconditioned equilibria under collision, and the pre- and post-collision mapping matrices involve a simpler moment basis and account for the grid aspect ratios only via certain diagonal scaling matrices. This results in a modular implementation of the PRC-LBM, and its formulation involves simpler expressions for the necessary corrections and the transport coefficients. Moreover, we note while all the prior rectangular LB schemes (e.g.,~\cite{peng2016hydrodynamically,yahia2021central}) indicated that the speed of sound should be adjusted to accommodate for its variations with the grid aspect ratio when compared to the speed sound for the square lattice, they did not provide any rationale or explicit formulas to accomplish this other than providing some tabulated data. In this work, we provide some physical arguments to consistently obtain the speed of sound for the rectangular lattice, with its explicit parametrization by the grid aspect ratio. Finally, we will perform some numerical studies to demonstrate the accuracy of the PRC-LBM and reductions in the number of steps for convergence to the steady state for simulations of selected cases of inhomogeneous and anisotropic flows for different choices of the preconditioning parameter and the grid aspect ratio.

This paper is organized as follows. The following section (Sec.~\ref{sec:2}) discusses a consistent approach for the selection of the speed of sound for lattice schemes based on rectangular lattice grids. Next, in Sec.~\ref{sec:3}, we present a Chapman-Enskog analysis of the preconditioned non-orthogonal moment LB formulation on a rectangular D2Q9 lattice and derive the correction terms necessary to eliminate the truncation errors due to grid anisotropy and non-Galilean invariant velocity terms arising from aliasing effects. Such corrections are shown to be parameterized by the grid aspect ratio, preconditioning parameter and the velocity gradients, where the latter are obtained locally from non-equilibrium moments. The construction of the preconditioned rectangular central moment LBM for an efficient implementation is discussed in Sec.~\ref{sec:4}, with the attendant algorithmic details of the PRC-LBM provided in~\ref{sec:algorithmic-details-PRC-LBM}. Then, in Sec.~\ref{sec:5}, we present numerical results for some case studies involving anisotropic and inhomogeneous shear flows, validating the PRC-LBM for accuracy and demonstrating convergence acceleration via preconditioning on rectangular lattice grids for various characteristic parameters. Moreover, comparisons between the preconditioned rectangular central moment LBM (PRC-LBM) another formulation involving the preconditioned rectangular raw moment LBM are made in Sec.~\ref{sec:comparisonsbetweenmodels}. The conclusions of this work are highlighted in Sec.~\ref{sec:6}.

\section{Selection of Speed of Sound on Rectangular Lattice Grid for Physical Consistency} \label{sec:2}\par
\begin{figure}[H]
\centering
 \includegraphics[width=0.8\textwidth] {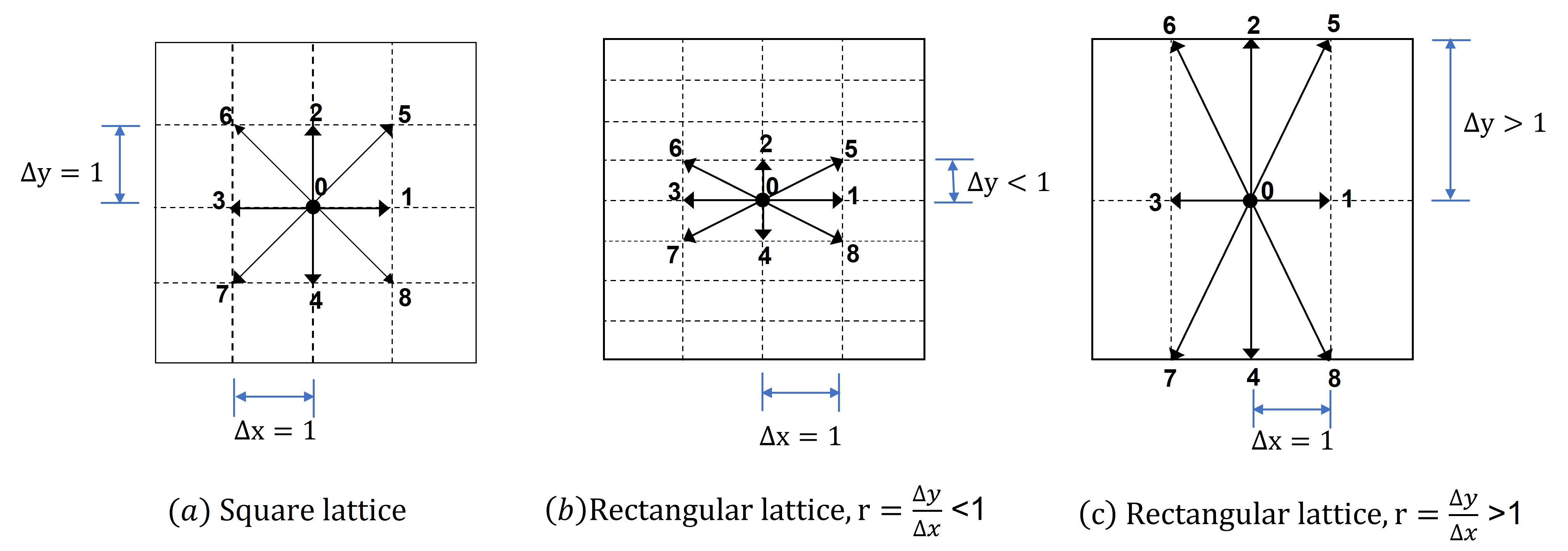}
  \caption{Two-dimensional, nine velocities (D2Q9) lattice with different possible arrangements based on the grid aspect ratio $r$.}
 \label{fig:lattice}  
\end{figure}

Before discussing the preconditioned rectangular central moment LB scheme and its analysis, we will now present a general physical consideration on the selection of the speed of sound for rectangular lattice grids and its relation to the sound speed of the usual square lattice. For the D2Q9 \emph{square} lattice shown in Fig.~\ref{fig:lattice}(a), with a lattice spacing $\Delta x$ and a time step $\Delta t$ resulting in the particle speed $c=\Delta x/\Delta t$, based on considerations of isotropy and Galilean invariance, it is well known that its optimal value of the speed of sound $c_{s*}$ is given by
\begin{equation}\label{Eq:1}
  c_{s*}= \frac{1}{\sqrt{3}}c = \frac{1}{\sqrt{3}}\frac{\Delta x}{\Delta t}= \frac{1}{\sqrt{3}}.
\end{equation}
Thus, $c_{s*}=1/\sqrt{3}$ in the usual lattice units (i.e., when $\Delta x=\Delta t=1.0$). For the two possible arrangements of the D2Q9 \emph{rectangular} lattice shown in Figs.~\ref{fig:lattice}(b) and ~\ref{fig:lattice}(c), we can generally define a \emph{grid aspect ratio} $r$ representing the ratio of the grid spacing in the $y$ direction with respect to that in the $x$ direction, i.e., $r=\Delta y/\Delta x$.
Denoting the particle speeds in the $x$ and $y$ directions, respectively, as $c_x=\Delta x/\Delta t = c$ and $c_y=\Delta y/\Delta t$, it then follows that $c_x=c$ and $c_y=r \;c$. From these two particle speeds, one may introduce two possible values of the speed of sound $c_{sx}$ and $c_{sy}$ in the two coordinate directions: $c_{sx}=c_{s*}$ and $c_{sy}=r \;c_{s*}$. However, it can be readily established from a Chapman-Enskog analysis that the pressure field is related to the local density times the square of the speed of sound, which reflects the equation of state of a weakly compressible athermal fluid motion represented by the LB method. Since the pressure field is an isotropic and a scalar quantity, the rectangular LB schemes need to use only one among the possible values of the speed of sound to satisfy physical consistency. Moreover, given that the speed of sound is a fraction of the particle speed, and in order to be consistent with the Courant-Friedrichs-Lewy condition, we prescribe that the effective speed of sound on the rectangular lattice be chosen as the one that provides the \emph{minimum} value among the two possibilities, i.e., $c_s= \mbox{min}(c_{sx}, c_{sy})$, so that it picks the more limiting case. In other words, our selection procedure for the speed of sound on the rectangular lattice $c_s$ relative to that for the corresponding square lattice $c_{s*}$ can be written as
\begin{equation}\label{eq:2}
  c_s= q \; c_{s*},  \quad  q= \mbox{min}(1,r).
\end{equation}
Thus, if $r<1$, $c_s= r c_{s*}$ (see Fig.~\ref{fig:lattice}(b)) and when $r>1$, $c_s= c_{s*}$ (see Fig.~\ref{fig:lattice}(c)). Moreover, when $r=1$, it naturally recovers the optimal value of $1/\sqrt{3}$ used for the square lattice. Thus, Eq.~\eqref{eq:2} automatically adapts the sound speed according to the grid aspect ratio, $r$, unlike previous LB schemes based on the rectangular lattice (e.g.,~\cite{peng2016hydrodynamically,yahia2021central}), where no such expressions are provided. Also, noting that a Chapman-Enksog analysis relates the kinematic viscosity of the fluid $\nu$, to a relaxation parameter and the square of the speed of sound $c_s^2$ (see the next section), it follows that $\nu$ is then parameterized by $q^2$, which facilitates maintaining numerical stability self consistently as the grid aspect ratio $r$ is varied with the use of rectangular lattice grids. Finally, we mention here that the specification of the Mach number $\mbox{Ma}$ for the rectangular LB simulations should be based on Eq.~\eqref{eq:2}, i.e., for any characteristic flow speed $U$, $\mbox{Ma}=U/c_s= U/(q c_{s*})= \mbox{Ma}_*/q$, where $\mbox{Ma}_*$ is the Mach number used for the square lattice.

We will formulate our central moment LB scheme on a $D2Q9$ rectangular lattice grid, where the particle velocity components $e_x$ and $e_y$ in the directions $x$ and $y$ (see Fig.~\ref{fig:lattice}) following the definition of the grid aspect ratio $r$ can be written as
\begin{subequations}
\begin{eqnarray}
\ket{e_{x}} &=& (0,1,0,-1,0,1,-1,-1,1)^\dag, \label{eq:3a}\\
\ket{e_{y}} &=& (0, 0, r, 0, -r, r, r, -r, -r)^\dag,\label{eq:3b}
\end{eqnarray}
\end{subequations}
where $\ket{\cdot}$ denotes a column vector based on the standard `ket' notation and $\dag$ refers to taking the transpose. We will also need the following 9-dimensional vector in defining the moment basis in the next section:
\begin{eqnarray}
\ket{1}  = (1,1,1,1,1,1,1,1,1)^\dag. \label{eq:4}
\end{eqnarray}

\section{Chapman-Enskog Analysis of Preconditioned LBE a Rectangular Lattice Grid: Isotropy Corrections, Macroscopic
Flow Equations, and Local Expressions for Velocity Gradients}\label{sec:3}
In the following, we will construct a preconditioned rectangular central moment LBM on the D2Q9 rectangular lattice with grid aspect ratio-adapted speed of sound to solve the following preconditioned NS equations:
\begin{subequations}\label{eq:pNSE}
\begin{eqnarray}\label{eq:5}
&\partial_t \rho + \bm{\nabla}\cdot \left( \rho \bm{u}\right) = 0,
\end{eqnarray}
\begin{eqnarray}\label{eq:6}
&\partial_t \left( \rho \bm{u}\right)+\bm{\nabla}\cdot\left( \cfrac{\rho \bm{u} \bm{u}}{\gamma}\right)= -\cfrac{1}{\gamma}\bm{\nabla}{p}+\cfrac{1}{\gamma}\; \bm{\nabla}\cdot \tensor{\tau}+\cfrac{\bm{F}}{\gamma},
\end{eqnarray}
\end{subequations}
where $\gamma$ is the preconditioned parameter used to achieve convergence acceleration to the steady state, $\bm{u}$ and $\rho$ are the fluid velocity and density, respectively, $p=\gamma c_s^2 \rho $ is the pressure field and $\tensor{\tau}$ is the viscous stress tensor $\tensor{\tau}=\rho \nu (\bm{\nabla}\bm{u}+ (\bm{\nabla}\bm{u})^\dag)$, and $\bm{F}$ is the body force.

\subsection{Moment basis, and definitions of central moments and raw moments}
In this regard, as in our previous work~\cite{yahia2021central,yahia2021three}, we employ a linearly independent set of non-orthogonal basis vectors for moments, by noting that they are chosen to especially allow for the separation of the isotropic parts from the non-isotropic parts of the second order moments for independent specification of the transport coefficients (i.e., the shear and bulk viscosities). For the D2Q9 lattice, such basis vectors are defined using a combination of the monomials of the type $\ket{e_x^m e_y^n}$, where $m$ and $n$ are integer exponents, as follows:
\begin{equation} \label{eq:7}
\tensor{T}= \Big[\ket{T_{0}},\ket{T_{1}},\ket{T_{2}},\ldots,\ket{T_{8}} \Big]^{\dag},
 \end{equation}
where
\begin{align}\label{eq:8}
&\ket{T_0}=\ket{1}, \qquad \ket{T_1}=\ket{e_x},\qquad \ket{T_2}=\ket{e_y}, \qquad \ket{T_3}=\ket{e_x^2 +e_y^2},\qquad  \ket{T_4}=\ket{e_x^2-e_y^2}, \nonumber\\
&\ket{T_5}=\ket{{e_x} {e_y}},\qquad\ket{T_6}=\ket{e_x^2 {e_y}},\qquad \ket{T_7}=\ket{{e_x} e_y^2},\qquad \ket{T_8}=\ket{e_x^2 e_y^2}.
\end{align}
Then, defining the sets of discrete distribution functions $\mathbf{f}$, their equilibria $\mathbf{f}^{eq}$, and the source terms $\mathbf{S}$, which represent the effect of the body force $\bm{F}=(F_x,F_y)$ on the fluid motion, respectively, as
\begin{subequations}
\begin{eqnarray}\label{eq:8A}
&\mathbf{f}=\left(f_{0},f_{1},f_{2},\ldots,f_{8}\right)^{\dag}, \quad
\mathbf{{f}}^{eq}=\left({f}_{0}^{eq},{f}_{1}^{eq},{f}_{2}^{eq},\ldots,{f}_{8}^{eq}\right)^{\dag},\quad &\mathbf{S}=\left({S}_{0},{S}_{1},{S}_{2},\ldots,{S}_{8}\right)^{\dag},
\end{eqnarray}
\end{subequations}
we can then express their \emph{raw moments} of order ($m+n$), $k_{mn}^\prime$, $k_{mn}^{eq\prime}$, and $\sigma_{mn}^\prime$, respectively,
\begin{subequations}\label{eq:9A}
\begin{eqnarray}
  &k_{mn}^\prime= \sum_{\alpha=0}^{8} f_{\alpha} \;e_{\alpha x}^m  e_{\alpha y}^n,\\
  &k_{mn}^{eq\prime}= \sum_{\alpha=0}^{8} f_{\alpha}^{eq} \;e_{\alpha x}^m  e_{\alpha y}^n,\\
  &\sigma_{mn}^\prime= \sum_{\alpha=0}^{8} S_{\alpha} \;e_{\alpha x}^m  e_{\alpha y}^n.
\end{eqnarray}
\end{subequations}
Similarly, we can write the corresponding \emph{central moments} $k_{mn}$, $k_{mn}^{eq}$ and $\sigma_{mn}$, respectively, by subtracting the particle velocities ($e_{\alpha x}, e_{\alpha y}$) by the fluid velocity ($u_x,u_y$) as follows:
\begin{subequations}\label{eq:9B}
\begin{eqnarray}
  &k_{mn}= \sum_{\alpha=0}^{8} f_{\alpha} \;(e_{\alpha x} -u_x)^m  (e_{\alpha y} -u_y)^n, \\
  &k_{mn}^{eq}= \sum_{\alpha=0}^{8} f_{\alpha}^{eq} \;(e_{\alpha x} -u_x)^m  (e_{\alpha y} -u_y)^n\\
  &\sigma_{mn}= \sum_{\alpha=0}^{8} S_{\alpha} \;(e_{\alpha x} -u_x)^m  (e_{\alpha y} -u_y)^n.
\end{eqnarray}
\end{subequations}
For convenience, we collect the various raw moments supported by the lattice set in view of the moment basis given in Eq.~\eqref{eq:8} in the form of the following 9-dimensional vectors as
\begin{subequations} \label{eq:9AA}
\begin{eqnarray}
&\mathbf{n}=\left(k_{00}^\prime, k_{10}^\prime, k_{01}^\prime, k_{20}^\prime+ k_{02}^\prime, k_{20}^\prime - k_{02}^\prime, k_{11}^\prime, k_{21}^\prime, k_{12}^\prime, k_{22}^\prime\right)^{\dag},\\
&\mathbf{n}^{eq}=\left(k_{00}^{eq \prime}, k_{10}^{eq \prime}, k_{01}^{eq \prime}, k_{20}^{eq \prime}+ k_{02}^{eq\prime}, k_{20}^{eq\prime} - k_{02}^{eq\prime}, k_{11}^{eq\prime}, k_{21}^{eq\prime}, k_{12}^{eq\prime}, k_{22}^{eq\prime}\right)^{\dag},\label{eq:basemomenteqvector}\\
&\mathbf{\Psi}=\Big(\sigma_{00}^\prime, \sigma_{10}^\prime, \sigma_{01}^\prime, \sigma_{20}^\prime+ \sigma_{02}^\prime, \sigma_{20}^\prime- \sigma_{02}^\prime, \sigma_{11}^\prime, \sigma_{21}^\prime, \sigma_{12}^\prime, \sigma_{22}^\prime\Big)^{\dag},
\end{eqnarray}
\end{subequations}
Then, the mappings between the various raw moments and the distribution functions can be compactly expressed via the matrix $\tensor{T}$ as
\begin{eqnarray} \label{eq:9}
&\mathbf{n}= \tensor{T} \mathbf{f}, \quad  \mathbf{n}^{eq}=  \tensor{T} \mathbf{f^{eq}}, \quad \mathbf{\Psi}= \tensor{T}\mathbf{S}.
\end{eqnarray}
Here, it should be mentioned that we use the combinations of the second order moment basis $\ket{e_x^2 +e_y^2}$ and $\ket{e_x^2 -e_y^2}$ to retain the flexibility of an independent specification of the bulk viscosity and shear viscosity, and the matrix $\tensor{T}$ as formulated above then facilitates in the demonstration of the consistency of our approach with the preconditioned NS equations and in the derivation of the required attendant corrections in the reminder of this section. However, in the actual implementation of the algorithm in the next section (see Sec.~\ref{sec:4}), we introduce the effects equivalent to the independent evolution of the moments related to $\ket{e_x^2 +e_y^2}$ and $\ket{e_x^2 -e_y^2}$ only within the sub-step involving the relaxations under collision and not for performing the mappings between the distribution functions and moments.

\subsection{Preconditioned Lattice Boltzmann Equation}
Next, it is important to note that the use of the rectangular lattice would result in an anisotropic form of the viscous stress tensor dependent on the grid aspect ratio $r$. Such spurious effects along with the truncation errors arising from the non-Galilean invariant aliasing effects on the D2Q9 lattice dependent on the cubic velocity terms and the preconditioning parameter need to be eliminated via certain counteracting corrections, which appear in the evolution of the non-equilibrium part of the second order moments (see Ref~\cite{yahia2021central}). Since by construction, the non-equilibrium second order central moments are identical to those of raw moments, it suffices to perform a consistency analysis and derive the necessary correction terms based on the simpler preconditioned rectangular raw moment MRT formulation of the lattice Boltzmann equation (MRT-LBE) written in a compact matrix-vector form given by
\begin{equation}\label{eq:10}
 \mathbf{f} (\bm{x}+\mathbf{e}\Delta t, t+\Delta t)- \mathbf{f} (\bm{x},t) =  \tensor{T^{-1}}
 \Big[  \tensor{\Lambda} \;  \left(\; \mathbf{n}^{eq}-\mathbf{n} \;\right) + \left(\tensor{I} - \frac{\Lambda}{2}\right)  \mathbf{\Psi}\Delta t \Big],
\end{equation}
where $\tensor{I}$ is an identity matrix of dimension $9\times 9$ and $\tensor{\Lambda}$ is a diagonal matrix holding the relaxation parameters given by
\begin{equation}\label{eq:11}
\tensor{\Lambda} = \mbox{diag} \; \big( 0, 0, 0,\omega_3, \omega_4,\omega_5, \omega_6, \omega_7, \omega_8 \big).
\end{equation}
The solution of this LBE (Eq.~\eqref{eq:10}) yields the distribution functions $f_\alpha=f_\alpha(\bm{x},t+\Delta t)$, whose leading order moments then provide the hydrodynamic fields as
\begin{equation}\label{eq:hydrodynamicfields}
\rho =\sum_{\alpha=0}^{8} f_{\alpha}, \qquad \rho \bm{u} =\sum_{\alpha=0}^{8} f_{\alpha} \bm{e}_{\alpha} + \frac{\bm{F}}{2\gamma}\Delta t, \qquad p= \gamma c_s^2 \rho.
\end{equation}
The key issue here is the specification of the moment equilibria components appearing in $\mathbf{n}^{eq}$ in Eq.~\eqref{eq:10} so that the preconditioned NS equations (Eq.~\eqref{eq:pNSE}) can be recovered consistently on a rectangular lattice grid.

\subsection{Preconditioned Equilibria and Sources: Raw Moments and Central Moments} \par
In this regard, our starting point is the matching of the components of the discrete raw moment equilibria supported by the D2Q9 lattice with those following from the continuous Maxwell distribution, where the speed of sound is based on Eq.~\eqref{eq:2}. Then, we account for the modifications needed for recovering the preconditioned NS equations, which was obtained in an earlier analysis performed in Ref.~\cite{Hajabdollahi201897} in the case of the \emph{square} lattice. Using these as our initial formulation, we can then write the leading terms of the raw moment equilibria as
\begin{align}\label{eq:12}
&k_{00}^{eq\prime}=\rho, \quad\quad k_{10}^{eq\prime}=\rho u_x,\quad\quad k_{01}^{eq\prime}=\rho u_y,\nonumber\\
&k_{20}^{eq\prime}=q^2 c_{s*}^2\rho+\frac{\rho u_x^2}{\gamma} ,\quad\quad\qquad k_{02}^{eq\prime}=q^2c_{s*}^2\rho+\frac{\rho u_y^2}{\gamma}, \quad\quad\quad k_{11}^{eq\prime}=\frac{\rho u_x u_y}{\gamma},\nonumber \\
&k_{21}^{eq\prime}=\rho \left(q^2 c_{s*}^2+ \frac{u_x^2}{\gamma^2} \right)u_y,\quad\quad k_{12}^{eq\prime}=\rho \left(q^2 c_{s*}^2+ \frac{u_y^2}{\gamma^2}\right)u_x,\quad k_{22}^{eq\prime}=\rho q^4 c_{s*}^4 + \rho q^2 c_{s*}^2\left( u_x^2+ u_y^2 \right)+ \rho u_x^2 u_y^2.
\end{align}
The expressions in Eq.~\eqref{eq:12} need to be corrected further to consistently recover the preconditioned NS equations on \emph{rectangular} lattice grids, which will be accomplished later in this section. Moreover, the raw moment equilibria of the source terms also need to be scaled appropriately by the preconditioning parameter $\gamma$ as follows~\cite{premnath2009steady,Hajabdollahi201897}:
\begin{align}\label{eq:13}
&\sigma_{00}^{\prime}=0,\quad \sigma_{10}^{\prime}=\frac{F_x}{\gamma},\quad \sigma_{01}^{\prime}=\frac{F_y}{\gamma},\qquad \sigma_{20}^{\prime}= \frac{2F_x u_x}{\gamma^2}, \quad \sigma_{02}^{\prime}=\frac{2F_y u_y}{\gamma^2},\quad \sigma_{11}^{\prime}=\frac{ \left(F_xu_y+F_yu_x \right)}{\gamma^2},
\end{align}
and $\sg_{mn}^\prime=0,$ \;if $(m+n)\ge 3$.

Then, the countable set of preconditioned discrete central moment equilibria on the D2Q9 lattice can be obtained from the corresponding the raw moment equilibria given in Eq.~\eqref{eq:12} via the binomial transformations as
\begin{align}\label{eq:14}
&k_{00}^{eq}=\rho, \qquad k_{10}^{eq}=0,\qquad k_{01}^{eq}=0,\nonumber\\
&k_{20}^{eq}=q^2 c_{s*}^2\rho+\left(\frac{1}{\gamma}-1\right)\rho u_x^2 ,\qquad k_{02}^{eq}=q^2 c_{s*}^2\rho+\left(\frac{1}{\gamma}-1\right)\rho u_y^2,\qquad k_{11}^{eq}=\left(\frac{1}{\gamma}-1\right)\rho u_x u_y,,\nonumber\\
&k_{21}^{eq}=\left( \frac{1}{\gamma^2}-\frac{3}{\gamma}+2 \right)\rho u_x^2 u_y, \qquad
k_{12}^{eq}=\left( \frac{1}{\gamma^2}-\frac{3}{\gamma}+2 \right)\rho u_x u_y^2 ,\qquad k_{22}^{eq}=q^4 c_{s*}^4\rho,
\end{align}
Note that since the fourth order component of the equilibrium central moment $k_{22}^{eq}$ does not appear at the leading order in Chapman-Enskog analysis of the preconditioned NS equations, for simplicity, we set it as $k_{22}^{eq}=q^4 c_{s*}^4\rho$ following our previous work~\cite{yahia2021central}. Moreover, similarly the central moment components of the source terms follow from the corresponding raw moments Eq.~\eqref{eq:13} using binomial expansions as
\begin{align}\label{eq:13-central}
&\sigma_{00}=0,\quad \sigma_{10}=\frac{F_x}{\gamma},\quad \sigma_{01}=\frac{ F_y}{\gamma},\nonumber\\
&\sigma_{20} = \left(\frac{1}{\gamma^2}-\frac{1}{\gamma}\right)2F_x u_x, \quad \sigma_{02}=\left(\frac{1}{\gamma^2}-\frac{1}{\gamma}\right)2F_y u_y,\quad \sigma_{11}=\left(\frac{1}{\gamma^2}-\frac{1}{\gamma}\right)\left(F_xu_y+F_yu_x \right),
\end{align}
and $\sigma_{mn}=0,$ \;if $(m+n)\ge 3$.
An alternative approach to implementing body forces in central moment LB schemes has been proposed by Fei and Luo~\cite{fei2017consistent,fei2018three}. It has been used for various applications, including thermal flows and multiphase flows, and involves including the effect of the body force on the higher order central moments. By contrast, the forcing scheme given above and others such as in~\cite{geier2015cumulant,hajabdollahi2018symmetrized} involve the effect of body forces up to the second order moments, and are constructed to recover the hydrodynamics (Navier-Stokes equations) as prescribed by the Chapman-Enskog analysis.

\subsection{Chapman-Enskog Analysis: Identification of Truncation Errors due to Grid Anisotropy, Preconditioning and Non-Galilean
Invariance from Aliasing Effects on the D2Q9 Rectangular Lattice}\par
Next, we will perform a Chapman-Enskog (C-E) analysis~\cite{chapman1990mathematical} in order to determine the truncation errors arising from grid anisotropy with the use of the rectangular lattice and the non-Galilean invariant (GI) cubic velocity terms due to the aliasing effects manifesting as a result of the discreteness of the D2Q9 lattice. This would be carried out following the approach taken in our previous works~\cite{premnath2009incorporating,Hajabdollahi201897,yahia2021central}. First, expanding the moments about their equilibria and the time derivative by means of a multiple time expansion, we write
\begin{equation}\label{eq:C-Eexpansion}
\mathbf{n}= \sum_{j=0}^{\infty} \epsilon^{j} \mathbf{n}^{(j)}, \quad \partial_t= \sum_{j=0}^{\infty} {\epsilon}^{j} {\partial_{t_j}},
\end{equation}
where $\epsilon= \Delta t$ represents the perturbation parameter serving in what follows to delineating the terms of different orders. Substituting the above equation in Eq.~\eqref{eq:10} and rewriting its left side via a Taylor series expansion and converting the resulting expression in terms of moments using $\mathbf{f}=  \tensor{T}^{-1} \mathbf{n}$, we obtain the evolution equations of the moments of different orders of $\epsilon$, i.e., $O(\epsilon^k)$, where $k=0,1$, and $2$ as
\begin{subequations}\label{eq:16}
\begin{eqnarray}
\centering
&O (\epsilon^0 ):  \mathbf{n}^{(0)} =  \mathbf{n}^{eq},  \label{eq:16a}\\
&O (\epsilon^1 ): \left(\partial_{t_0} + \bm{E}_i \partial_i\right)  \mathbf{n}^{(0)} =  - \tensor{\Lambda} \; \mathbf{n}^{(1)}+\mathbf{\Psi},  \label{eq:16b} \\
&O (\epsilon^2 ): \partial_{t_1} \; \mathbf{n}^{(0)} +  \left(\partial_{t_0} + \bm{E}_i \partial_i\right) \;\left[ \tensor{I} - \frac{\tensor{\Lambda}} {2} \right] \mathbf{n}^{(1)} =  - \tensor{\Lambda} \; \mathbf{n}^{(2)}, \label{eq:16c}
\end{eqnarray}
\end{subequations}
where $\tensor{E}_i= \tensor{T} \;( \mathbf{e}_i \; \tensor{I})\tensor{T}^{-1}$ and $ \mathbf{e}_i=\ket{e_{i}}$, $ i \in (x,y)$. Then, substituting the raw moments and the source terms shown in Eqs.~\eqref{eq:12} and \eqref{eq:13}, respectively, into Eq.~\eqref{eq:16b}, the relevant moment system $O(\epsilon)$ which are relevant in recovering the preconditioned hydrodynamics can be written as
\begin{subequations}
\begin{eqnarray}\label{eq:17a}
 &\partial_{t_0}\rho + \partial_x \rho u_x + \partial_y \rho u_y = 0,
\end{eqnarray}
\begin{eqnarray}\label{eq:17b}
 &\partial_{t_0}\rho u_x + \partial_x (\rho q^2 c_{s*}^2  +\rho u_x^2/\gamma) + \partial_y (\rho u_x u_y/\gamma) = F_x/\gamma,
\end{eqnarray}
\begin{eqnarray}\label{eq:17c}
&\partial_{t_0}\rho u_y + \partial_x (\rho u_x u_y/\gamma)+ \partial_y (\rho q^2 c_{s*}^2 + \rho u_y^2/\gamma) = F_y/\gamma,
\end{eqnarray}
\begin{eqnarray}\label{eq:17d}
& \partial_{t_0}\left[2 \rho q^2 c_{s*}^2 + \rho( u_x^2+ u_y^2)/\gamma\right]+  \partial_x \left[(1 + q^2 c_{s*}^2)\rho u_x + \rho u_x u_y^2/\gamma^2 \right] + \partial_y \left[(r^2 + q^2 c_{s*}^2)\rho u_y + \rho u_x^2 u_y/\gamma^2 \right] = \nonumber \\
& - \omega_3\;  n_3^{(1)}+  2 \left( F_x u_x + F_y u_y \right)/\gamma^2,
\end{eqnarray}
\begin{eqnarray}\label{eq:17e}
& \partial_{t_0}\left[\rho (u_x^2 - u_y^2)/\gamma\right]+  \partial_x \left[(1 - q^2 c_{s*}^2)\rho u_x - \rho u_x u_y^2/\gamma^2\right] + \partial_y \left[(-r^2 + q^2 c_{s*}^2)\rho u_y +\rho u_x^2 u_y/\gamma^2\right] = \nonumber \\
&- \omega_4\;  n_4^{(1)}+ 2 \left(F_x u_x - F_y u_y\right)/\gamma^2,
\end{eqnarray}
\begin{eqnarray}\label{eq:17f}
& \partial_{t_0}(\rho u_x u_y/\gamma)+  \partial_x \left[ q^2 c_{s*}^2 \rho u_y + \rho u_x^2 u_y/\gamma^2 \right] + \partial_y \left[q^2 c_{s*}^2 \rho u_x +\rho u_x u_y^2/\gamma^2 \right] = \nonumber \\
&-\omega_5\;  n_5^{(1)}+ \left(F_x u_y+ F_y u_x \right)/\gamma^2.
\end{eqnarray}
\end{subequations}
Similarly, the $O(\epsilon^2)$ evolution equations for the conserved moments at the slower time scale $t_1$ reads from  Eq.~\eqref{eq:16c} as
\begin{subequations}
\begin{eqnarray}
&\partial_{t_1}\rho=0,
\label{eq:18a}\\&
\partial_{t_1}\left(\rho u_x\right)+\partial_x \left[\dfrac{1}{2}\left(1-\dfrac{\omega_3}{2}\right) n_3^{(1)}+\dfrac{1}{2}\left(1-\dfrac{\omega_4}{2}\right)n_4^{(1)}\right]
+\partial_y \left[\left(1-\dfrac{\omega_5}{2}\right)n_5^{(1)}\right]=0, \label{eq:18b}\\
&\partial_{t_1}\left(\rho u_y\right)+\partial_x \left[\left(1-\dfrac{\omega_5}{2}\right) n_5^{(1)}\right]+\partial_y \left[\dfrac{1}{2}\left(1-\dfrac{\omega_3}{2}\right)n_3^{(1)}-\dfrac{1}{2}\left(1-\dfrac{\omega_4}{2}\right)n_4^{(1)}\right]=0. \label{eq:18c}
\end{eqnarray}
\end{subequations}
The above Eqs.~\eqref{eq:18a}-\eqref{eq:18c} depend on the components of the non-equilibrium moments $n_3^{(1)}$, $n_4^{(1)}$ and $n_5^{(1)}$, which can be obtained from Eqs.~(\ref{eq:17d})-(\ref{eq:17f}). Hence, rewriting Eqs.~(\ref{eq:17d})-(\ref{eq:17f}) to express the non-equilibrium moments as
\begin{subequations}
\begin{align}
&n_3^{(1)}= \frac{1}{\omega_3} \Big\{-\partial_{t_0}\left[ 2 q^2 c_{s*}^2 \rho + \rho u_x^2/\gamma + \rho u_y^2/\gamma \right] - \partial_x \left[(1+q^2 c_{s*}^2)\rho u_x\right]- \partial_x (\rho u_x u_y^2/\gamma^2)- \partial_y \left[(r^2 +q^2 c_{s*}^2)\rho u_y\right]\nonumber\\
& \; \; \; \; \; \; \; \; \; \; \; \; \; \; \; \; \; \quad - \partial_y (\rho u_x^2 u_y/\gamma^2)+ 2(F_x u_x + F_y u_y)/\gamma^2 \Big\}, \label{eq:20a}\\
&n_4^{(1)}= \frac{1}{\omega_4} \Big\{-\partial_{t_0}[(\rho u_x^2-\rho u_y^2)/\gamma] - \partial_x \left[(1- q^2 c_{s*}^2)\rho u_x\right]+ \partial_x(\rho u_x u_y^2/\gamma^2) + \partial_y \left[(r^2-q^2 c_{s*}^2)\rho u_y\right]\nonumber\\
&\; \; \; \; \; \; \; \; \; \; \; \; \; \; \; \; \; \quad - \partial_y(\rho u_x^2 u_y/\gamma^2) + 2(F_x u_x - F_y u_y)/\gamma^2 \Big\}, \label{eq:20b}\\
&n_5^{(1)}= \frac{1}{\omega_5} \Big\{-\partial_{t_0}(\rho u_x u_y/\gamma) - \partial_x (q^2 c_{s*}^2 \rho u_y)- \partial_x (\rho u_x^2 u_y/\gamma^2)- \partial_y (q^2 c_{s*}^2\rho u_x)- \partial_y(\rho u_x u_y^2/\gamma^2) + (F_x u_y + F_x u_y)/\gamma^2 \Big\}.\label{eq:20c}
\end{align}
\end{subequations}
Clearly, the second order non-equilibrium moments $n_3^{(1)}$, $n_4^{(1)}$ and $n_5^{(1)}$ involve terms related to the non-GI cubic velocity errors, whose prefactors are dependent on the preconditioning parameter $\gamma$ and the grid-anisotropy error terms dependent on the grid aspect ratio $r$, in addition to those are related to the physically consistent terms that contribute towards the viscous stress tensor. Denoting the truncation errors related to the grid anisotropy as $E_{js}$ and the non-GI cubic velocity terms as $E_{jg}$ for $j=3,4$ and $5$, and after replacing the time derivatives appearing in Eqs.~\eqref{eq:20a}-\eqref{eq:20c} in terms of the spatial derivatives of the conserved moments via Eqs.~\eqref{eq:17a}-\eqref{eq:17c}, we can then simplify the resulting equations by retaining terms up to $O(u_i^3)$ (see Refs.~\cite{Hajabdollahi201897} and~
\cite{yahia2021central} for details). Then, the final expressions for the second order non-equilibrium moment components on the rectangular lattice resulting for our LB formulation can be written as follows:
\begin{subequations}\label{eq:2ndordernoneqmmomentswitherrorterms}
\begin{eqnarray}
n_3^{(1)} &=& -\frac{2 q^2 c_{s*}^2}{\omega_3}  \rho \bm{\nabla}\cdot \bm{u}+ E_{3g}+ E_{3s} ,\\
n_4^{(1)} &=&-\frac{2 q^2 c_{s*}^2 }{\omega_4} \rho \left(\partial_x u_x - \partial_y u_y \right)+ E_{4g}+ E_{4s},\\
n_5^{(1)} &=& -\frac{q^2 c_{s*}^2\rho}{\omega_5} \left(\partial_x u_y + \partial_y u_x \right)+ E_{5g},
\end{eqnarray}
\end{subequations}
where the expressions for the truncation errors due to grid anisotropy $E_{3s}$ and $E_{4s}$, and the non-GI cubic velocity aliasing errors $E_{3g}$, $ E_{4g}$,a nd $ E_{5g}$ read as
\begin{subequations}\label{eq:truncationerrors-gridanisotropy}
\begin{eqnarray}
E_{3s}&=& \cfrac{1}{\omega_3} (3 q^2 c_{s*}^2 -1)\rho \partial_x u_x+ \cfrac{1}{\omega_3} (3 q^2 c_{s*}^2 -r^2)\rho \partial_y u_y, \label{eq:truncationerrors3s}\\
E_{4s}&=& \dfrac{1}{\omega_4} (3 q^2 c_{s*}^2 -1)\rho \partial_x u_x- \dfrac{1}{\omega_4} (3 q^2 c_{s*}^2 -r^2)\rho \partial_y u_y,\label{eq:truncationerrors4s}
\end{eqnarray}
\end{subequations}
and
\begin{subequations}\label{eq:truncationerrors-nonGI}
\begin{eqnarray}
E_{3g} &=& \cfrac{1}{\omega_3} \left[(2/\gamma+1)q^2 c_{s*}^2 -1 \right] u_x \partial_x \rho+ \cfrac{1}{\omega_3} \left[(2/\gamma+1) q^2 c_{s*}^2 -r^2 \right] u_y \partial_y \rho + \cfrac{M_3}{\omega_3} \; \partial_x u_x+ \cfrac{N_3}{\omega_3}\; \partial_y u_y,\label{eq:truncationerrors3g}\\
E_{4g} &=& \dfrac{1}{\omega_4} \left[ (2/\gamma+1)q^2 c_{s*}^2 -1 \right] u_x \partial_x \rho- \dfrac{1}{\omega_4} \left[(2/\gamma+1)q^2 c_{s*}^2 -r^2 \right] u_y \partial_y \rho + \dfrac{M_4}{\omega_4}\partial_x u_x+ \dfrac{N_4}{\omega_4}\partial_y u_y,\label{eq:truncationerrors4g}\\
E_{5g} &=& \frac{1}{\omega_5}(1/\gamma-1) q^2 c_{s*}^2 \left(u_x \partial_y \rho + u_y \partial_x \rho\right) + \frac{1}{\omega_5} (1/\gamma^2-1/\gamma) \rho u_x u_y \left( \partial_x u_x+ \partial_y u_y\right),\label{eq:truncationerrors5g}
\end{eqnarray}
\end{subequations}
where the prefactors $M_3$, $N_3$, $M_4$ and $N_4$ appearing in Eqs.~\eqref{eq:truncationerrors3g} and \eqref{eq:truncationerrors4g} can be expressed as
\begin{subequations}\label{eq:prefactors}
\begin{eqnarray}
M_3  &=& \rho \left[(4/\gamma^2-1/\gamma) u_x^2 + (1/\gamma^2-1/\gamma) u_y^2\right],\label{eq:prefactors1}\\
N_3  &=& \rho \left[(4/\gamma^2-1/\gamma) u_y^2 + (1/\gamma^2-1/\gamma) u_x^2\right],\label{eq:prefactors2}\\
M_4  &=& \rho \left[(4/\gamma^2-1/\gamma) u_x^2 - (1/\gamma^2-1/\gamma) u_y^2\right],\label{eq:prefactors3}\\
N_4  &=& \rho \left[-(4/\gamma^2-1/\gamma)u_y^2 + (1/\gamma^2-1/\gamma) u_x^2\right].\label{eq:prefactors4}
\end{eqnarray}
\end{subequations}
Recognizing $q=\mbox{min}\{r,1\}$, it is evident that the various truncation errors given above are dependent on the preconditioning parameter $\gamma$ and the grid aspect ratio $r$, which need to be eliminated.

\subsection{Corrections via Extended Moment Equilibria for Elimination of Truncation Errors due to Grid Anisotropy, Preconditioning,
and Aliasing Effects}\label{subsec:correctionsextendemomentequilibria}
In this regard, we propose an extended moment equilibria $\mathbf{n}^\mathit{eq,eff}$
\begin{eqnarray}\label{eq:30}
\mathbf{n}^\mathit{eq,eff}= \mathbf{n}^{eq}+ \Delta t \mathbf{n}^{eq(1)},
\end{eqnarray}
where $\mathbf{n}^{eq(1)}$ are the corrections made to the base moment equilibria $\mathbf{n}^{eq}$ introduced in Eqs.~\eqref{eq:basemomenteqvector} and~\eqref{eq:12}. As shown in Eqs.~\eqref{eq:2ndordernoneqmmomentswitherrorterms}, \eqref{eq:truncationerrors-gridanisotropy}, and \eqref{eq:truncationerrors-nonGI}, the truncation
errors exist in the evolution of the second order moments $n_3$, $n_4$ and $n_5$, and involve spatial derivatives of the velocity field and the density field. Recognizing this fact, for the purpose of consistently recovering the preconditioned NS equations in a rectangular lattice grid, we thus write the following expressions for the corrections to the moment equilibria, where it suffices to introduce them to only the second order components:
\begin{eqnarray}\label{eq:31}
n^{eq(1)}_j =
\begin{cases}
\theta_{3x} \partial_x u_x  +  \theta_{3y} \partial_y u_y + \lambda_{3x} \partial_x \rho + \lambda_{3y} \partial_y \rho  & \quad j =3\\
\theta_{4x} \partial_x u_x  -  \theta_{4y} \partial_y u_y + \lambda_{4y} \partial_x \rho + \lambda_{4y} \partial_y \rho  & \quad j =4\\
\theta_{5x} \partial_x u_x  +  \theta_{5y} \partial_y u_y + \lambda_{5y} \partial_x \rho + \lambda_{5y} \partial_y \rho  & \quad j =5\\
0 & \quad \mbox{otherwise},\\
\end{cases}
\end{eqnarray}
Here, $\theta_{jx}$, $\theta_{jy}$, $\lambda_{jx}$, $\lambda_{jy}$, where $j=3,4,$ and $5$, are the unknown coefficients, which will be determined by carrying out a modified C-E expansion that includes the extended moment equilibria given in Eq.~\eqref{eq:30}. Thus, replacing the expansions appearing in Eq.~\eqref{eq:C-Eexpansion} with
\begin{eqnarray} \label{eq:C-Eexpansion1}
\mathbf{n} &=& \mathbf{n}^\mathit{eq,eff}+\epsilon \mathbf{n}^{(1)} + \epsilon^2 \mathbf{n}^{(2)}+ \ldots = \mathbf{n}^{(0)}+\underline{\epsilon \mathbf{n}^{eq(1)}} +\epsilon \mathbf{n}^{(1)}+ \epsilon^2 \mathbf{n}^{(2)}+ \ldots \nonumber\\
\partial_t&=&\partial_{t_0} +\epsilon \partial_{t_1} + \epsilon^2 \partial_{t_2}+ \ldots,
\end{eqnarray}
and performing the same steps that follow Eq.~\eqref{eq:C-Eexpansion} with using Eq.~\eqref{eq:C-Eexpansion1}, then the evolution of the moment systems at various orders of $\epsilon$ in Eq.~\eqref{eq:16} modify to the following by accounting for the presence of the corrections $\mathbf{n}^{eq(1)}$:
\begin{subequations}
\begin{eqnarray}\label{eq:16-extendedequilibria}
 \centering
&O (\epsilon^0 ):  \mathbf{n}^{(0)} =  \mathbf{n}^\mathit{eq},  \label{eq:16-extendedequilibria-a}\\
&O (\epsilon^1 ): \left({\partial_{t_0}} + \tensor{E}_i \partial_i\right)  \mathbf{n}^{(0)} =  - \tensor{\Lambda} \left[ \mathbf{n}^{(1)}- \underline{\mathbf{n}^{eq(1)}}\right]  +\mathbf{\Psi},  \label{eq:16-extendedequilibria-b} \\
&O (\epsilon^2 ): {\partial_{t_1}} \mathbf{n}^{(0)} +  \left({\partial_{t_0}} + \tensor{E}_i \partial_i\right) \left[\left( \tensor{I} - \frac{\tensor{\Lambda}} {2} \right)\mathbf{n}^{(1)}\right] +  \left({\partial_{t_0}} + \tensor{E}_i \partial_i\right) \underline{\left[ \frac{\tensor{\Lambda}} {2} \mathbf{n}^{eq(1)}\right]}=  - \tensor{\Lambda}\mathbf{n}^{(2)}. \label{eq:16-extendedequilibria-c}
\end{eqnarray}
\end{subequations}
In view of the derivation given in the previous section and the changes appearing in Eq.~\eqref{eq:16-extendedequilibria-b} relative to Eq.~\eqref{eq:16b}, the second order non-equilibrium moments for the rectangular lattice with preconditioning in Eq.~\eqref{eq:2ndordernoneqmmomentswitherrorterms} modify to
\begin{subequations}\label{eq:2ndordernoneqmmomentswitherrorterms-extended}
\begin{eqnarray}
n_3^{(1)} &=& -\frac{2 q^2 c_{s*}^2}{\omega_3}  \rho \bm{\nabla}\cdot \bm{u}+ E_{3g}+ E_{3s}+\underline{n_3^{eq(1)}}, \label{eq:2ndordernoneqmmomentswitherrorterms-extended1}\\
n_4^{(1)} &=&-\frac{2 q^2 c_{s*}^2 }{\omega_4} \rho \left(\partial_x u_x - \partial_y u_y \right)+ E_{4g}+ E_{4s}+\underline{n_4^{eq(1)}}, \label{eq:2ndordernoneqmmomentswitherrorterms-extended2}\\
n_5^{(1)} &=& -\frac{q^2 c_{s*}^2\rho}{\omega_5} \left(\partial_x u_y + \partial_y u_x \right)+ E_{5g}+\underline{n_5^{eq(1)}}, \label{eq:2ndordernoneqmmomentswitherrorterms-extended3}
\end{eqnarray}
\end{subequations}
where the error terms $E_{3g}$, $E_{3s}$, $E_{4g}$, $E_{4s}$, and $E_{5g}$ are given in the previous section in Eqs.~\eqref{eq:truncationerrors-gridanisotropy} and~\eqref{eq:truncationerrors-nonGI}.

Now, in order to derive explicit formulas for the corrections $n_3^{eq(1)}$, $n_4^{eq(1)}$ and $n_5^{eq(1)}$, we need certain constraint relationships between them and the error terms. These follow from combining Eq.~\eqref{eq:16-extendedequilibria-b} and $\epsilon$ times Eq.~\eqref{eq:16-extendedequilibria-c} and using the expressions for the non-equilibrium moments in Eq.~\eqref{eq:2ndordernoneqmmomentswitherrorterms-extended} along with $\partial_t=\partial_{t_0} + \epsilon \partial_{t_1}$ to obtain the effective evolution equations for the conserved moments, which would include both the truncation error terms identified earlier and the unknown corrections whose combined effects are set to zero so that the evolution equations correspond to the preconditioned NS equations. See e.g., Refs.~\cite{Hajabdollahi201897,yahia2021central,yahia2021three} for details of these steps. Writing the truncation error terms compactly in the form a vector $\mathbf{\Xi}$
\begin{equation}\label{eq:errorvector}
\mathbf{\Xi}=\left(\varphi_{0},\varphi_{1},\varphi_{2},\dots,\varphi_{8}\right)^{\dag},
\end{equation}
where
\begin{eqnarray}\label{eq:errorvector1}
\varphi_j =
\begin{cases}
 E_{3s}+ E_{3g}& \quad j =3\\
 E_{4s}+ E_{4g}& \quad j =4\\
 E_{5g}& \quad j =5\\
 0 & \quad \mbox{otherwise},
\end{cases}
\end{eqnarray}
then the necessary constraint equation between the vector of moment corrections $\mathbf{n}^{eq(1)}$ identified whose functional forms with unknown coefficients are given in Eq.~\eqref{eq:31} and the above vector holding the truncation errors $\mathbf{\Xi}$ (see Eqs.~\eqref{eq:errorvector} and~\eqref{eq:errorvector1}) reads as
\begin{equation}\label{eq:constraint-errors-corrections}
\mathbf{n}^{eq(1)}+ \left( \tensor{I} - \frac{\tensor{\Lambda}} {2} \right)\mathbf{\Xi}=0,
\end{equation}
which, in component form, becomes
\begin{equation}\label{eq:constraint-errors-corrections-component-form}
n_j^{eq(1)}+ \left( 1 - \frac{\omega_j} {2} \right)(E_{js}+ E_{jg})=0, \quad j=3,4,5.
\end{equation}
Evaluating Eq.~\eqref{eq:constraint-errors-corrections-component-form} and using Eqs.~\eqref{eq:31},~\eqref{eq:errorvector} and~\eqref{eq:errorvector1} for $j=3, 4$ and $5$, respectively, we get
\begin{eqnarray*}
&&\theta_{3x} \partial_x u_x  +  \theta_{3y} \partial_y u_y + \lambda_{3x} \partial_x \rho + \lambda_{3y} \partial_y \rho  = -\left( 1 - \dfrac{\omega_3} {2} \right)E_{3g} -\left( 1 - \dfrac{\omega_3} {2} \right) E_{3s} \\
&&= -\left(\dfrac{1}{\omega_3}- \dfrac{1}{2}\right) \left[M_3 +(q^2 c_{s*}^2 -1)\rho \right] \partial_x u_x -\left(\dfrac{1}{\omega_3}- \dfrac{1}{2}\right) \left[N_3 +(q^2 c_{s*}^2 -r^2)\rho \right]\partial_y u_y \\
&&- \left(\dfrac{1}{\omega_3}- \dfrac{1}{2}\right) \left[\left(\dfrac{2}{\gamma}+1 \right) q^2 c_{s*}^2 -1 \right] u_x \partial_x \rho- \left(\dfrac{1}{\omega_3}- \dfrac{1}{2}\right) \left[\left(\dfrac{2}{\gamma}+1 \right) q^2 c_{s*}^2 -r^2 \right]u_y \partial_y \rho.
\end{eqnarray*}
\begin{eqnarray*}
&&\theta_{4x} \partial_x u_x -  \theta_{4y} \partial_y u_y + \lambda_{4x} \partial_x \rho + \lambda_{4y} \partial_y \rho  = -\left( 1 - \dfrac{\omega_4} {2} \right)E_{4g} -\left( 1 - \dfrac{\omega_4} {2} \right) E_{4s} \\
&&= -\left(\dfrac{1}{\omega_4}- \dfrac{1}{2}\right) \left[M_4 +(q^2 c_{s*}^2 -1)\rho \right] \partial_x u_x -\left(\dfrac{1}{\omega_4}- \dfrac{1}{2}\right) \left[M_4 +(q^2 c_{s*}^2 -r^2)\rho \right]\partial_y u_y \\
&&- \left(\dfrac{1}{\omega_4}- \dfrac{1}{2}\right) \left[\left(\dfrac{2}{\gamma}+1 \right) q^2 c_{s*}^2 -1 \right] u_x \partial_x \rho+ \left(\dfrac{1}{\omega_4}- \dfrac{1}{2}\right) \left[\left(\dfrac{2}{\gamma}+1 \right) q^2 c_{s*}^2 -r^2 \right]u_y \partial_y \rho,
\end{eqnarray*}
\begin{eqnarray*}
&&\theta_{5x} \partial_x u_x  +  \theta_{5y} \partial_y u_y + \lambda_{5x} \partial_x \rho + \lambda_{5y} \partial_y \rho  = -\left( 1 - \dfrac{\omega_5} {2} \right)E_{5g} \\
&&= -\left(\dfrac{1}{\omega_5}- \dfrac{1}{2}\right) \left(\dfrac{1}{\gamma^2}-\dfrac{1}{\gamma} \right) \rho u_x u_y (\partial_x u_x +\partial_y u_y) -\left(\dfrac{1}{\omega_5}- \dfrac{1}{2}\right) \left(\dfrac{1}{\gamma}-1 \right) q^2 c_{s*}^2 \left(u_x \partial_x \rho +u_y \partial_y \rho\right),
\end{eqnarray*}
Comparing the terms involving the spatial gradients of the same type of quantity in each side of the above three equations, we finally get the coefficients for the correction terms in the second order moment equilibria as
\begin{subequations}\label{eq:38}
\begin{eqnarray}
\theta_{3x}&=& -\Big[ M_3 + \left(3 q^2 c_{s*}^2- 1\right)\rho \Big] \left( \frac{1}{\omega_3}- \frac{1}{2}\right),\\
\theta_{3y}&=& -\Big[ N_3 + \left(3 q^2 c_{s*}^2- r^2\right)\rho \Big] \left(\frac{1}{\omega_3}- \frac{1}{2}\right) ,\\
\lambda_{3x}&=& - \left[\left(\frac{2}{\gamma}+1 \right) q^2 c_{s*}^2 -1 \right] \left(\frac{1}{\omega_3}- \frac{1}{2}\right) u_x,\\
\lambda_{3y}&=& - \left[\left(\frac{2}{\gamma}+1 \right) q^2 c_{s*}^2 -r^2 \right]\left(\frac{1}{\omega_3}- \frac{1}{2}\right) u_y,
\end{eqnarray}
\end{subequations}
\begin{subequations}\label{eq:39}
\begin{eqnarray}
\theta_{4x}&=& -\Big[ M_4 + \left(3 q^2 c_{s*}^2- 1\right)\rho \Big] \left( \frac{1}{\omega_4}- \frac{1}{2}\right),\\
\theta_{4y}&=& +\Big[ N_4 - \left(3 q^2 c_{s*}^2- r^2\right)\rho \Big] \left(\frac{1}{\omega_4}- \frac{1}{2}\right) ,\\
\lambda_{4x}&=& - \left[\left(\frac{2}{\gamma}+1 \right) q^2 c_{s*}^2 -1 \right] \left(\frac{1}{\omega_4}- \frac{1}{2}\right) u_x,\\
\lambda_{4y}&=& + \left[\left(\frac{2}{\gamma}+1 \right) q^2 c_{s*}^2 -r^2 \right]\left(\frac{1}{\omega_4}- \frac{1}{2}\right) u_y,
\end{eqnarray}
\end{subequations}
\begin{subequations}\label{eq:40}
\begin{eqnarray}
\theta_{5x}&=& -\left(\frac{1}{\gamma^2}-\frac{1}{\gamma} \right) \left( \frac{1}{\omega_5}- \frac{1}{2}\right)\rho u_x u_y ,\\
\theta_{5y}&=& -\left(\frac{1}{\gamma^2}-\frac{1}{\gamma} \right)  \left( \frac{1}{\omega_5}- \frac{1}{2}\right)\rho u_x u_y,\\
\lambda_{5x}&=& - \left(\frac{1}{\gamma}-1 \right) \left(\frac{1}{\omega_5}- \frac{1}{2}\right) q^2 c_{s*}^2 u_y,\\
\lambda_{5y}&=& - \left(\frac{1}{\gamma}-1 \right) \left(\frac{1}{\omega_5}- \frac{1}{2}\right) q^2 c_{s*}^2 u_x.
\end{eqnarray}
\end{subequations}
These expressions (Eqs.~\eqref{eq:38}-\eqref{eq:40}) together with Eqs.~\eqref{eq:30} and~\eqref{eq:31} are among the main results of this work that contribute towards formulating a new preconditioned LB approach on a rectangular lattice grid. The above choices for the moment equilibria corrections, which depend on both the grid aspect ratio $r$ and the preconditioning parameter $\gamma$, ensures that the resulting algorithm using a rectangular lattice represents the preconditioned NS equations with the shear viscosity $\nu$ and bulk viscosity $\xi$ satisfying the following relationships among the various model parameters:
\begin{equation}\label{eq:transport-coefficients}
  \nu =\gamma q^2 c_{s*}^2 \left( \frac{1}{\omega_j}- \frac{1}{2} \right)\Delta t,\quad j=4,5, \;\; \quad \xi = \gamma q^2 c_{s*}^2 \left( \frac{1}{\omega_3}- \frac{1}{2} \right)\Delta t,
\end{equation}
where the optimal value of $c_{s*}^2$ is $1/3$, and the emergent pressure field $p$ is given by $p=\gamma q^2 c_{s*}^2 \rho$. We emphasize here that the simple expressions given in Eq.~\eqref{eq:transport-coefficients} self-consistently parameterize the transport coefficients in terms of $q$ which is given in Eq.~\eqref{eq:2} and maintains desired numerical stability in LB simulations using rectangular lattice grids. Unlike in previous works (see e.g.,~\cite{peng2016lattice,peng2016hydrodynamically,zong2016designing}), there is no need to rely on trial and error to adjust the speed of sound when a rectangular lattice is used and the grid aspect ratio is varied to any desired value.

\subsection{Strain rate tensor components based on non-equilibrium moments}
We will now show the diagonal components of the strain rate tensor  $\partial_x u_x$ and $\partial_y u_y$, which appear in the moment equilibria corrections given in Eqs.~\eqref{eq:30},~\eqref{eq:31} and Eqs.~\eqref{eq:38}-\eqref{eq:40} can be computed locally via second-order non-equilibrium moments. First, using Eqs.~\eqref{eq:2ndordernoneqmmomentswitherrorterms-extended1} and~\eqref{eq:2ndordernoneqmmomentswitherrorterms-extended2}, and simplifying via Eq.~\eqref{eq:constraint-errors-corrections-component-form}, we obtain
\begin{eqnarray*}
n_3^{(1)}&=& -\cfrac{ 2 q^2 c_{s*}^2}{\omega_3}  \rho \left(\partial_x u_x + \partial_y u_y \right)+ \cfrac{\omega_3}{2} E_{3g}+ \cfrac{\omega_3}{2} E_{3s},\\
n_4^{(1)}&=& -\dfrac{ 2 q^2 c_{s*}^2}{\omega_4}  \rho \left(\partial_x u_x - \partial_y u_y \right)+ \dfrac{\omega_4}{2} E_{4g}+ \dfrac{\omega_4}{2} E_{4s}
\end{eqnarray*}
Then, substituting for $E_{3s}$, $E_{4s}$, $E_{3g}$, and $E_{4g}$ using Eqs.~\eqref{eq:truncationerrors-gridanisotropy} and \eqref{eq:truncationerrors-nonGI} in the last two equations and rearranging them leads to
\begin{eqnarray}\label{eq:41}
&\left[-\dfrac{2 q^2 c_{s*}^2}{\omega_3}\rho+ \dfrac{M_3}{2} + \dfrac{1}{2}(3 q^2 c_{s*}^2-1) \rho \right] \partial_x u_x + \left[-\dfrac{2 q^2c_{s*}^2}{\omega_3}\rho + \dfrac{N_3}{2}+ \dfrac{1}{2}(3 q^2 c_{s*}^2-r^2) \rho \right] \partial_y u_y \nonumber \\
&=n_3^{(1)}-\dfrac{1}{2} \left[ \left(\dfrac{2}{\gamma}+1\right) q^2 c_{s*}^2-1 \right]
u_x \partial_x \rho - \dfrac{1}{2} \left[ \left(\dfrac{2}{\gamma}+1\right) q^2 c_{s*}^2-r^2 \right] u_y \partial_y \rho.
\end{eqnarray}
\begin{eqnarray}\label{eq:42}
&\left[-\dfrac{2 q^2 c_{s*}^2}{\omega_4}\rho+ \dfrac{M_4}{2} + \dfrac{1}{2}(3 q^2 c_{s*}^2-1) \rho \right] \partial_x u_x + \left[+\dfrac{2 q^2c_{s*}^2}{\omega_4}\rho + \dfrac{N_4}{2}- \dfrac{1}{2}(3 q^2 c_{s*}^2-r^2) \rho \right] \partial_y u_y \nonumber \\
&=n_4^{(1)}-\dfrac{1}{2} \left[ \left(\dfrac{2}{\gamma}+1\right) q^2 c_{s*}^2-1 \right]
u_x \partial_x \rho + \dfrac{1}{2} \left[ \left(\dfrac{2}{\gamma}+1\right) q^2 c_{s*}^2-r^2 \right] u_y \partial_y \rho.
\end{eqnarray}
Based on Eqs.~\eqref{eq:41} and \eqref{eq:42}, the required local expressions for the diagonal components of the strain rate tensor $\partial_x u_x$ and $\partial_y u_y$ can be obtained. In this regard, we first introduce the following intermediate variables
\begin{subequations}\label{eq:43}
\begin{eqnarray}
 A&=& \cfrac{1}{2} \left[\left(\frac{2}{\gamma}+1\right) q^2 c_{s*}^2 - 1 \right] u_x,\qquad B = \cfrac{1}{2} \left[\left(\frac{2}{\gamma}+1\right) q^2 c_{s*}^2 - r^2 \right] u_y,\\
 e_{3\rho}&=& -A \partial_x \rho - B \partial_y \rho,\qquad\qquad\qquad e_{4\rho}= -A \partial_x \rho + B \partial_y \rho,
 \end{eqnarray}
\end{subequations}
where the density gradients $\partial_x \rho$ and $\partial_y \rho$ may be obtained via an isotropic finite-difference scheme. The non-equilibrium moments $n_3^{(1)}$ and $n_4^{(1)}$ appearing in Eqs.~\eqref{eq:41} and \eqref{eq:42} can be computed using either raw moments or central moments as
\begin{eqnarray*}
n_3^{(1)}&=& \left(k_{20}^\prime + k_{02}^\prime \right)- \left(k_{20}^{eq \prime} + k_{02}^{eq\prime} \right)= \left(k_{20} + k_{02} \right)- \left(k_{20}^{eq} + k_{02}^{eq} \right) \nonumber\\
\quad &=& \left(k_{20} + k_{02} \right)- 2 q^2 c_{s*}^2\rho-  \left(\frac{1}{\gamma}-1\right) (u_x^2 +u_y^2)\rho,\\
n_4^{(1)}&=& \left(k_{20}^\prime - k_{02}^\prime \right)- \left(k_{20}^{eq \prime} - k_{02}^{eq\prime} \right)= \left(k_{20} - k_{02} \right)- \left(k_{20}^{eq} - k_{02}^{eq} \right) \nonumber\\
\quad &=& \left(k_{20} - k_{02} \right)- \left(\frac{1}{\gamma}-1\right) (u_x^2 -u_y^2)\rho.
\end{eqnarray*}
Based on these considerations, we can then identify the right sides and the left sides of Eqs.~\eqref{eq:41} and \eqref{eq:42}, respectively, conveniently by further introducing the following additional intermediate variables
\begin{subequations}\label{eq:45}
\begin{eqnarray}
R_3 &=& n_3^{(1)}+ e_{3\rho} = k_{20} + k_{02}- 2 q^2 c_{s*}^2\rho- \left(\frac{1}{\gamma}-1\right) (u_x^2 +u_y^2)\rho +e_{3\rho},\\
R_4 &=& n_4^{(1)}+ e_{4\rho} = k_{20} - k_{02} - \left(\frac{1}{\gamma}-1\right) (u_x^2 -u_y^2)\rho + e_{4\rho},
\end{eqnarray}
\end{subequations}
and
\begin{subequations}\label{eq:46}
\begin{eqnarray}
&&C_{3x}= \left[-\frac{2 q^2 c_{s*}^2}{\omega_3}\rho + \frac{M_3}{2} + \frac{1}{2}(3 q^2 c_{s*}^2-1)\rho \right],\quad C_{3y} = \left[-\frac{2 q^2 c_{s*}^2}{\omega_3}\rho + \frac{N_3}{2} + \frac{1}{2}(3 q^2 c_{s*}^2-r^2)\rho \right],\\
&&C_{4x}= \left[-\frac{2 q^2 c_{s*}^2}{\omega_4} \rho + \frac{M_4}{2} + \frac{1}{2}(3 q^2 c_{s*}^2-1)\rho \right],\quad C_{4y} = \left[+\frac{2 q^2 c_{s*}^2}{\omega_4} \rho + \frac{N_4}{2} - \frac{1}{2}(3 q^2 c_{s*}^2-r^2)\rho \right],
\end{eqnarray}
\end{subequations}
where $M_3$, $N_3$, $M_4$ and $N_4$ are given in Eq.~\eqref{eq:prefactors}. Then, Eqs.~\eqref{eq:41} and \eqref{eq:42} can be more compactly written as
\begin{subequations}\label{eq:47}
\begin{eqnarray}
C_{3x}\partial_x u_x  + C_{3y}\partial_y u_y &=& R_3,\\
C_{4x}\partial_x u_x  + C_{4y}\partial_y u_y &=& R_4.
\end{eqnarray}
\end{subequations}
Solving the last two equations, we finally get the required local expressions for the diagonal parts of the strain rate tensor as follows:
\begin{equation}\label{eq:48}
\partial_x u_x = \frac{\left[ C_{4y}R_3 - C_{3y}R_4 \right]}{\left[ C_{3x} C_{4y}  - C_{4x} C_{3y} \right]},\qquad\qquad \partial_y u_y = \frac{1}{C_{3y}} \left[ R_3 - C_{3x} \partial_x u_x \right].
\end{equation}
For completeness, we note that a similar relation for the off-diagonal component ($\partial_x u_y+ \partial_y u_x$) follows from combining Eqs.~\eqref{eq:2ndordernoneqmmomentswitherrorterms-extended3} and~\eqref{eq:constraint-errors-corrections-component-form} and then simplifying via Eq.~\eqref{eq:truncationerrors-nonGI}.


\section{Preconditioned Rectangular Central Moment Lattice Boltzmann Method (PRC-LBM)}\label{sec:4}
In this section, we will present a robust and efficient implementation of a LB algorithm on rectangular lattice grids for solving preconditioned NS equations using and extending the results of the C-E analysis performed in the last section. In this regard, we note that the effect of the moment basis $\tensor{T}$ as defined in Eqs.~\eqref{eq:7} and~\eqref{eq:8} will be equivalently utilized in our implementation in a more modular fashion so the LB schemes based on the square lattice can be readily extended for utilizing rectangular lattice grids along with preconditioning and the necessary corrections. This involves using a simpler re-defined moment basis in conjunction with diagonal scaling matrices based on the grid aspect ratio for performing the pre- and post-collision transformations between the raw moments and distribution functions, and segregate the evolution of the trace of the diagonal parts of the second order moments from the others for achieving independent variations of bulk and shear viscosities~\cite{geier2015cumulant}, along with accounting for the corrections to eliminate the grid-anisotropy and non-GI truncation errors, only within the collision step under moment relaxations (see Ref.~\cite{yahia2021three}). In other words, the linear combinations of moments as required are considered only for performing the collision step and not for any mappings. This represents an improvement over the implementation discussed over all the previous 2D rectangular LB schemes for the solution of the NS equations, including our recent work~\cite{yahia2021central}, and is consistent with our more recent 3D formulation~\cite{yahia2021three}), but extended here with a preconditioning strategy for convergence acceleration. Similar approach based on a natural independent moment set without involving the mixed moments has also been used in previous work~\cite{fei2018modeling,fei2018cascaded} by using a block diagonal relaxation matrix~\cite{asinari2008generalized} in the context of LB formulations using a square lattice.

\subsection{Reformulation of the Preconditioned Rectangular Raw Moment LBE}
Thus, we first introduce a moment basis $\tensor{Q}$, which unlike $\tensor{T}$ in Eqs.~\eqref{eq:7} and~\eqref{eq:8}, does not contain any combinations of the basis vectors, but only a set of bare basis vectors for the D2Q9 lattice:
\begin{eqnarray}\label{eq:Q-momentbasis}
\tensor{Q}=\Big[ \; \ket{1}, \ket{e_x}, \ket{e_y},\ket{e_x^2},\ket{e_y^2},\ket{{e_x} {e_y}}, \ket{e_x^2 e_y}, \ket{e_x e_y^2}, \ket{e_x^2 e_y^2}\; \Big] ^{\dag},
\end{eqnarray}
where $\ket{e_x}$, $\ket{e_y}$ and $\ket{1}$ are given in Eqs.\eqref{eq:3a}-\eqref{eq:3b} and \eqref{eq:4}, respectively. Hence, $\tensor{Q}$ depends on the grid aspect ratio $r$. We can relate this moment basis for a \emph{rectangular} lattice $\tensor{Q}$ to an equivalent moment basis for a \emph{square} lattice $\tensor{P}$ given by
\begin{eqnarray}\label{eq:P-momentbasis}
\tensor{P}=\Big[ \; \ket{1}, \ket{\bar{e}_x}, \ket{\bar{e}_y},\ket{\bar{e}_x^2},\ket{\bar{e}_y^2},\ket{{\bar{e}_x} {\bar{e}_y}}, \ket{\bar{e}_x^2 \bar{e}_y}, \ket{\bar{e}_x \bar{e}_y^2}, \ket{\bar{e}_x^2 \bar{e}_y^2}\; \Big] ^{\dag},
\end{eqnarray}
where the particle velocity components of the square lattice $\ket{\bar{e}_x}$ and $\ket{\bar{e}_y}$ are given as
\begin{subequations}
\begin{eqnarray*}
\ket{\bar{e}_{x}} &=& (0,1,0,-1,0,1,-1,-1,1)^\dag, \label{eq:3a-square}\\
\ket{\bar{e}_{y}} &=& (0, 0, 1, 0, -1, 1, 1, -1, -1)^\dag.\label{eq:3b-square}
\end{eqnarray*}
\end{subequations}
Evidently, the two moment basis matrices $\tensor{Q}$ and $\tensor{P}$ can be readily related via a diagonal scaling matrix $\tensor{S}$
\begin{equation}\label{eq:QandPrelation}
\tensor{Q}= \tensor{S} \tensor{P},
\end{equation}
where $\tensor{S}$ reads as
\begin{equation}\label{eq:S-matrix}
\tensor{S} = \mbox{diag}{\begin{bmatrix}\;1  &  1  &  r  &   1  &  r^2  &  r  &   r  &  r^2  &  r^2 \;  \end{bmatrix}}.
\end{equation}
Importantly, from Eq.~\eqref{eq:QandPrelation}, the matrix inverse of $\tensor{Q}$ follows directly from the inverse of $\tensor{P}$ for the square lattice, which is quite straightforward to perform, and the inverse of the scaling matrix $\tensor{S}$ using
\begin{equation}\label{eq:QandPrelation-inverse}
\tensor{Q}^{-1}= \tensor{P}^{-1}\tensor{S}^{-1},
\end{equation}
where $\tensor{S}^{-1}$ is obtained from Eq.~\eqref{eq:S-matrix}, which being a diagonal matrix, by simply taking the reciprocal of each of its elements, i.e.,
\begin{equation}\label{eq:S-matrix-inverse}
\tensor{S}^{-1} = \mbox{diag}{\begin{bmatrix}\;1  &  1  &  r^{-1}  &   1  &  r^{-2}  &  r^{-1}  &   r^{-1}  &  r^{-2}  &  r^{-2} \;  \end{bmatrix}}.
\end{equation}
In other words, $\tensor{Q}^{-1}$ for the rectangular lattice is easy to perform knowing the corresponding $\tensor{P}^{-1}$ by appropriate scalings of the latter's elements based on the grid aspect ratio $r$. By contrast, since $\tensor{T}$ is defined using combinations of the basis vectors, its inverse $\tensor{T}^{-1}$, involves cumbersome expressions with parameterizations based on $r$. This fact confers a significant advantage of using $\tensor{Q}$ (and its inverse) rather than $\tensor{T}$ in performing mappings between moments and distribution functions~\cite{yahia2021three} and is thus adopted in designing our LB algorithm in what follows. However, as mentioned earlier, the effect of such combinations should still be accounted for in the evolution of the moments, which we accomplish by formally introducing a matrix $\tensor{B}$ in
\begin{equation}\label{eq:TandQrelation}
\tensor{T}= \tensor{B} \tensor{Q}.
\end{equation}
Thus, $\tensor{B}$ expresses the combinations of the moments (for the second order components $\ket{e_x^2+e_y^2}$ and $\ket{e_x^2-e_y^2}$ in the case of the D2Q9 lattice), which will be effectively introduced in the LB scheme in the evolution of the corresponding combinations of moments under collision and not for mappings. For this purpose, using the moment basis defined by $\tensor{Q}$, we can then define a set of bare moments $\mathbf{m}$ from the distribution functions $\mathbf{f}$ (and vice versa) using
\begin{eqnarray}\label{eq:f-Q-mappings}
\mathbf{m}= \tensor{Q} \mathbf{f}, \qquad \mathbf{f}={\tensor{Q}}^{-1} \mathbf{m},
\end{eqnarray}
where $\mathbf{m}$ is given by
\begin{equation}\label{baremoments-m}
\mathbf{m}=\left(k_{00}^\prime,k_{10}^\prime,k_{01}^\prime,k_{20}^\prime,k_{02}^\prime, k_{11}^\prime, k_{21}^\prime, k_{12}^\prime, k_{22}^\prime\right)^{\dag},
\end{equation}
and similarly for the sets of raw moment equilibria and the source terms, respectively, via $\mathbf{m}^{eq}= \tensor{Q} \mathbf{f}^{eq}$ and $\mathbf{\Phi}= \tensor{Q} \mathbf{S}$ as
\begin{subequations}
\begin{eqnarray}\label{baremoments-meq-phi}
\mathbf{m}^{eq}&=&\left(k_{00}^{eq\prime},k_{10}^{eq\prime},k_{01}^{eq\prime},k_{20}^{eq\prime},k_{02}^{eq\prime}, k_{11}^{eq\prime}, k_{21}^{eq\prime}, k_{12}^{eq\prime}, k_{22}^{eq\prime}\right)^{\dag},\\
\mathbf{\Phi}&=&\left(\sigma_{00}^\prime,\sigma_{10}^\prime,\sigma_{01}^\prime,\sigma_{20}^\prime,\sigma_{02}^\prime, \sigma_{11}^\prime, \sigma_{21}^\prime, \sigma_{12}^\prime, \sigma_{22}^\prime\right)^{\dag}.
\end{eqnarray}
\end{subequations}
These represent the simpler bare moments versions of those given in Eq.~\eqref{eq:9AA} for the combined moments of various quantities (which followed from Eq.~\eqref{eq:9}).

From the above developments by exploiting the properties of the various matrices introduced and rearranging, the preconditioned raw moment based MRT-LBE in Eq.~\eqref{eq:10} can be rewritten in the following equivalent form (see Ref.~\cite{yahia2021three} for details):
\begin{equation}\label{eq:10-equivalent}
 \mathbf{f} (\bm{x}+\mathbf{e}\Delta t, t+\Delta t) = \tensor{P}^{-1}\tensor{S}^{-1}
 \Big[\mathbf{m} + \tensor{B}^{-1}\tensor{\Lambda}\;\left(\; \tensor{B}\mathbf{m}^{eq}-\tensor{B}\mathbf{m} \;\right) + \tensor{B}^{-1}\left(\tensor{I} - \frac{\tensor{\Lambda}}{2}\right)  \tensor{B}\mathbf{\Phi}\Delta t \Big].
\end{equation}
This equation (Eq.~\eqref{eq:10-equivalent}) can be more conveniently represented by splitting them in the form of the following sequence of sub-steps that are amenable for implementation:
\begin{eqnarray}
\mathbf{m}&=&\tensor{S}\tensor{P}\mathbf{f},\nonumber\\
\tilde{\mathbf{m}}&=&\mathbf{m} + \tensor{B}^{-1}\left\{\tensor{\Lambda}\;\left(\; \tensor{B}\mathbf{m}^{eq}-\tensor{B}\mathbf{m} \;\right) + \left(\tensor{I} - \frac{\tensor{\Lambda}}{2}\right)  \tensor{B}\mathbf{\Phi}\Delta t\right\},\nonumber\\
\tilde{\mathbf{f}} (\bm{x},t) &=&  \tensor{P}^{-1}\tensor{S}^{-1}\tilde{\mathbf{m}},\nonumber\\
 \mathbf{f} (\bm{x}+\mathbf{e}\Delta t, t+\Delta t)&=&\tilde{\mathbf{f}} (\bm{x},t).\label{eq:LBErawmomentrectangularlattice}
\end{eqnarray}
Here, we emphasize that $\tensor{P}$ and $\tensor{P}^{-1}$ perform transformations between the distribution functions and raw moments in a way as done for the usual \emph{square} lattice using the non-orthogonal moment basis, $\tensor{S}$ and $\tensor{S}^{-1}$ reflect the simple scalings of the raw moments by factors based on grid aspect ratio and the order of the moment before and after collision, respectively, and $\tensor{B}$ and $\tensor{B}^{-1}$ represent combining moments prior to their relaxations under collision with the addition of the source terms, and their subsequent segregation, respectively. Equation~\eqref{eq:LBErawmomentrectangularlattice} expresses the preconditioned rectangular LBM based on \emph{raw moments}. As such, this scheme, by including both the numerical enhancement features, viz., preconditioning and rectangular lattice grids together, even in the context of raw moments, is new and suitable for implementation. Nevertheless, a number of prior studies (see e.g.,~\cite{ning2016numerical,chavez2018improving,hajabdollahi2019cascaded,hajabdollahi2020local,hajabdollahi2021central,hajabdollahi2021central,adam2019numerical,adam2021cascaded}), including those involving rectangular/cuboid lattices~\cite{yahia2021central,yahia2021three}, have demonstrated that constructing LB schemes involving the relaxations of \emph{central moments} under collision offer significant improvements in numerical stability over those based on raw moments. Hence, in this work we will only implement and perform a numerical study on the generalization of the above developments to central moments, viz., the preconditioned rectangular central moment LBE, which will be discussed next and followed by a summary of its algorithmic steps.

\subsection{Formulation of the Preconditioned Rectangular Central Moment LBE}
For this purpose, we will utilize the independently supported bare central moments defined in Eq.~\eqref{eq:9B} for the D2Q9 lattice and collect them in the form of the following vectors:
\begin{subequations}\label{eq:centralmoments-mc}
\begin{eqnarray}
\mathbf{m}^c&=&\left(k_{00},k_{10},k_{01},k_{20},k_{02}, k_{11}, k_{21}, k_{12}, k_{22}\right)^{\dag}, \\
\mathbf{m}^{c,eq}&=&\left(k_{00}^{eq},k_{10}^{eq},k_{01}^{eq},k_{20}^{eq},k_{02}^{eq}, k_{11}^{eq}, k_{21}^{eq}, k_{12}^{eq}, k_{22}^{eq}\right)^{\dag},\\
\mathbf{\Phi}^c&=&\left(\sigma_{00},\sigma_{10},\sigma_{01},\sigma_{20},\sigma_{02}, \sigma_{11}, \sigma_{21}, \sigma_{12}, \sigma_{22}\right)^{\dag},
\end{eqnarray}
\end{subequations}
Now, the raw moments defined in Eq.~\eqref{eq:9A} can be related to the central moments in Eq.~\eqref{eq:9B} via straightforward binomial expansions involving the former in combinations with monomials of the fluid velocity components at different order (of the form $u_x^pu_y^q$). Thus, the mappings from the raw moments to central moments (and vice versa) can be formally expressed as
\begin{eqnarray}\label{eq:m-mc-F-mappings}
\mathbf{m}^c= \tensor{F} \mathbf{m}, \qquad \mathbf{m}={\tensor{F}}^{-1} \mathbf{m}^c,
\end{eqnarray}
where $\tensor{F}$ is referred to as the frame transformation matrix reflecting the binomial transforms of moments at different orders supported by the D2Q9 lattice. Such a formulation to represent the transformations between the raw moments and central moments for the complete set supported by the lattice in the form of a shift matrix was first introduced by Fei and Luo in~\cite{fei2017consistent,fei2018three}. It is given by
\begin{equation} \label{eq:Fmatrix}
  \tensor{F} =
  \begin{bmatrix}
   1 &  0 &  0 &  0 &  0 &  0 &  0  &  0  &  0\\[4pt]
  -u_x  &  1 &  0 &  0 &  0 &  0 &  0  &  0  &  0\\[4pt]
  -u_y  &  0  & 1  & 0  &  0   & 0 & 0 & 0 &   0 \\[4pt]
  u_x ^2 +u_y ^2  &  -2 u_x  & -2 u_y  & 1   &  0   & 0 & 0 & 0 &   0 \\[4pt]
  u_x ^2 -u_y ^2   &  -2 u_x  & 2 u_y & 0   & 1   & 0 & 0 & 0 &   0
  \\[4pt]
  u_x u_y  &  -u_y  & -u_x  & 0  &  0   & 1 & 0 & 0 &  0 \\[4pt]
  -u_x ^2 u_y    &  2 u_x u_y  & u_x ^2   & - \frac{1}{2} u_y  &  - \frac{1}{2} u_y  & -2 u_x & 1 & 0 &  0 \\[4pt]
  -u_x u_y ^2   & u_y ^2   & 2 u_x u_y  & - \frac{1}{2} u_x  &  \frac{1}{2} u_x  & -2 u_y & 0 & 1 &  0 \\[4pt]
  u_x^2 u_y ^2   & -2 u_x u_y ^2   & -2 u_x^2 u_y  &  \frac{1}{2} (u_x^2+u_y^2)  &  \frac{1}{2} (u_y^2 -u_x^2)  & 4 u_x u_y & -2 u_y & -2 u_x &  1\\.
  \end{bmatrix}
\end{equation}
Also, as noted in Ref.~\cite{yahia2021central}, its inverse ${\tensor{F}}^{-1}$ can be read off directly from the elements of ${\tensor{F}}={\tensor{F}}(u_x,u_y)$ with minor changes by exploiting the following property that exists for such transforms: ${\tensor{F}}^{-1}={\tensor{F}}(-u_x,-u_y)$. Thus, and thus naturally both of them are lower triangular matrices. Then, by an analogy with Eq.~\eqref{eq:10-equivalent}, we can write the following preconditioned rectangular central moment LBE by involving relaxations of central moments $\mathbf{m}^c$ under collision (rather than raw moments $\mathbf{m}$) and including the additional transforms between them and the raw moments (via $\tensor{F}$ and $\tensor{F}^{-1}$)~\cite{yahia2021three}:
\begin{equation}\label{eq:10-equivalent-centralmoment}
 \mathbf{f} (\bm{x}+\mathbf{e}\Delta t, t+\Delta t) = \tensor{P}^{-1}\tensor{S}^{-1}\tensor{F}^{-1}
 \Big[\mathbf{m}^c + \tensor{B}^{-1}\tensor{\Lambda}\;\left(\; \tensor{B}\mathbf{m}^{eq,c}-\tensor{B}\mathbf{m}^c \;\right) + \tensor{B}^{-1}\left(\tensor{I} - \frac{\tensor{\Lambda}}{2}\right)  \tensor{B}\mathbf{\Phi}^c\Delta t \Big].
\end{equation}
It may be noted that this Eq.~(\ref{eq:10-equivalent-centralmoment}) has some similarities with those presented by Luo and collaborators~\cite{fei2017consistent,fei2018three}, where they presented the cascaded central moment LB method in a generalized MRT framework on a square lattice. Our notations follow from those presented in an earlier work of the second author~\cite{premnath2009incorporating} for the frame transformation matrix $\tensor{F}$. This matrix as shown in Eq.~(\ref{eq:Fmatrix}) is identical to the shift matrix used in~\cite{fei2017consistent,fei2018three}. Nevertheless, there are some key differences: Eq.~(\ref{eq:10-equivalent-centralmoment}) also involves the forward and inverse scaling transformations (via diagonal matrices) related to the grid aspect ratio $r$ to accommodate the use of a rectangular lattice in a modular fashion. Moreover, for the collision step, Refs.~\cite{fei2017consistent,fei2018three} combine the relaxation parameters of the second order moments similar to that in~\cite{asinari2008generalized}. By contrast, here, to execute the collision step, the second order moments are combined prior to collision, which are then relaxed at independent rates to their equilibria (with appropriate corrections based on $r$ and $\gamma$) and then segregated post collision. Such a strategy for performing collision was presented by Geier \emph{et al.}~\cite{geier2015cumulant}, and present work can be considered as an extension of such an approach for performing flow simulations on rectangular lattices with preconditioning. Also, it should be noted that not all collision models admit interpretations based on matrices. For example, the highly sophisticated and nonlinear cumulant LB scheme cannot be represented in the form of matrices. Hence, the algorithms (including the special cases for raw moments and central moments) presented in~\cite{geier2015cumulant} are given only as series of substeps and no matrices are utilized in this regard. Thus, to maintain generality of our approach, in what follows, we will represent our PRC-LBM in the form of a sequence of operations that conveys the essence of Eq.~(\ref{eq:10-equivalent-centralmoment}). This last equation (Eq.~\eqref{eq:10-equivalent-centralmoment}) can then be more conveniently split up into various sub-steps, which then results into the following preconditioned rectangular central moment LBM or PRC-LBM:
\begin{eqnarray}
\mathbf{m}^c&=&\tensor{F}\tensor{S}\tensor{P}\mathbf{f},\nonumber\\
\tilde{\mathbf{m}}^c&=&\mathbf{m}^c + \tensor{B}^{-1}\left\{\tensor{\Lambda}\;\left(\; \tensor{B}\mathbf{m}^{c,eq}-\tensor{B}\mathbf{m}^c \;\right) + \left(\tensor{I} - \frac{\tensor{\Lambda}}{2}\right)  \tensor{B}\mathbf{\Phi}^c\Delta t\right\},\nonumber\\
\tilde{\mathbf{f}} (\bm{x},t) &=&  \tensor{P}^{-1}\tensor{S}^{-1}\tensor{F}^{-1}\tilde{\mathbf{m}}^c,\nonumber\\
 \mathbf{f} (\bm{x}+\mathbf{e}\Delta t, t+\Delta t)&=&\tilde{\mathbf{f}} (\bm{x},t).\label{eq:LBEcentralmomentrectangularlattice}
\end{eqnarray}
This PRC-LBM is a numerically more robust formulation than its raw moment counterpart given earlier in Eq.~\eqref{eq:LBErawmomentrectangularlattice}. The algorithmic details of the PRC-LBM to facilitate its implementation are discussed in~\ref{sec:algorithmic-details-PRC-LBM}.

\section{Results and discussion} \label{sec:5}
We will now discuss some case studies based on the PRC-LB algorithm for simulations of shear flows at various characteristic parameters that show its numerical validation against certain benchmark problems and the significant advantages of combining the rectangular lattice grid and preconditioning over the LB scheme based on the square lattice and without preconditioning. In this regard, as noted at the end of~\ref{sec:algorithmic-details-PRC-LBM}, in what follows, the no-slip boundary condition for the moving walls, which generate shear flows, are accounted for via the momentum augmented half-way bounce back approach and including the parametrization for the grid aspect ratio for the rectangular lattice given in our recent work~\cite{yahia2021central}.

\subsection{2D Shear Flows in Lid-driven Square Cavity using PRC-LBM: Validation}
\begin{figure}[H]
\centering
\advance\leftskip-1.7cm
    \subfloat[$\mbox{Re}=100$ ] {
        \includegraphics[width=.4\textwidth] {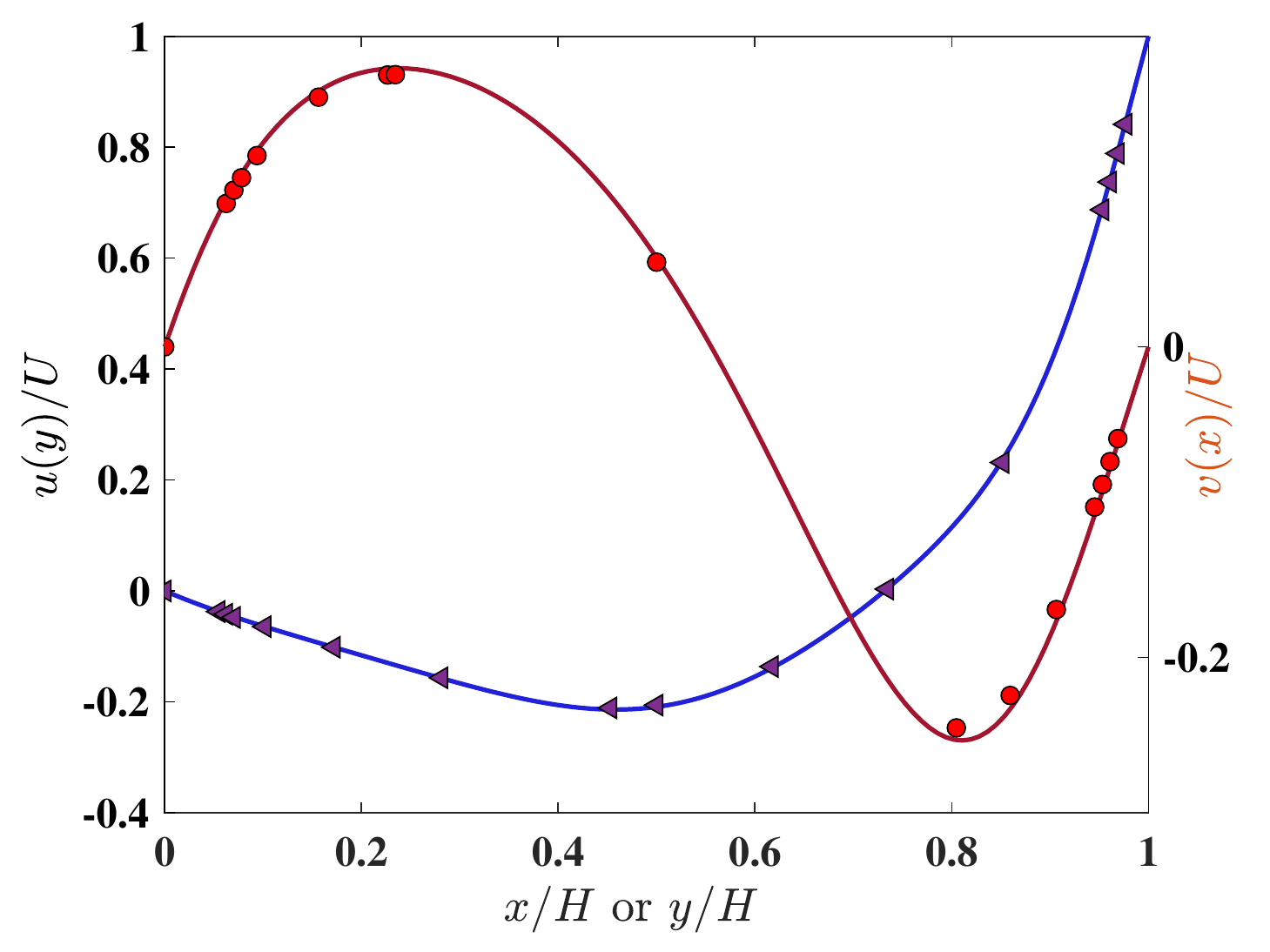}
        \label{fig:1a} } 
    \subfloat[$\mbox{Re}=1000$] {
        \includegraphics[width=.4\textwidth] {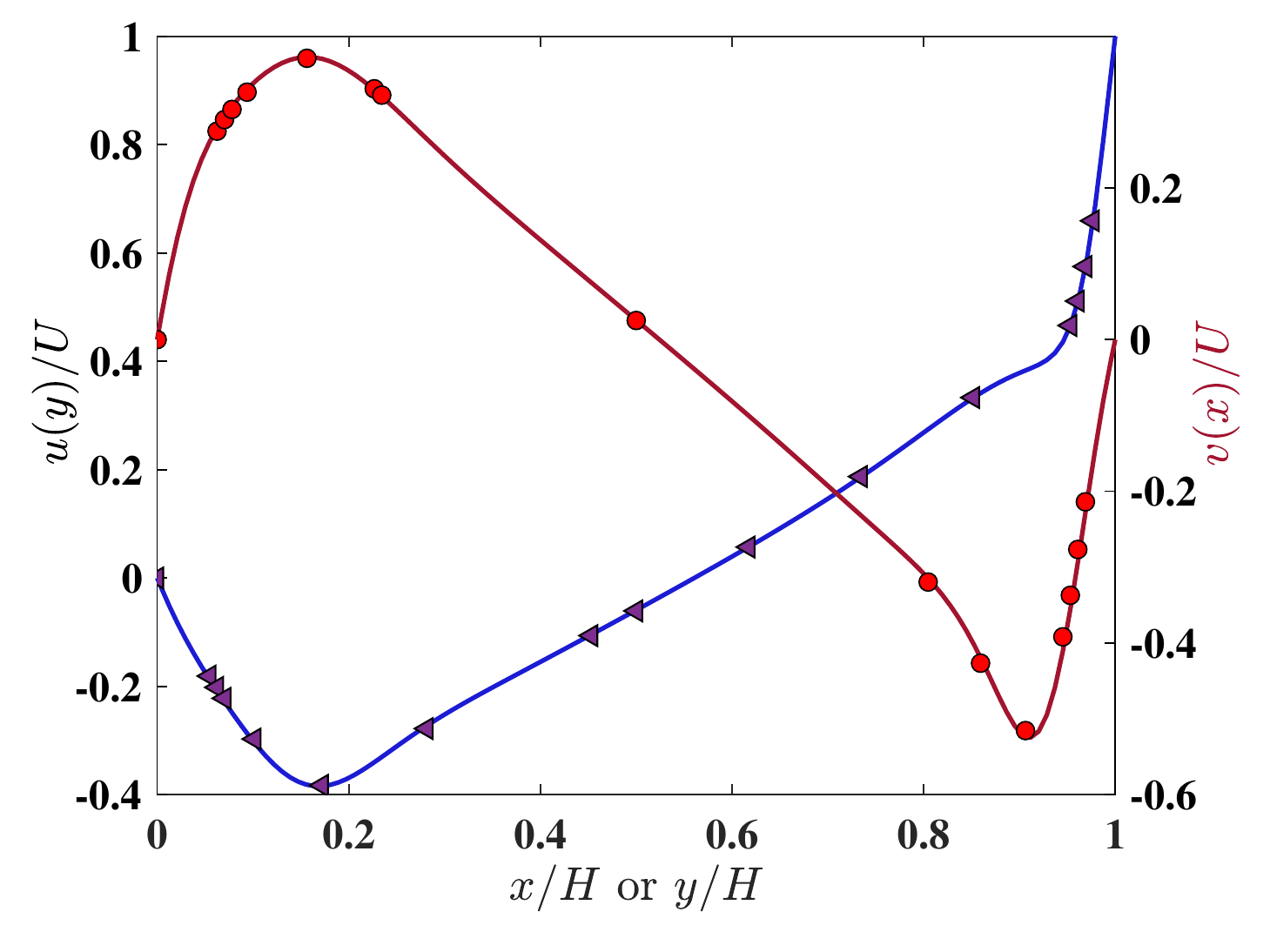}
        \label{fig:1b} } 
         \\
        \subfloat[$\mbox{Re}=3200$ ] {
        \includegraphics[width=.4\textwidth] {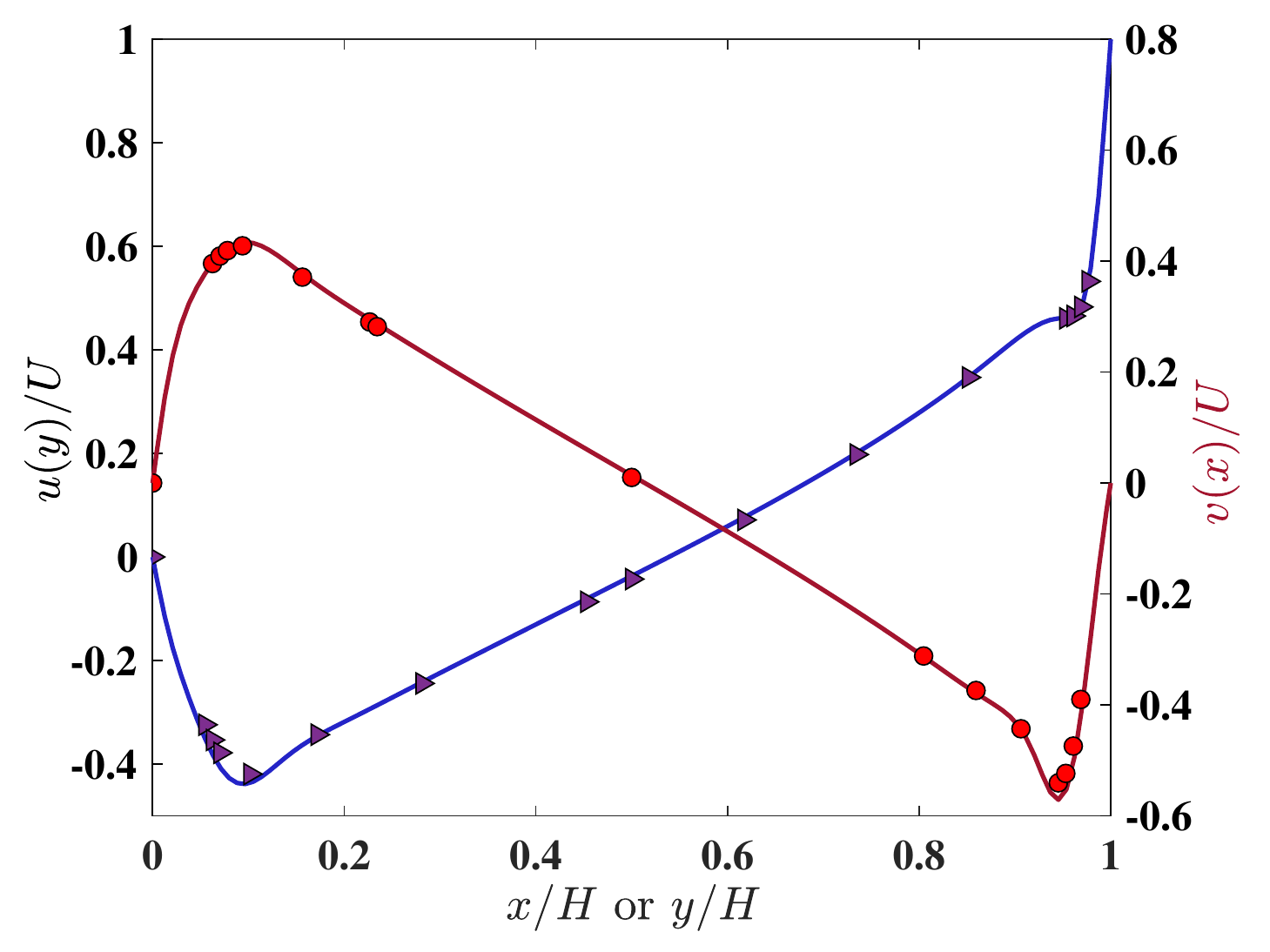}
        \label{fig:1c} } 
    \advance\leftskip0cm
    \caption{The components of the velocity profiles $u(y)$ and $v(x)$ along the vertical and horizontal centerlines of a square cavity, i.e., $x=H/2$ and $y=H/2$ respectively, computed using the PRC-LBM on a rectangular lattice grid of aspect ratio of $r=0.5$ with the preconditioning parameter $\gamma=0.1$ at \mbox{Ma}$=0.05$ for Reynolds numbers of (a)~\mbox{Re}$=100$, (b)~\mbox{Re}$=1000$, (c)~\mbox{Re}$=3200$ and compared with the benchmark numerical solutions of Ref.~\cite{ghia1982high} (symbols).}
    \label{fig:Restreams}
\end{figure}

First, we will assess the accuracy of the PRC-LBM for the simulation of the classical flow within a \emph{square} cavity of side $H$, whose top surface moves at a constant velocity $U$ setting up flow patterns that depend on the Reynolds number given by $\mbox{Re}=UH/\nu$. We performed simulations at Reynolds numbers of~$\mbox{Re}=100,1000$ and $3200$ at a fixed Mach number~$\mbox{Ma}=0.05$. A rectangular lattice with grid aspect ratio of $r=0.5$ using a grid resolution of $N_x\times N_y=200\times 400$ is employed by setting the preconditioning parameter $\gamma=0.1$ in our algorithm given in~\ref{sec:algorithmic-details-PRC-LBM}. The numerical results of the horizontal and vertical components of the velocity profiles along the centerlines of the cavity predicted by the PRC-LBM at the above three choices of $\mbox{Re}$ are compared against the benchmark numerical solutions given by Ghia \emph{et al.}~\cite{ghia1982high} in Fig.~\ref{fig:Restreams}. It is evident that the PRC-LBM results are in very good agreement with the benchmark data.

\subsection{2D Shear Flows in Lid-driven Shallow and Deep Cavities using PRC-LBM: Validation and Convergence Acceleration}
\begin{figure}[H]
\centering
\advance\leftskip-1.7cm
  \subfloat[Shallow cavity ($\mbox{AR}<1$)] {
        \includegraphics[width=0.4\textwidth] {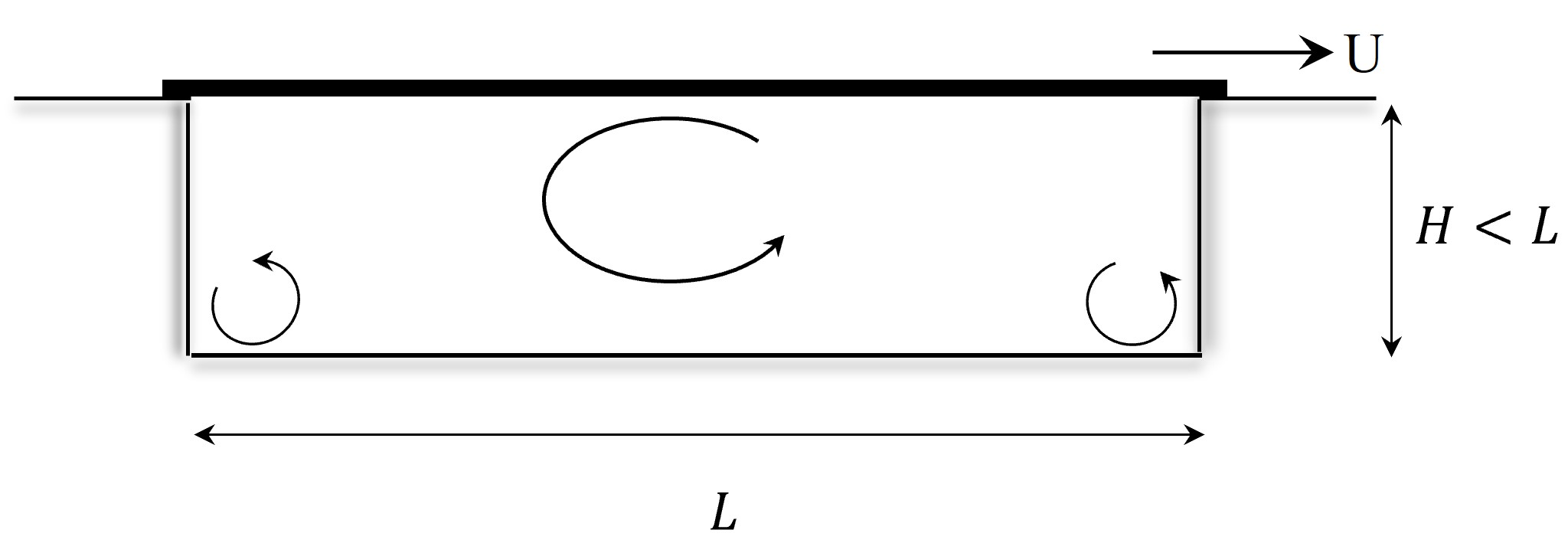}
        \label{fig:schematicshallowcavity} } 
    \subfloat[Deep cavity ($\mbox{AR}>1$)] {
        \includegraphics[width=0.25\textwidth] {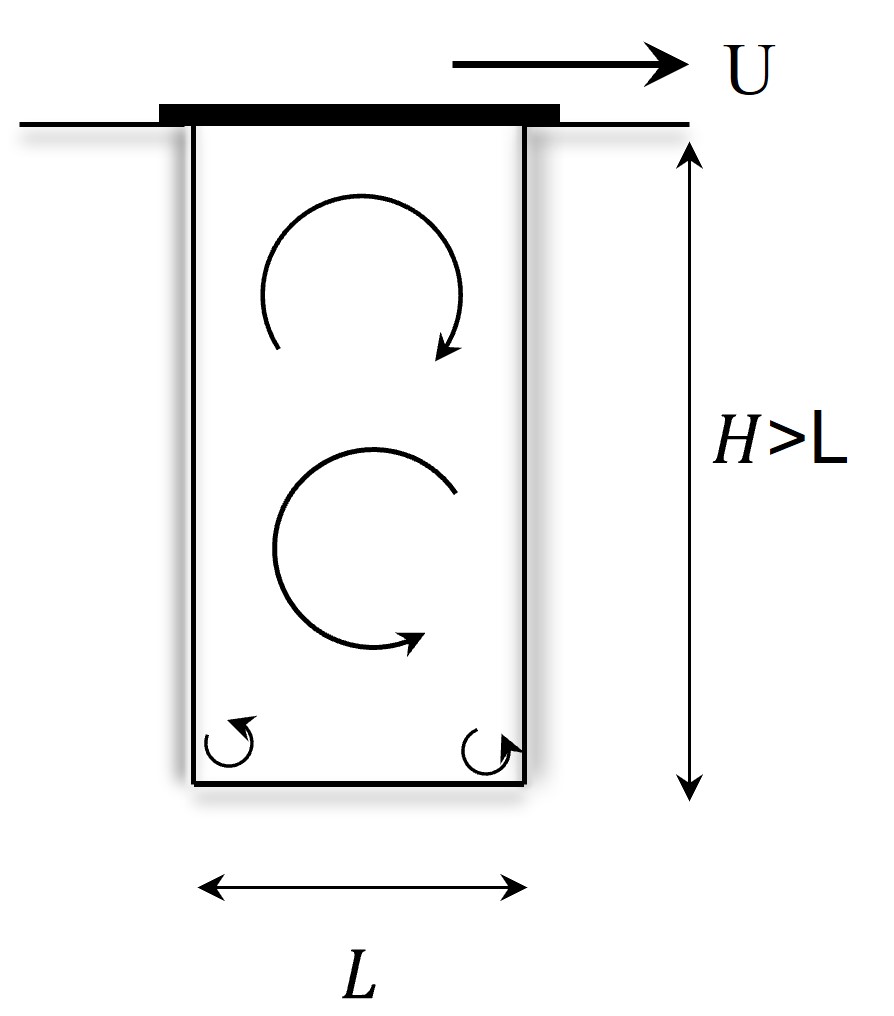}
        \label{fig:schematicdeepcavity} } \\
                \advance\leftskip0cm
 \caption{Schematic arrangements of the flows inside a 2D (a) shallow and (b) deep cavities of dimensions $L\times H$.}
 \label{fig:schematicavity}
\end{figure}

Next, we will demonstrate the accuracy and computational advantages of using the PRC-LBM for computing anisotropic and inhomogeneous shear flows inside \emph{rectangular} cavities of length $L$ and height $H$, and characterized by the geometric aspect ratio $\mbox{AR}=H/L$. As shown in Fig~\ref{fig:schematicavity}, we consider two cases: (a) shallow cavity with aspect ratio $\mbox{AR}<1$ and (b) deep cavity with aspect ratio $\mbox{AR}>1$. In each case, the flow is set up by the motion of the upper lid with a velocity $U$ in the positive $x$ direction which generates vortices that are different in size and shape based on the Reynolds number specified by $\mbox{Re}=UH/\nu$. The confinement effect characterized by the aspect ratio $\mbox{AR}$ results in the characteristic flow scales or the spatial gradients in velocities that can be different in different coordinate directions, which can be more naturally and efficiently resolved by using a rectangular lattice grid. In our previous work, we illustrated the benefits of using the rectangular lattice over that based on the square lattice for simulating such flows within shallow cavities with the use of fewer grid nodes for the former when compared to the latter~\cite{yahia2021central}. In the current study, we aim to show further improvements of utilizing preconditioning with rectangular lattice grids for convergence acceleration of flows to their steady states, resulting in dramatic cumulative advantages of simulating such flows using the square lattice and without preconditioning.

\begin{figure}[H]
\centering
\advance\leftskip-1.7cm
    \subfloat[$\mbox{AR}=0.25$, \mbox{Re}=100, $r=0.2$] {
        \includegraphics[width=0.45\textwidth] {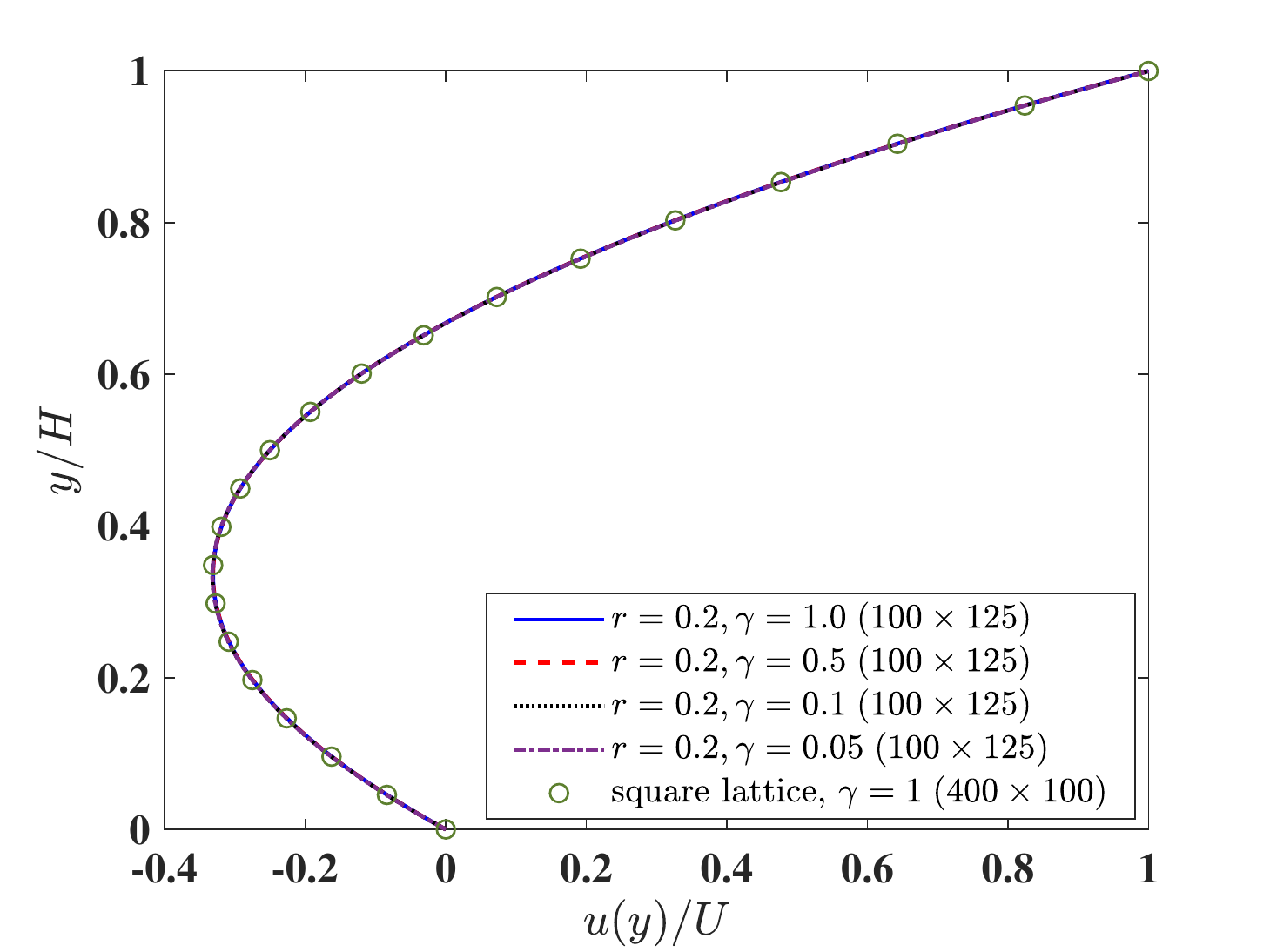}
        \label{ushallow_Re100} } 
    \subfloat[$\mbox{AR}=0.25$, \mbox{Re}=100, $r=0.2$] {
        \includegraphics[width=0.45\textwidth] {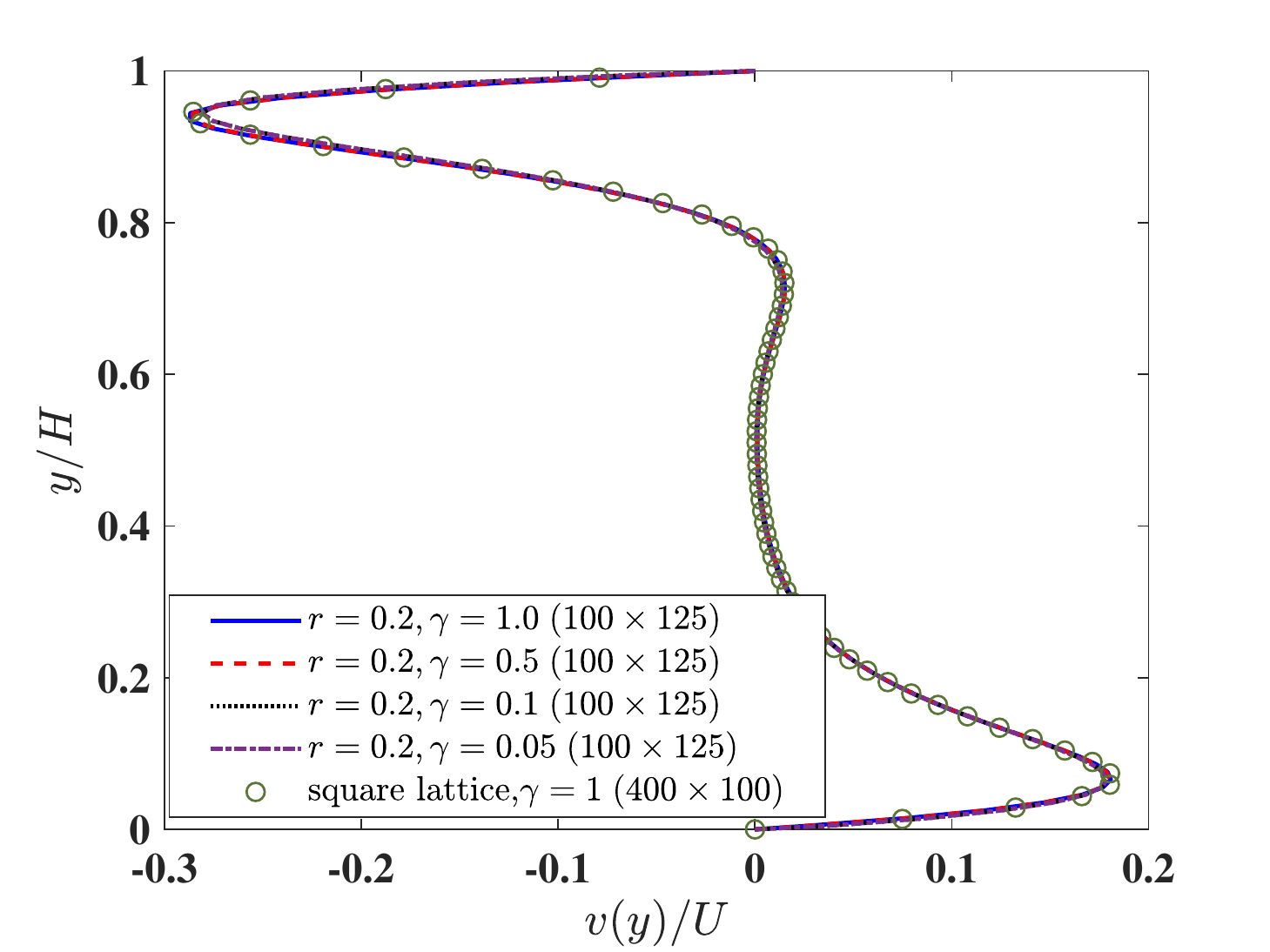}
        \label{vshallow_Re100} } \\
                \advance\leftskip0cm
    \caption{The velocity profiles along the centerlines of a shallow rectangular cavity of aspect ratio $\mbox{AR}=0.25$ at a Reynolds number~$\mbox{Re}=100$ and Mach number~$\mbox{Ma}=0.06$ computed using the PRC-LBM with a grid resolution $100\times 125$ using the rectangular lattice of grid aspect ratio $r=0.2$ with different levels of preconditioning, i.e., $\gamma=1.0, 0.5, 0.1$ and $0.05$, and compared with the results obtained using a square lattice ($r=1.0$) with a grid resolution of $400\times 100$ at $\gamma=1.0$. (a) $u$ component of the velocity along the vertical centerline, and (b) $v$ component of the velocity along the vertical centerline.}
    \label{fig:velocityprofile1}
\end{figure}

\begin{figure}[H]
\centering
\advance\leftskip-1.7cm
    \subfloat[$\mbox{Ma}=0.06$, $r=0.2$] {
        \includegraphics[width=0.45\textwidth] {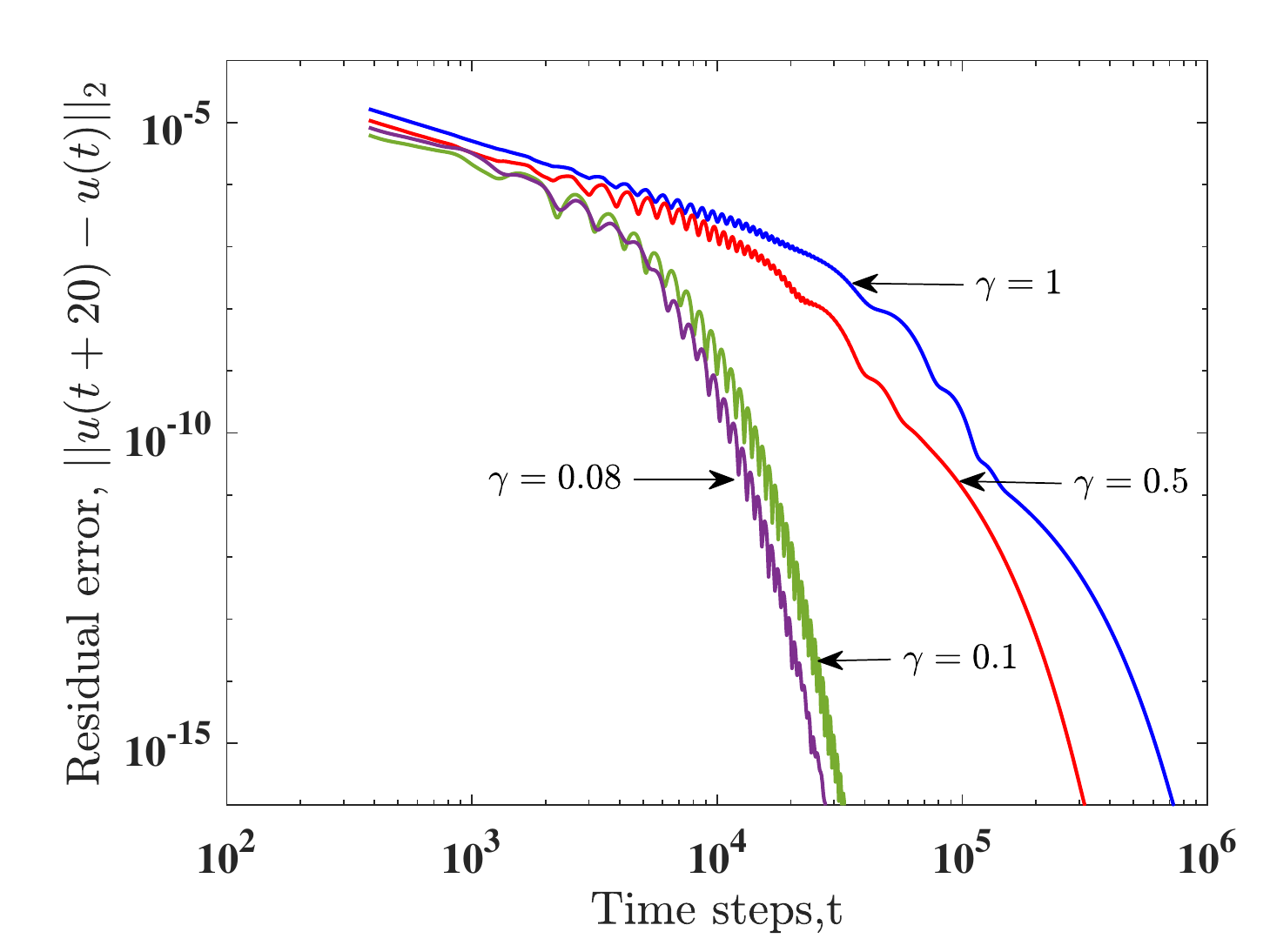}
        \label{histconverge} } 
    \subfloat[$\mbox{Ma}=0.01$, $r=0.2$] {
        \includegraphics[width=0.45\textwidth] {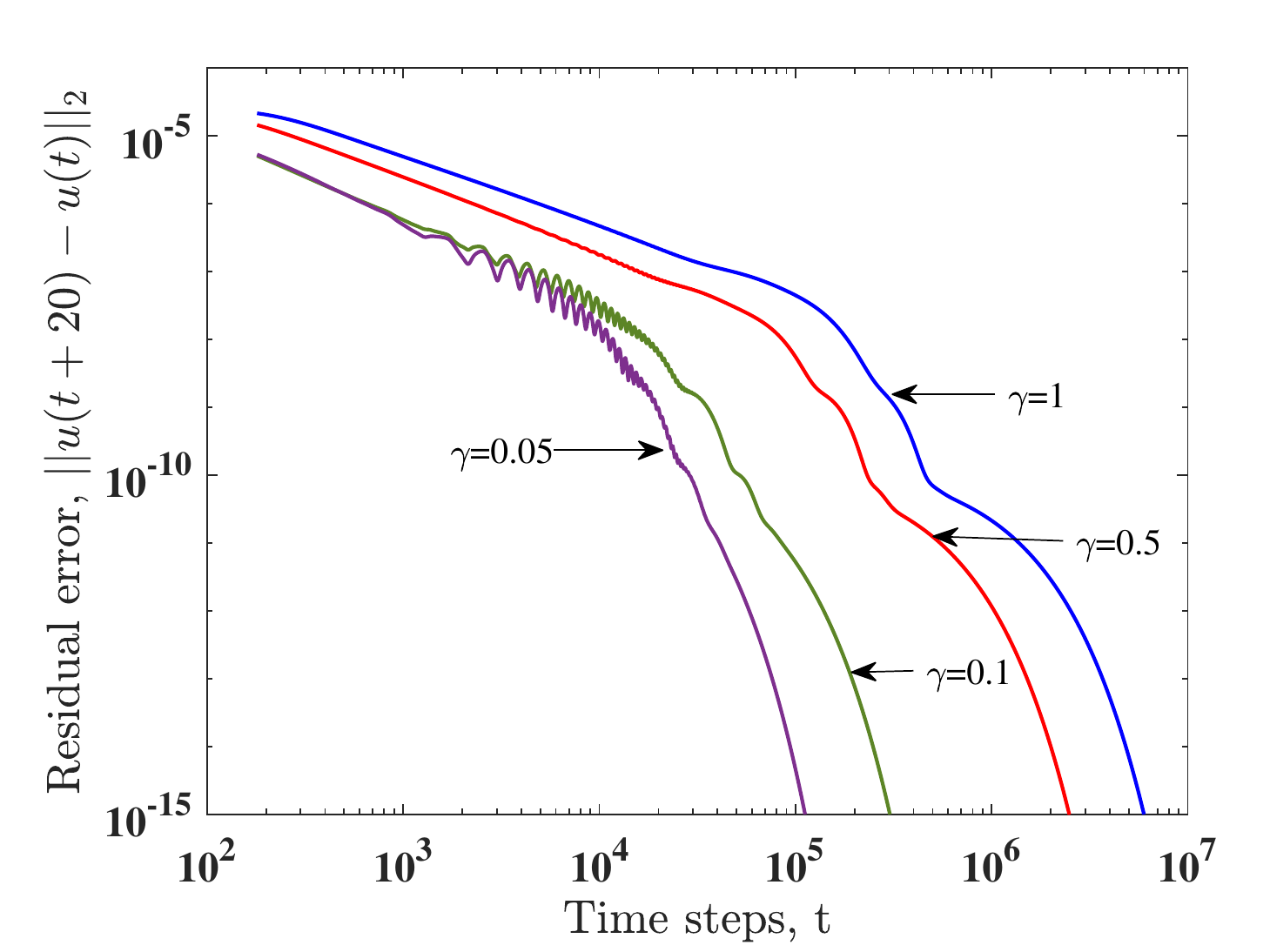}
        \label{Mahistconverge} } \\
                \advance\leftskip0cm
    \caption{Convergence histories to the steady state for simulations of flows within a shallow rectangular cavity of aspect ratio $\mbox{AR}=0.25$ at a Reynolds number~\mbox{Re}$=100$ using the PRC-LBM with a grid resolution of $100\times 125$ using a rectangular lattice of grid aspect ratio $r=0.2$ with different values of the preconditioned parameter $\gamma$ at (a) Mach number~\mbox{Ma}$=0.06$, and (b) Mach number~\mbox{Ma}$=0.01$.}
    \label{fig:2}
\end{figure}

In this regard, first, simulations of flow inside a shallow rectangular cavity of aspect ratio $\mbox{AR}=0.25$ at a Reynolds number~$\mbox{Re}=100$ and Mach number~$\mbox{Ma}=0.06$ are performed using rectangular lattice with grid aspect ratio $r=\Delta y/\Delta x=0.2$. If $N_x$ and $N_y$ are the number of grid nodes resolving the cavity in $x$ and $y$ directions, respectively, the grid spacings in the respective directions satisfy $\Delta x=L/N_x$ and $\Delta y=H/N_y$, or $r=\mbox{AR}N_x/N_y$. Thus, $N_x=(r/\mbox{AR})N_y$ in the case of the rectangular lattice and $N_x=(1/\mbox{AR})N_y$ for the square lattice, where $r=1$. Choosing $N_y=125$, this results in $100\times 125$ as the total number of grid nodes for the rectangular lattice, while taking even somewhat smaller value of $N_y=100$, however, requires a total $400\times 100$ grid nodes for the square lattice case, which is significantly more when compared to the former. Thus, if the computed results in each case are in agrement with one another, this, by itself, is a saving in the memory storage and computational cost by a factor of over $3$ in using the rectangular lattice. Then, the effect of different levels of preconditioning in PRC-LBM is studied by considering $\gamma=1.0, 0.5, 0.1$ and $0.05$. Figures~\ref{ushallow_Re100} and \ref{vshallow_Re100} show comparisons of the centerline velocity components $u$ and $v$ computed using PRC-LBM with $r=0.2$ and $100\times 125$ for different $\gamma$ with the results obtained using the square lattice ($r=1$) with $400\times 100$ and without preconditioning ($\gamma=1$). It can be seen the PRC-LBM results are in remarkably good agreement for the entire range of the choice of $\gamma$, when compared to the corresponding velocity profiles for the square lattice case.
Then, in order to verify the benefits of utilizing the preconditioning procedure with the rectangular lattice, we study the convergence histories to the steady state in using the PRC-LBM at $r=0.2$ for two values of the Mach number and different $\gamma$ in each case. Figure~\ref{histconverge} shows the convergence histories at $\mbox{Ma}=0.06$ with $\gamma=1, 0.5,0.1$, and $0.08$ and Fig.~\ref{Mahistconverge} shows that at smaller Mach number $\mbox{Ma}=0.01$ with $\gamma=1, 0.5,0.1$, and $0.05$, where the residual global error of the $u$ velocity component is estimated under the second norm as $||u(t+20)-u(t)||_2$. At $\mbox{Ma}=0.06$, it takes about $610,000$ steps to reach the steady state without preconditioning ($\gamma=1$), while the PRC-LBM with $\gamma=0.08$ requires only about $24,000$ steps to reach similar residual error as the previous case, leading to a dramatic reduction in the number of steps for convergence by a factor of about $25$ in this case. On the other hand, at a lower $\mbox{Ma}=0.01$, the PRC-LBM takes about $5,000,000$ without preconditioning, while only about $97,000$ with preconditioning (using $\gamma=0.05$), with an even larger improvement corresponding to a reduction factor of about $51$. Clearly, at lower Mach numbers, the disparities between the flow speed and the sound speed are larger, and the associated higher stiffness is respectively alleviated to a greater degree with preconditioning, which is consistent with previous investigations on square lattice grids (see e.g.,~\cite{guo2004preconditioned,hajabdollahi2019improving}). Noting that we have already reduced the computational costs by involving the rectangular lattice grids when compared to that using the square lattice, preconditioning the rectangular central moment LBM provides a further, i.e., cumulative improvement in solving steady state flow problems more efficiently. However, it should be noted that while using smaller values of $\gamma$ does favor faster convergence speed, its smallest possible value is limited by the numerical stability considerations (see e.g.,~ \cite{guo2004preconditioned,premnath2009steady,izquierdo2009optimal}). The optimal value of the level of preconditioning is a compromise between convergence rate and stability. Typically, the minimum possible $\gamma$ is found to be proportional to the Mach number used in simulations.

\begin{figure}[H]
\centering
\advance\leftskip-1.7cm
    \subfloat[$AR=2$, $\mbox{Re}=100$, $r=1.6$] {
        \includegraphics[width=0.45\textwidth] {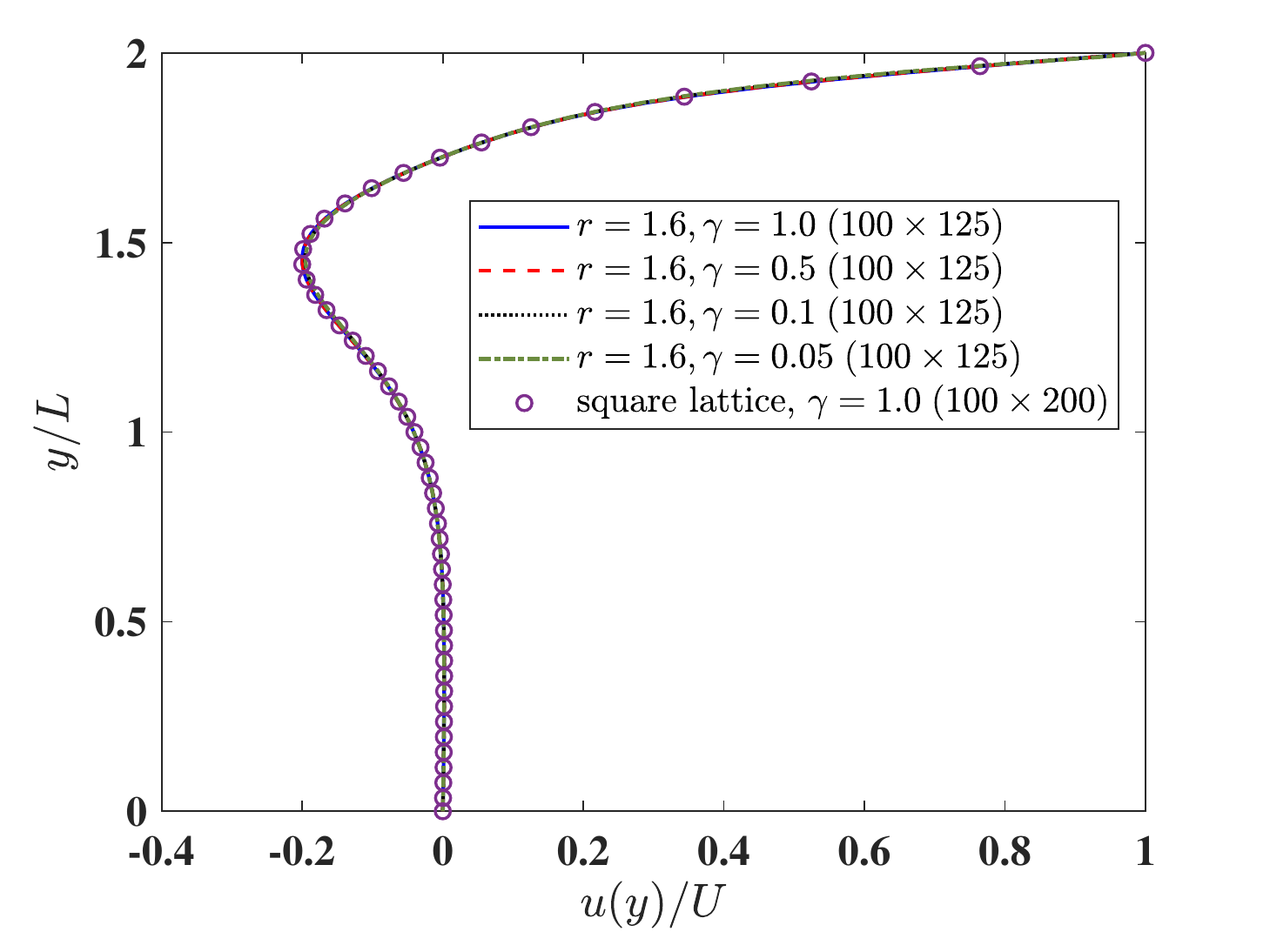}
        \label{udeep_Re100} } 
    \subfloat[$AR=2$, $\mbox{Re=100}$, $r=1.6$] {
        \includegraphics[width=0.45\textwidth] {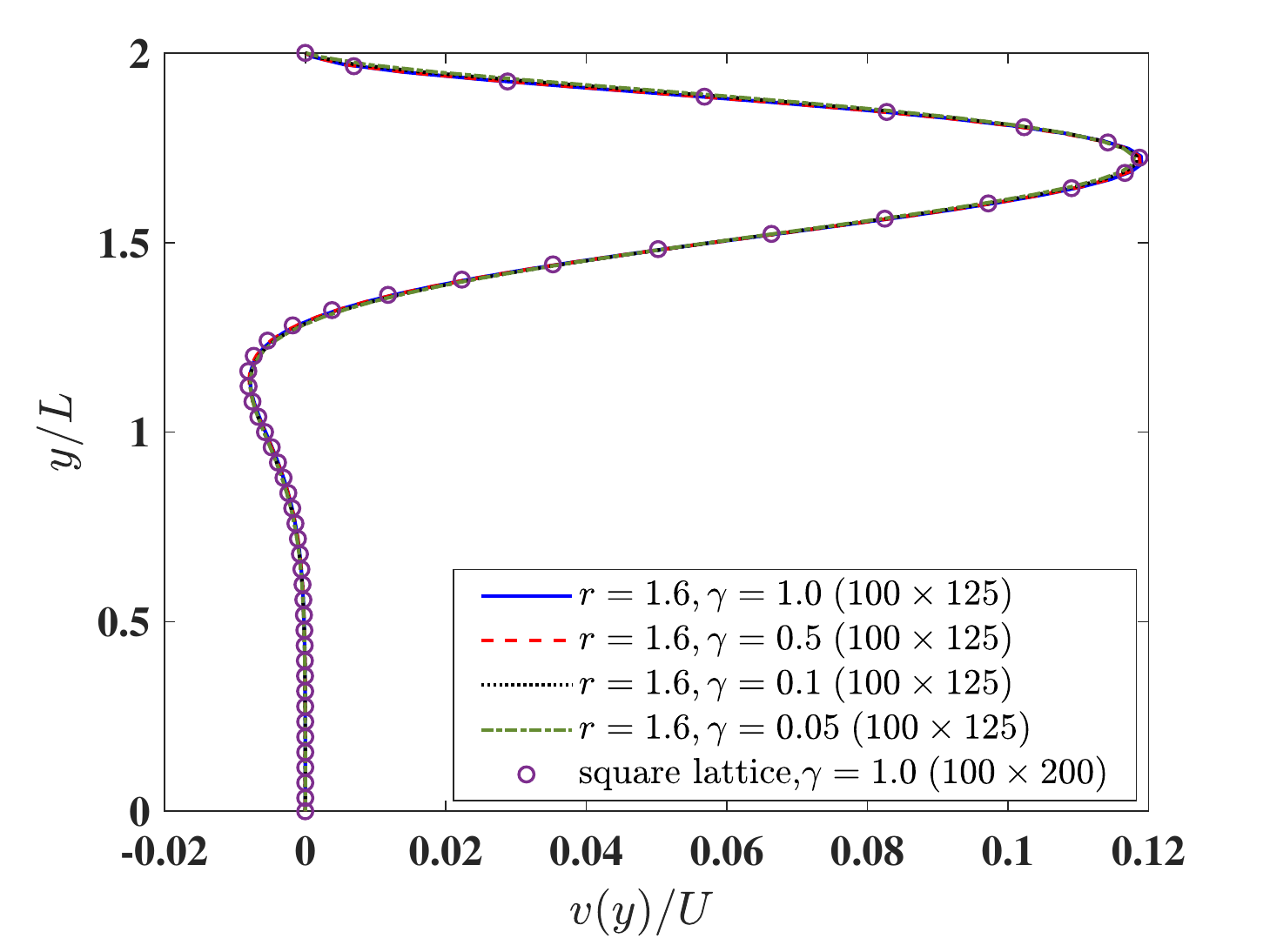}
        \label{vdeep_Re100} } \\
                \advance\leftskip0cm
    \caption{The velocity profiles along the centerlines of a deep rectangular cavity of aspect ratio $\mbox{AR}=2$ at a Reynolds number~$\mbox{Re}=100$ and Mach number~$\mbox{Ma}=0.06$ computed using the PRC-LBM with a grid resolution $100\times 125$ using the rectangular lattice of grid aspect ratio $r=1.6$ with different levels of preconditioning, i.e., $\gamma=1.0, 0.5, 0.1$ and $0.05$, and compared with the results obtained using a square lattice ($r=1.0$) with a grid resolution of $100\times 200$ at $\gamma=1.0$. (a) $u$ component of the velocity along the vertical centerline, and (b) $v$ component of the velocity along the vertical centerline.}
    \label{fig:deepvelocityprofile}
\end{figure}

\begin{figure}[H]
\centering
\advance\leftskip-1.7cm
    \subfloat[$\mbox{Ma}=0.06$, $r=1.6$] {
        \includegraphics[width=0.45\textwidth] {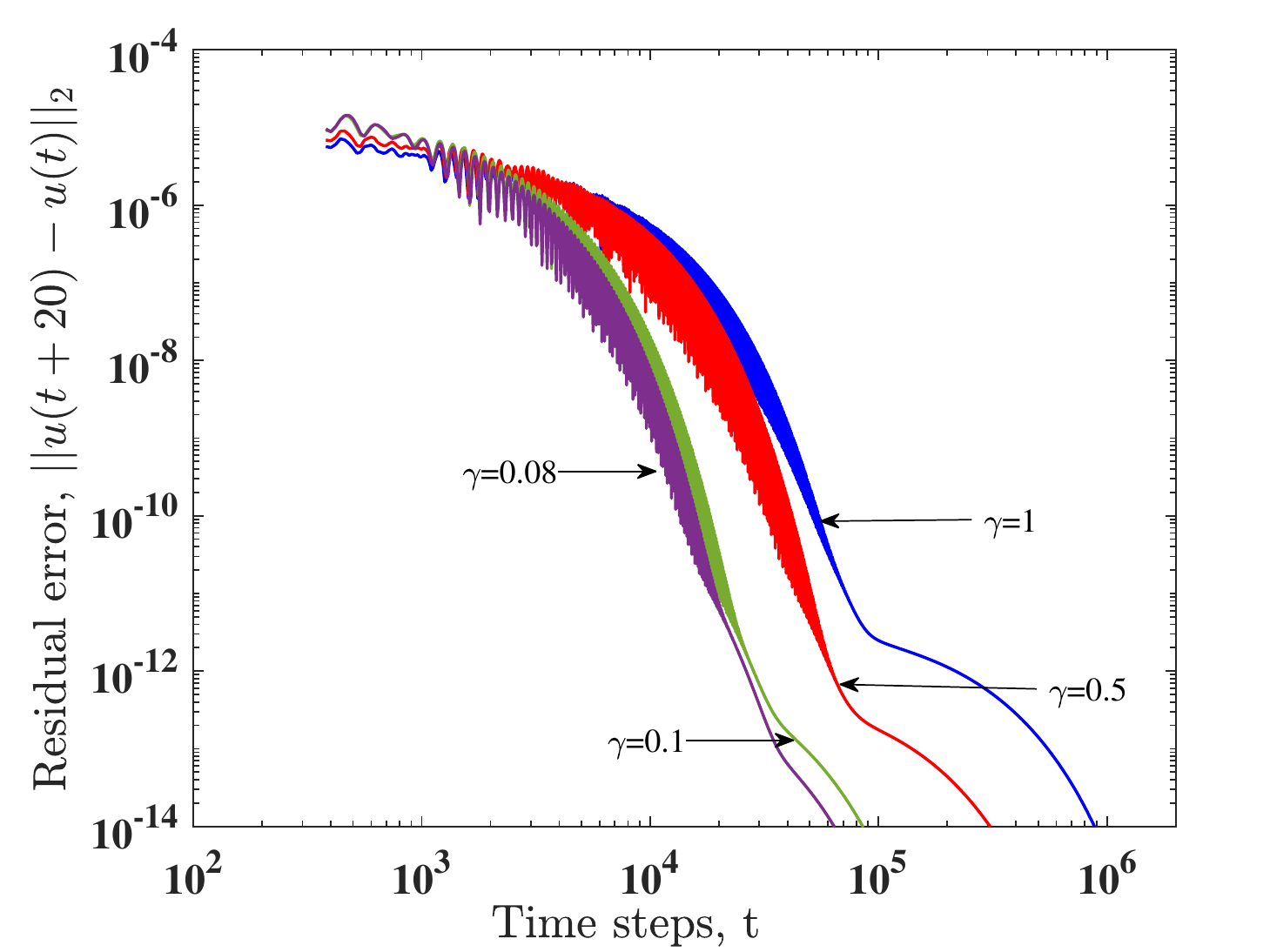}
        \label{udeephistMa06} } 
    \subfloat[$\mbox{Ma}=0.02$, $r=1.6$] {
        \includegraphics[width=0.45\textwidth] {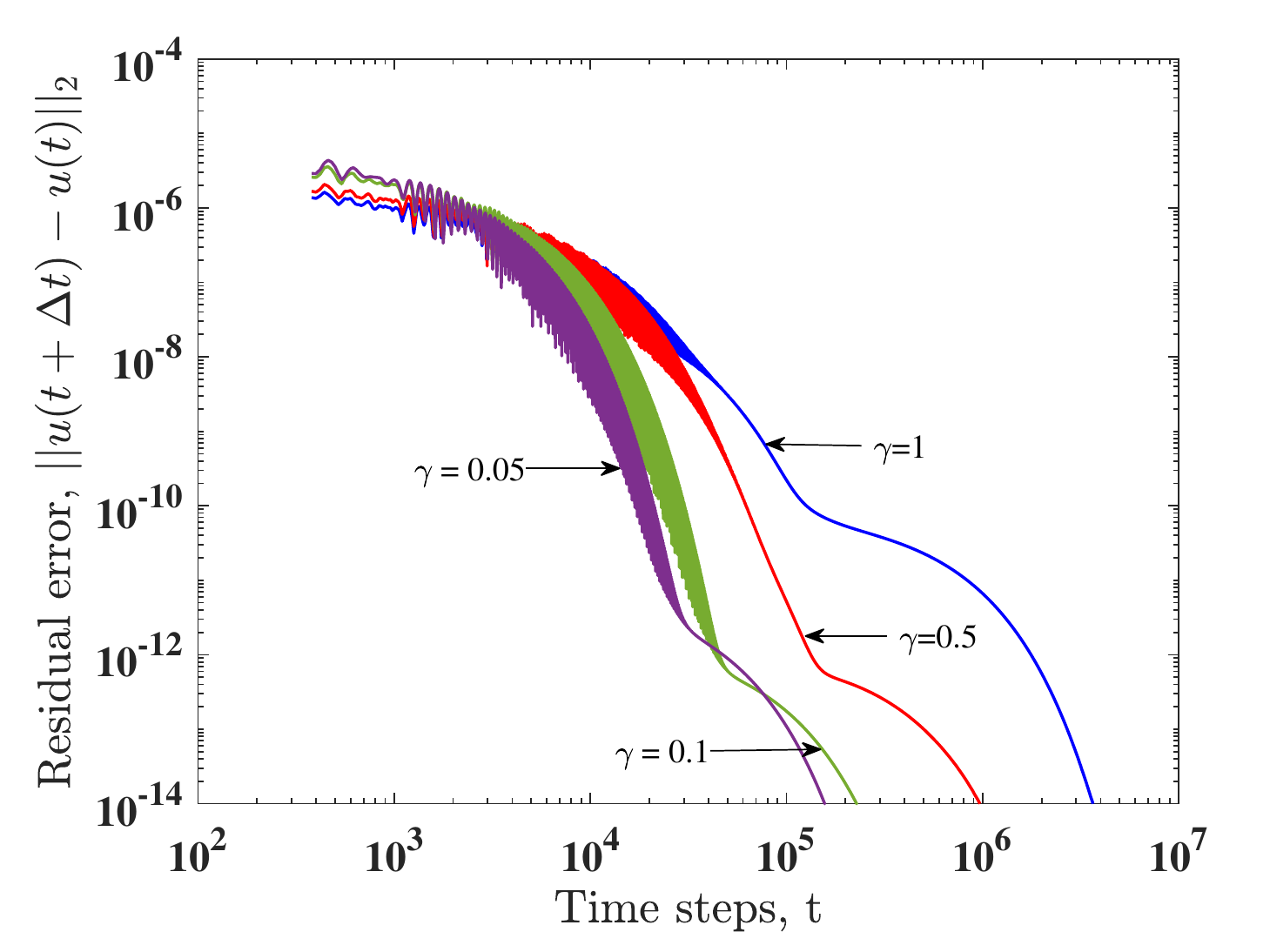}
        \label{vdeephistMa02} } \\
                \advance\leftskip0cm
    \caption{Convergence histories to the steady state for simulations of flows within a deep rectangular cavity of aspect ratio~\mbox{AR}$=2$ at a Reynolds number~\mbox{Re}$=100$ using the PRC-LBM with a grid resolution of $100\times 125$ using a rectangular lattice of grid aspect ratio $r=1.6$ with different values of the preconditioned parameter $\gamma$ at (a) Mach number~\mbox{Ma}$=0.06$, and (b) Mach number~\mbox{Ma}$=0.02$.}
    \label{fig:deepvelocityhist}
\end{figure}

Next, we simulate the flow inside a deep cavity ($H> L$) as shown in Fig.~\ref{fig:schematicdeepcavity} by considering $\mbox{AR}=2$ at $\mbox{Re}=100$ and $\mbox{Ma}=0.06$ computed using the PRC-LBM using $r=1.6$. Choosing $N_x=100$ for $r=1.6$, this leads to $N_y=125$ for the rectangular lattice case, while for the square lattice case, with $N_x=100$, we need $N_y=200$, i.e., more number of grid nodes in the $y$ direction. Results shown in Fig.~\ref{fig:deepvelocityprofile} present comparisons of the centerline profiles of the components of the velocity across the deep cavity computed using the PRC-LBM with $\gamma=1.0, 0.5, 0.1$ and $0.05$ using $100\times 125$ rectangular grids with $r=1.6$ against those based on the square lattice ($r=1$) with $100\times 200$ grid nodes and $\gamma=1.0$. As in the shallow cavity case, it can be seen that the preconditioned central moment LBM for all possible choice of $\gamma$ and with fewer number of grid nodes delivers solutions that are in very good agreement with those based on the square lattice.
Moreover, the convergence histories presented in Fig.~\ref{fig:deepvelocityhist} for deep cavity flow simulations at $\mbox{AR}=2$ and \mbox{Re}$=100$ using the PRC-LBM with $r=1.6$ with various levels of preconditioning again show a significant reduction in the number of steps for convergence -- for example, by a factor of $14$ with $\gamma=0.08$ for $\mbox{Ma}=0.06$, and a factor of $23$ with $\gamma=0.05$ for $\mbox{Ma}=0.02$ when compared to the corresponding cases without preconditioning. For the latter case with $\mbox{Ma}=0.02$, it may be noted that the choice $\gamma = 0.05$ is almost at the threshold of its smallest possible value dictated by stability considerations. As a result, it may be leading to a transient effect with the convergence history temporarily crossing over that for the case $\gamma = 0.1$ when the residual error is O($10^{-12}$). Nevertheless, when the residual error drops further down to O($10^{-13}$) or smaller, as expected, the convergence histories show that the case $\gamma = 0.05$ result in faster convergence to the steady state when compared to $\gamma = 0.1$.

\begin{figure}[H]
\centering
\advance\leftskip-1.7cm
    \subfloat[$AR=2$, $\mbox{Re}=1000$, $r=1.33$] {
        \includegraphics[width=0.45\textwidth] {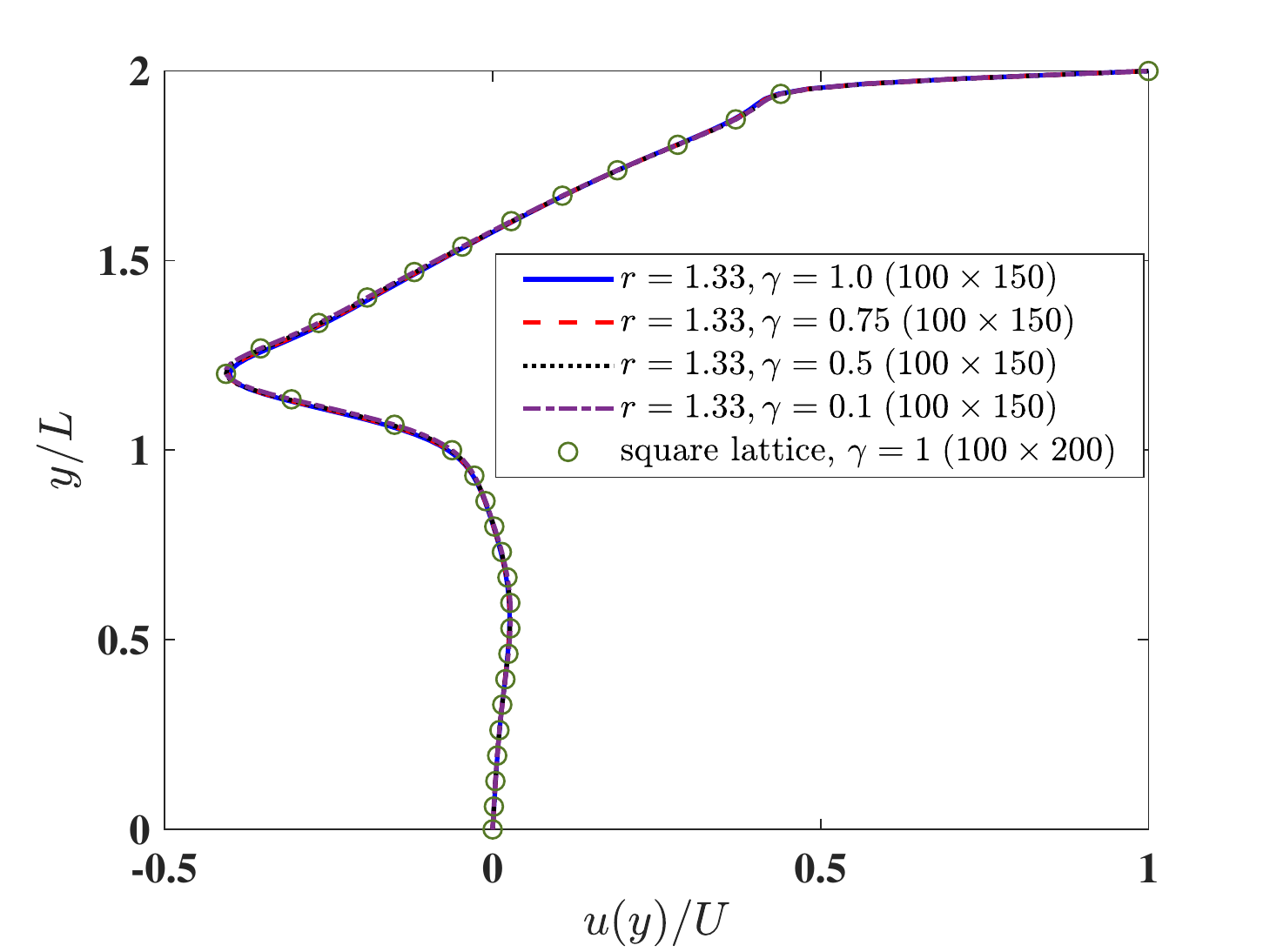}
        \label{udeepMa06Re1000} } 
    \subfloat[$AR=2$, $\mbox{Re}=1000$, $r=1.33$] {
        \includegraphics[width=0.45\textwidth] {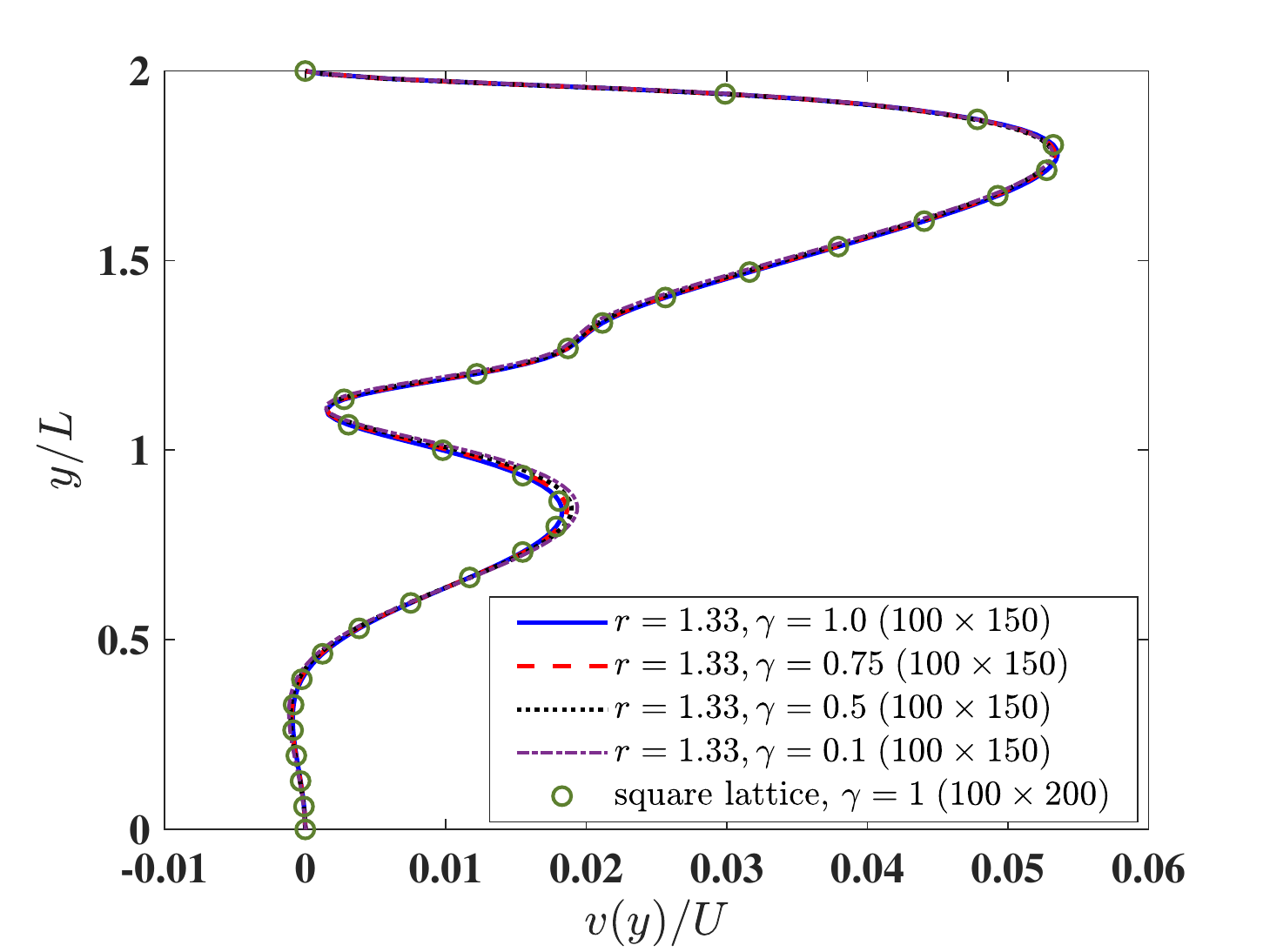}
        \label{vdeepMa06Re1000} } \\
                \advance\leftskip0cm
    \caption{The velocity profiles along the centerlines of a deep rectangular cavity of aspect ratio $\mbox{AR}=2$ at a Reynolds number~$\mbox{Re}=1000$ and Mach number~$\mbox{Ma}=0.06$ computed using the PRC-LBM with a grid resolution $100\times 150$ using the rectangular lattice of grid aspect ratio $r=1.33$ with different levels of preconditioning, i.e., $\gamma=1.0, 0.75, 0.5$ and $0.1$, and compared with the results obtained using a square lattice ($r=1.0$) with a grid resolution of $100\times 200$ at $\gamma=1.0$. (a) $u$ component of the velocity along the vertical centerline, and (b) $v$ component of the velocity along the vertical centerline.}
    \label{fig:deepvelocityRe1000}
\end{figure}

\begin{figure}[H]
\centering
\advance\leftskip-1.7cm
    \subfloat[$\mbox{Re}=100$, $AR=2$] {
        \includegraphics[width=0.45\textwidth] {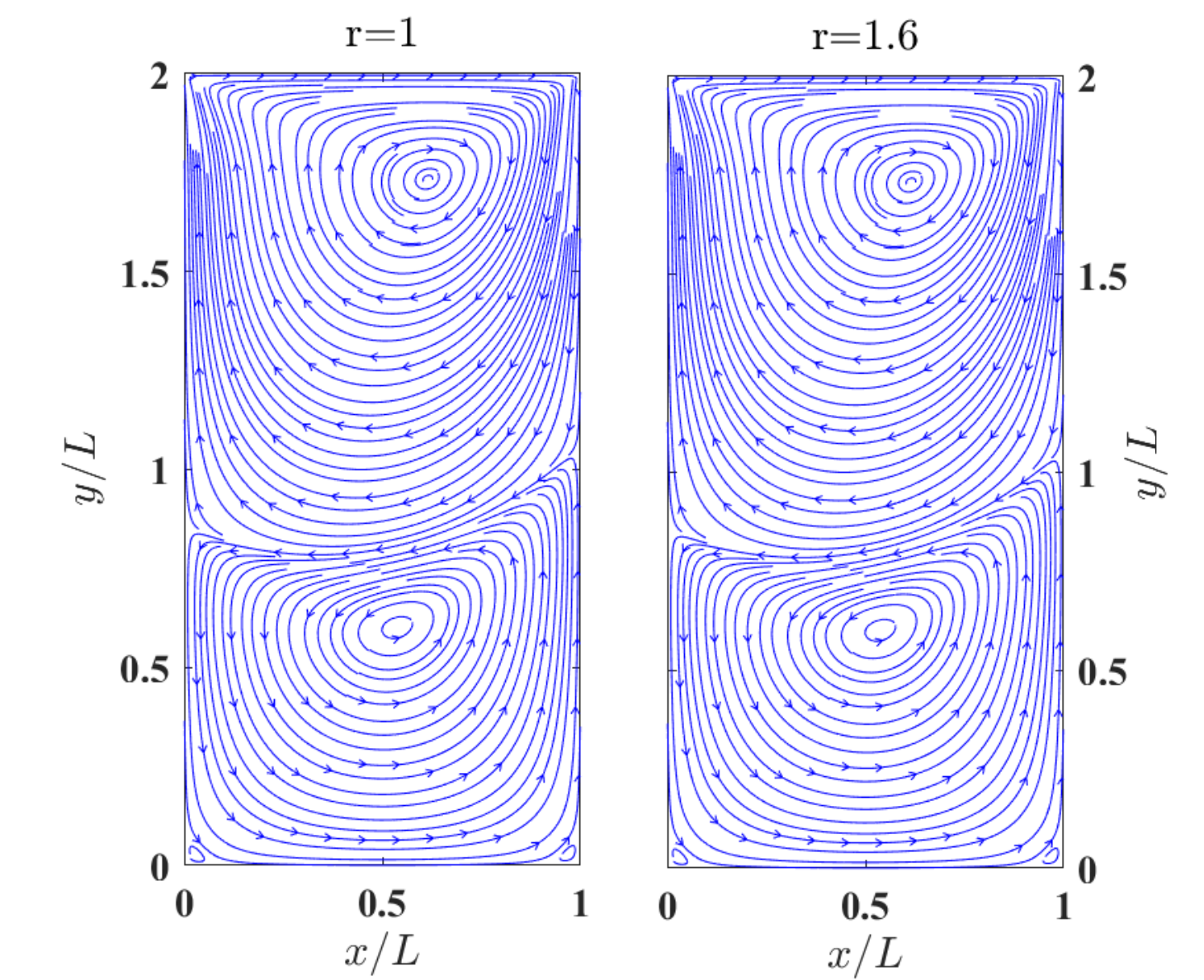}
        \label{streamlinesRe100} } 
    \subfloat[$\mbox{Re}=1000$, $AR=2$] {
        \includegraphics[width=0.45\textwidth] {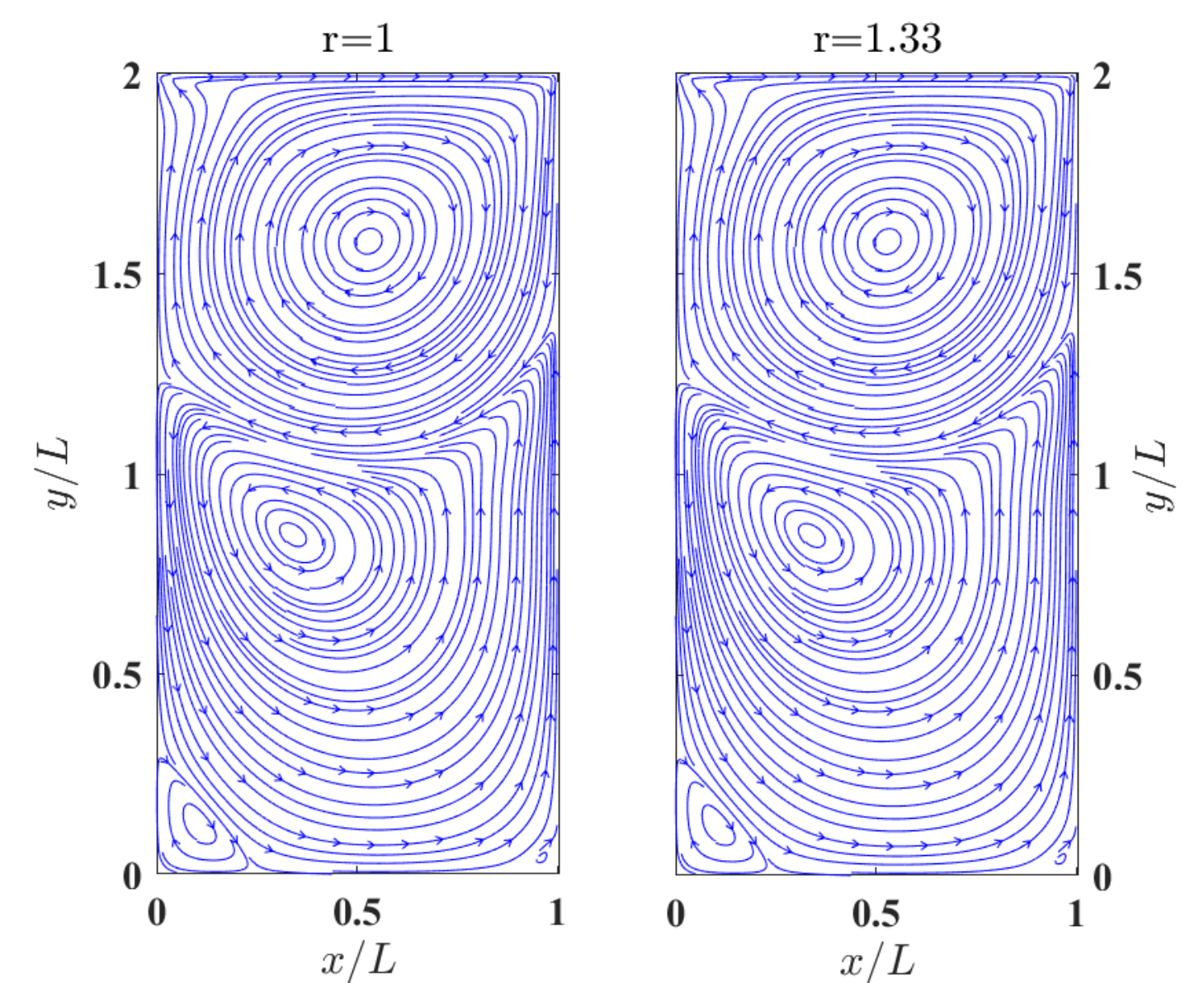}
        \label{streamsdeepRE1000} } \\
                \advance\leftskip0cm
    \caption{Streamline contours of the flow field in a 2D rectangular deep cavity of aspect ratio $\mbox{AR}=2$ computed using the PRC-LBM with $\gamma=0.1$ on a rectangular lattice at (a) \mbox{Re}=100 using $r=1.6$, and (b) \mbox{Re}=1000 using $r=1.33$ and, in each case, compared with the results of the non-preconditioned LBM using the square lattice ($\gamma=1.0$ and $r=1$).}
    \label{fig:streamlines}
\end{figure}

\begin{figure}[H]
\centering
\advance\leftskip-1.7cm
 \includegraphics[width=0.45\textwidth] {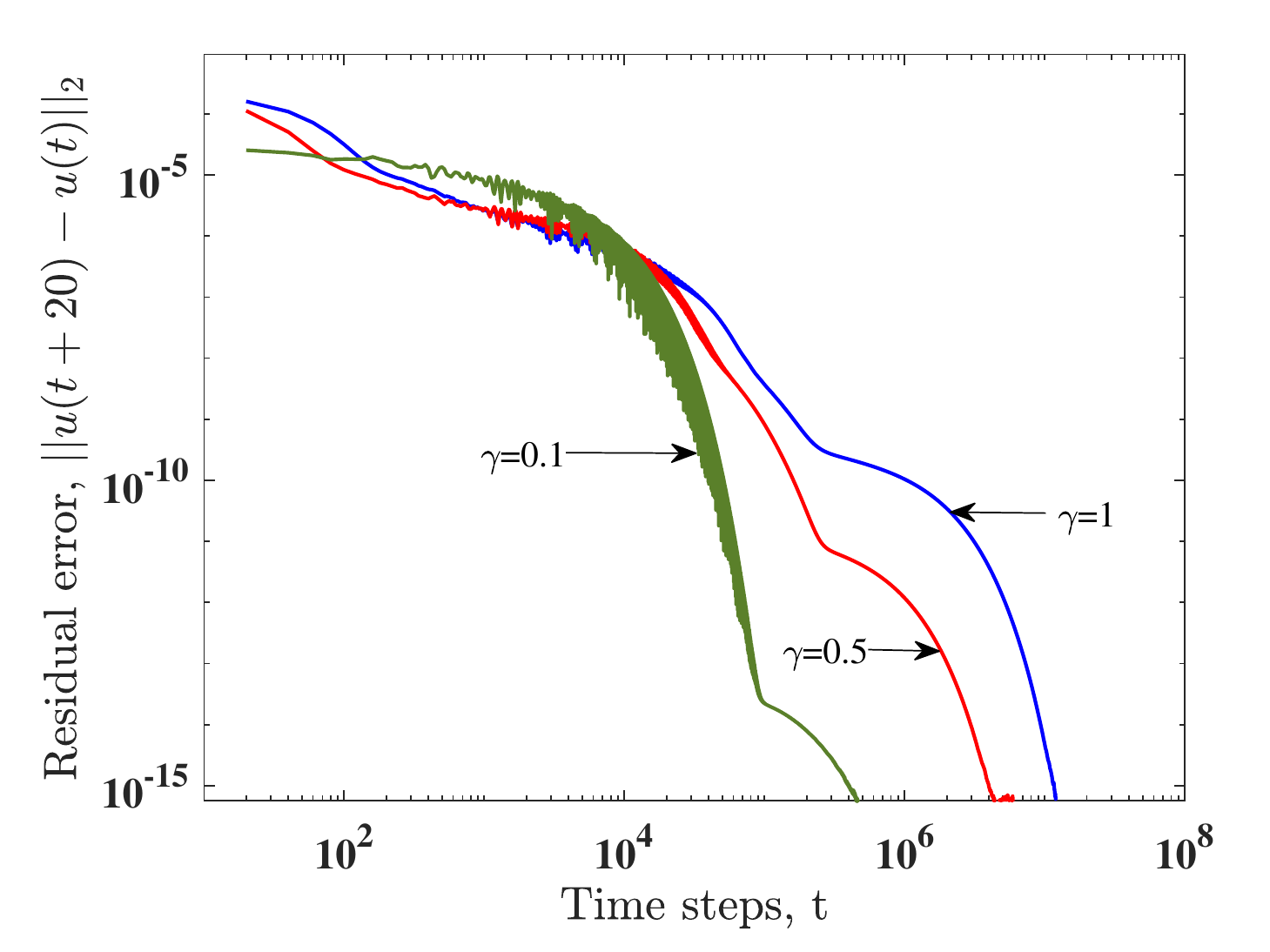}
        \label{vdeepMa02}  \\
                \advance\leftskip0cm
    \caption{Convergence histories to the steady states for simulations of flows within a deep rectangular cavity of aspect ratio $\mbox{AR}=2$ at a Reynolds number~$\mbox{Re}=1000$ using the PRC-LBM with a grid resolution of $100\times 150$ using a rectangular lattice of grid aspect ratio $r=1.33$ with different values of the preconditioned parameter $\gamma$ at Mach number~$\mbox{Ma}=0.06$.}
    \label{fig:vhistdeepRe1000}
\end{figure}

As a last case study, we perform an investigation on the efficacy of the PRC-LBM for simulations of flows inside a deep cavity ($\mbox{AR}=2$) at a higher Reynolds number of $\mbox{Re}=1000$ at $\mbox{Ma}=0.06$ using a grid resolution of $N_x\times N_y=100\times 150$ corresponding to the grid aspect ratio of $r=1.33$. The results of the centerline velocity component profiles obtained using the PRC-LBM at $\gamma=1.0, 0.75, 0.5$ and $0.1$ are reported in Fig.~\ref{fig:deepvelocityRe1000} and compared against the square lattice results using $100\times 200$ grid nodes at $\gamma=1.0$. Evidently, the rectangular LB scheme with all possible choices of $\gamma$ yields solutions that are again in very good agreement with those based on the square lattice. Moreover, these results are further corroborated by the plots of the streamline contours at two Reynolds numbers~$\mbox{Re}=100$ and $\mbox{Re}=1000$ simulated using the square lattice ($r=1$) and the rectangular lattice ($r=1.6$ for $\mbox{Re}=100$ and $r=1.33$ for $\mbox{Re}=1000$) and presented in Fig.~\ref{fig:streamlines}, which show the ability of the PRC-LBM to compute the flow patterns accurately.
Finally, the convergence histories presented in Fig.~\ref{fig:vhistdeepRe1000} for deep cavity flow simulations using a rectangular lattice grid ($r=2$) at $\mbox{Re}=1000$ and $\mbox{Ma}=0.06$ show that it takes about $11,154,000$ steps to reach the steady state without preconditioning, while the PRC-LBM with preconditioning ($\gamma=0.1$) requires significantly fewer steps of about $583,000$ to reach similar residual errors, delivering an improvement by a factor of about $19$. In general, the higher $\mbox{Re}=1000$ case takes longer to ready steady state when compared to the case at lower $\mbox{Re}=100$ (shown in Fig.~\ref{fig:deepvelocityhist}) since the former is set up by reducing the fluid viscosity when compared to the latter, resulting in a slower diffusion of momentum and its convergence. However, thanks to preconditioning, even at higher $\mbox{Re}$, the rectangular central moment LBM is able to achieve substantial savings in the overall computational effort.

%

\section{Comparisons between Preconditioned Rectangular LB Formulations based on Raw Moments and Central Moments} \label{sec:comparisonsbetweenmodels}
As noted in the introduction, no other formulation other than the current work that combines both preconditioning and rectangular lattice grids are available in the literature. Nevertheless, it should be noted that the derivation presented in Sec.~\ref{sec:3} can be utilized to construct a preconditioned rectangular \emph{raw} moment LBM (referred to as the PRNR-LBM in what follows) by performing the collision step in terms of relaxations involving the raw moments using the corrections to the equilibria based on $\gamma$ and $r$ given in Sec.~ \ref{subsec:correctionsextendemomentequilibria}. The implementation of this strategy is summarized in Eq.~(\ref{eq:LBErawmomentrectangularlattice}). As such this PRNR-LBM can be regarded as a special case of the PRC-LBM based on central moments. Here, we note that the single-relaxation-time (SRT) formulations are well-known to have serious deficiencies when compared other collision models. For example, they are significantly less stable in simulating flows at relatively low viscosities or large Reynolds numbers, even when compared to the approaches based on raw moments, and it is quite cumbersome to combine the use of both preconditioning and rectangular lattice grids. Given these issues, they are not given further considerations in this work, and we will focus on attention making comparisons between the PRC-LBM and PRNR-LBM in what follows.

\begin{figure}[H]
\centering
\advance\leftskip-1.7cm
 \includegraphics[width=0.45\textwidth] {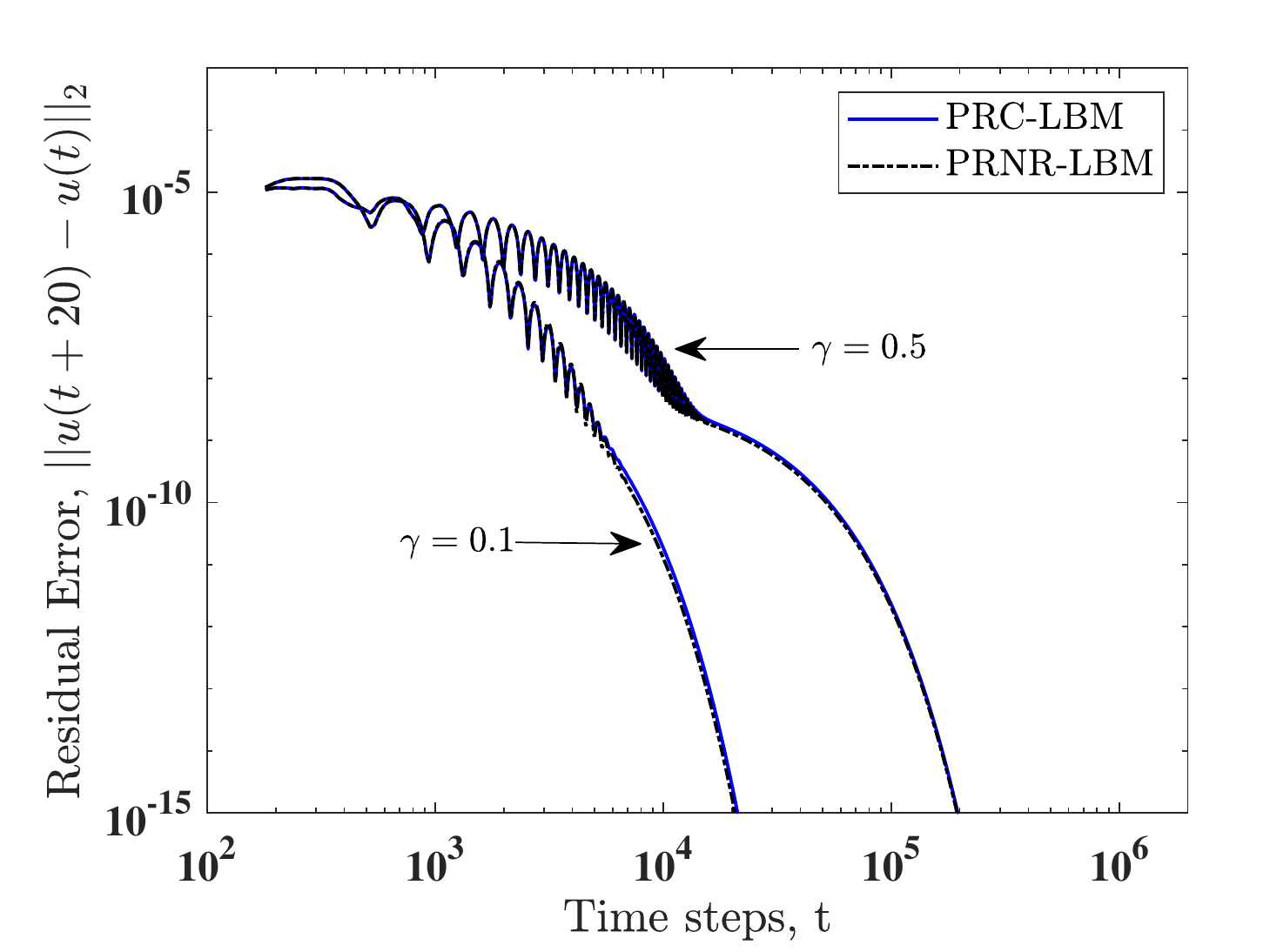}
        \label{vdeepMa02}  \\
                \advance\leftskip0cm
    \caption{Convergence histories to the steady states for simulations of flows within a shallow rectangular cavity of aspect ratio $\mbox{AR}=0.25$ at a Reynolds number~$\mbox{Re}=100$ using the PRC-LBM and PRNR-LBM with a grid resolution of $100\times 50$ using a rectangular lattice of grid aspect ratio $r=0.5$ with two different values of the preconditioned parameter $\gamma$ at Mach number~$\mbox{Ma}=0.06$.}
    \label{fig:vhistPRcPRNR}
\end{figure}

First, we point out that provided a given collision model yields numerically stable flow simulations, the convergence rate to the steady state is primarily influenced by the choice of the preconditioning parameter, and the use of two different collision models under otherwise similar conditions, such as the same choice of model parameters, is expected to result in a similar convergence speed. Figure~\ref{fig:vhistPRcPRNR} presents convergence histories for the PRC-LBM and PRNR-LBM for the simulations of flow with a shallow cavity with aspect ratio $\mbox{AR}=0.25$, $r=0.5$, $\mbox{Re}=100$, and $\mbox{Ma}=0.06$ for two different values of $\gamma$ (equal to $0.1$ and $0.5$). For these choices, both the collision models result in the same convergence rate to the steady state for a fixed preconditioning parameter.

\begin{figure}[H]
\centering
\advance\leftskip-1.7cm
    \subfloat[$N_x \times N_y=100 \times 100$, $AR=0.25$] {
        \includegraphics[width=0.45\textwidth] {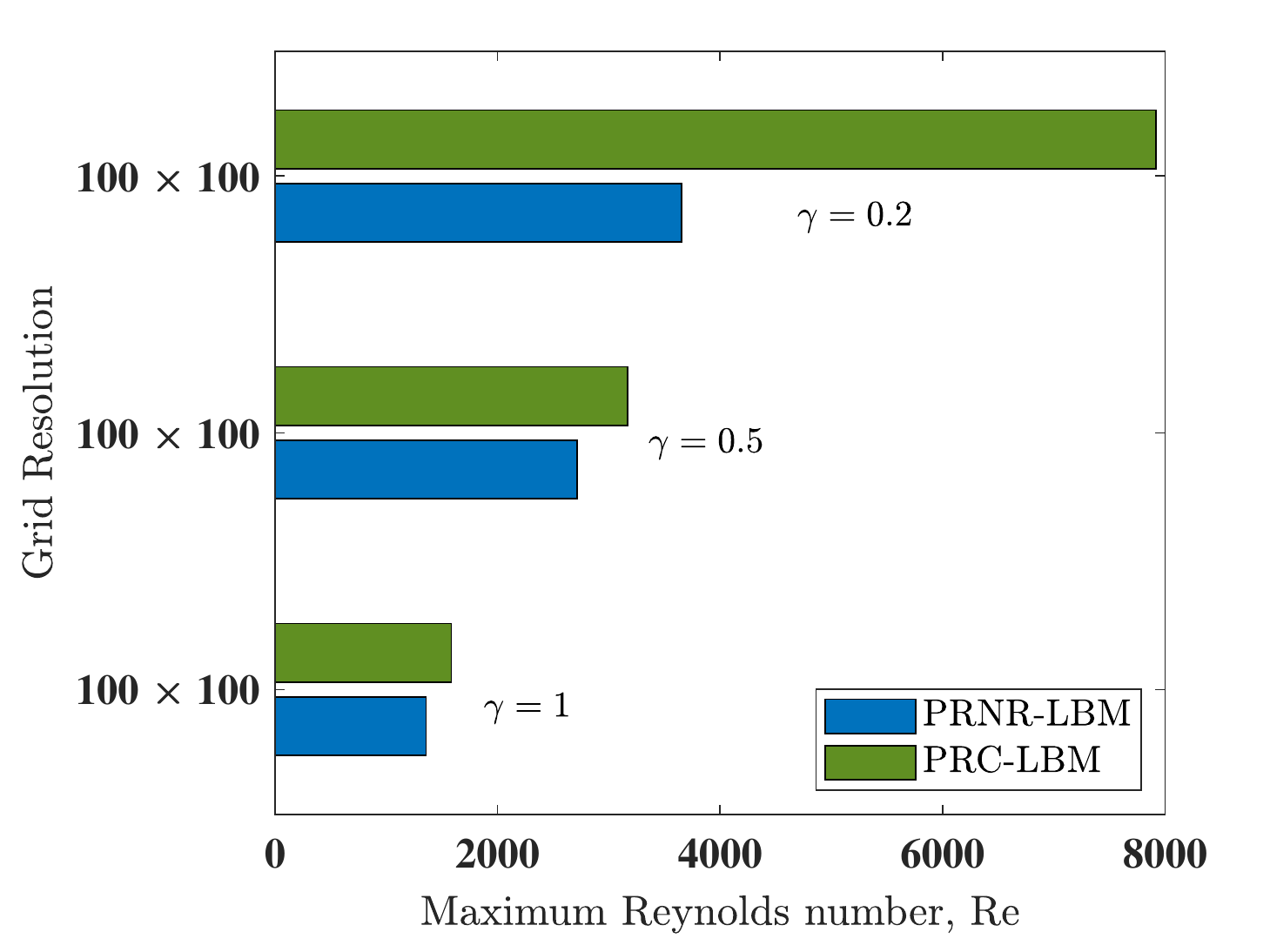}
        \label{numericalstabilitytest-lowresolution} } 
    \subfloat[$N_x \times N_y=200 \times 200$, $AR=0.25$] {
        \includegraphics[width=0.45\textwidth] {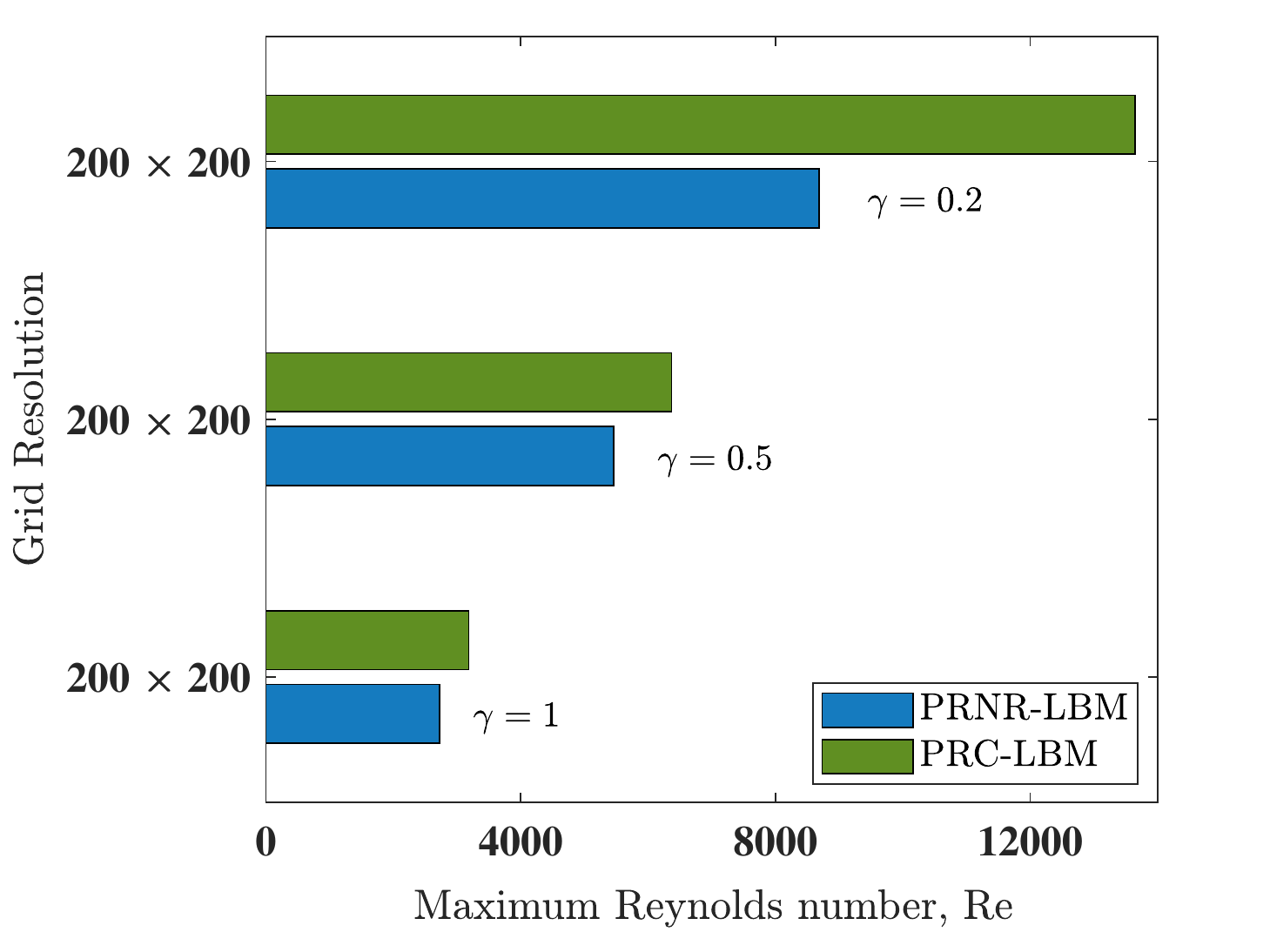}
        \label{numericalstabilitytest-highresolution} } \\
                \advance\leftskip0cm
   \caption{Comparison of the maximum Reynolds number $\mbox{Re}$ for numerical stability of PRC-LBM  and PRNR-LBM for simulations of flows within a shallow rectangular cavity of aspect ratio $\mbox{AR}=0.25$ with a fixed lid-velocity of $0.02$ at different mesh resolutions with a grid aspect ratio of $r=0.25$ for three different choices of the preconditioning parameter ($\gamma = 1.0$, $0.5$ and $0.2$).}
    \label{fig:numericalstabilitytest}
\end{figure}

On the other hand, we will now demonstrate the advantage of performing the collision step in a frame of reference based on the local fluid velocity in the PRC-LBM when compared to using the PRNR-LBM involving the rest or the lattice frame of reference is related to improving robustness or numerical stability of computations. In this regard, we investigate the maximum Reynolds number achieved by PRC-LBM and PRNR-LBM for simulating flows within a shallow cavity with an aspect ratio of  $\mbox{AR}=0.25$ with a rectangular grid using $r=0.25$ by maintaining the lid velocity at a relative small constant value of $U=0.02$ and reducing the shear viscosity of the fluid to the smallest possible value for which the computations remain numerically stable. The shear viscosity is varied by changing the relaxation parameters associated with the second order moments and the relaxation parameters for the higher order moments are set to unity for simplicity for both the collision models. Two different grid resolutions of $100\times100$, $200\times200$ are considered, and in each case, three different choices of the preconditioning parameter $\gamma = 1.0$, $0.5$ and $0.2$ are used. The results of these stability tests are presented in Fig.~\ref{fig:numericalstabilitytest}. Clearly, even for the relatively low lid velocity considered, the preconditioned rectangular LBM using central moments is consistently more stable when compared to the preconditioned rectangular LBM using raw moments, especially at smaller $\gamma$. These results extend those presented in Ref.~ \cite{yahia2021central} by including preconditioning effects. Further improvements in numerical stability are possible when simulating shear flows with larger characteristic velocities. Moreover, as emphasized by various studies involving central moments in LBM (see e.g.,~\cite{ hajabdollahi2019cascaded, hajabdollahi2021central}), the additional computational overhead of using central moments when compared to that of using raw moments is relatively small, by about 15\%, but with the benefit of stable simulations at significantly higher Reynolds numbers using the former when compared to the latter.

\section{Summary and Conclusions} \label{sec:6}
In this paper, we have developed a new LB algorithm based on central moments and using involving a preconditioning strategy and a rectangular lattice grid, viz., the PRC-LBM, for efficient simulations of inhomogeneous and anisotropic flows. By including a preconditioning parameter $\gamma$ in its moment equilibria and augmenting its second order components via corrections that eliminate the anisotropy effects associated with the rectangular lattice characterized by the grid aspect ratio $r$ and the non-Galilean invariant velocity terms due to the aliasing effects on the D2Q9 lattice, it can consistently represent the preconditioned Navier-Stokes equations. Such corrections to the equilibria, obtained via a Chapman-Enskog analysis, are related to the diagonal components of the velocity gradient tensor, which are expressed in terms of the non-equilibrium moments to facilitate local computations and their coefficients simultaneously depend on both $\gamma$ and $r$. In the construction of the PRC-LBM, unlike the prior rectangular LB formulations, we have used the natural, non-orthogonal moment basis with a physically consistent parametrization of the speed of sound based on $r$, and with a $\gamma$-adjusted equilibria obtained through a matching principle based on the continuous Maxwell distribution. These result in simpler correction terms for using rectangular lattice grids in conjunction with preconditioning, and which, while used here in terms of central moments for robustness, have general applicability for other collision models. Moreover, our algorithmic implementation is modular in nature with a clear interpretation based on special matrices, which also naturally extends to three dimensions using a cuboid lattice in solving the preconditioned Navier-Stokes equations. The PRC-LBM simulations of benchmark shear-driven flows within shallow and deep cavities with significant geometric anisotropy at various Reynolds numbers and Mach numbers and with different values of $r$ and $\gamma$, validate the method for accuracy and show improvements in stability when compared to another implementation based on raw moments. Furthermore, we demonstrate significant reductions in the computational cost with the use of PRC-LBM via convergence acceleration to the steady states and reduced memory storage when compared to the corresponding LB scheme using the square lattice and without involving preconditioning.

As final concluding remarks, we note the following. Given the ubiquity of inhomogeneous and anisotropic flows, classical schemes for CFD invariably use stretched grids that adapt to the local flow conditions and rarely utilize uniform square or cubic grids. Thus, the use of square/cubic lattices in LB algorithms is far from optimal in the use of overall computational resources in performing such flow simulations efficiently. Previous efforts in developing LB schemes based on rectangular lattices (e.g.,~\cite{bouzidi2001lattice, zhou2012mrt, peng2016lattice,peng2016hydrodynamically,zong2016designing}), which were generally based on orthogonal moment basis and non-optimal equilibria, were significantly restrictive in terms of complexity of implementation, limited stability ranges and cumbersome approach involved in choosing the various model parameters. All these issues have been circumvented in our present rectangular LB formulation involving a non-orthogonal moment basis and the construction of equilibria based on a matching principle. It is modular in construction in the sense that an LB code for the square lattice can be readily extended to the rectangular lattice with some minor efforts based on the former by making few simple changes. As shown in the appendix on the algorithmic implementation, such changes include the use of pre-collision and post-collision grid aspect ratio-based scalings of the raw moments and the use of extended second order moment equilibria adjusted suitably based on the grid aspect ratio to recover the Navier-Stokes equations. These simple additional changes to the existing square lattice-based LB codes to accommodate the rectangular grids have been numerically demonstrated to result in savings by an order of magnitude or more in terms of computational cost and memory when compared to that based on the square lattice (consistent with Refs.~\cite{yahia2021central, yahia2021three}). Furthermore, the use of preconditioning with rectangular lattice grids resulted in significant convergence acceleration to the steady states leading to further reduction in the overall computational efforts. Hence, the PRC-LBM represents an efficient approach for flow simulations.

\section*{Acknowledgements}
The first author (EY) thanks the Department of Mechanical Engineering at the University of Colorado Denver for financial support. The second author would like to acknowledge the support of the US National Science Foundation (NSF) under Grant CBET-1705630. The second author (KNP) would like to also thank the NSF for support of the development of a computer cluster infrastructure through the project ``CC* Compute: Accelerating Science and Education by Campus and Grid Computing'' under Award 2019089.

\appendix

\section{Algorithmic Details of the PRC-LBM}\label{sec:algorithmic-details-PRC-LBM}
The PRC-LB algorithm consists of the following sub-steps during a time step $\Delta t$, where the distribution function $f_\alpha(\bm{x},t+\Delta t)$ is updated from $f_\alpha(\bm{x},t)$ for $\alpha = 0, 1, 2,\ldots, 8$ on a rectangular lattice grid according to Eq.~\eqref{eq:LBEcentralmomentrectangularlattice} to simulate the preconditioned NS equations:

\begin{itemize}
\item Compute pre-collision raw moments
\newline
We perform
\begin{equation*}
\mathbf{m}=\tensor{P}\mathbf{f}
\end{equation*}
on the components of $\mathbf{f}(\bm{x},t)$, so that the pre-collision raw moment components $k_{mn}^\prime$ of $\mathbf{m}$  for the usual square lattice are obtained, where the mapping $\tensor{P}$ follows from Eq.~\eqref{eq:P-momentbasis}. In other words,
\begin{eqnarray}\label{eq:62}
  \Kps{00} &=& f_0 +f_1 +f_2 + f_3 +f_4 +f_5 +f_6 +f_7 +f_8, \nonumber\\
  \Kps{10} &=& f_1  -f_3  +f_5 -f_6 -f_7 +f_8, \nonumber\\
  \Kps{01} &=& f_2  -f_4  +f_5 +f_6 -f_7 -f_8, \nonumber\\
  \Kps{20} &=& f_1  +f_3  +f_5 +f_6 +f_7 +f_8, \nonumber\\
  \Kps{02} &=& f_2  +f_4  +f_5 +f_6 +f_7 +f_8, \nonumber\\
  \Kps{11} &=& f_5 -f_6 +f_7 -f_8, \nonumber\\
  \Kps{21} &=& f_5 +f_6 -f_7 -f_8, \nonumber\\
  \Kps{12} &=& f_5 -f_6 -f_7 +f_8, \nonumber\\
  \Kps{22} &=& f_5 +f_6 +f_7 +f_8.\nonumber
\end{eqnarray}

\item Scale pre-collision raw moments for rectangular lattice
\newline
Using  Eq.~\eqref{eq:S-matrix} for the scaling matrix $\tensor{S}$, implement
\begin{equation*}
\mathbf{m}\leftarrow\tensor{S}\mathbf{m},
\end{equation*}
which yields the pre-collision raw moments for the \emph{rectangular} lattice $k_{mn}^\prime$ from those computed for the square lattice above. Thus,
\begin{align*}
&\Kps{00}=\Kps{00}, \quad \Kps{10} =\Kps{10} ,\quad \Kps{01}= r  \Kps{01}, \nonumber\\
&\Kps{20}= \Kps{20},\quad \Kps{02}=r^2 \Kps{02},\quad
\Kps{11}=r \Kps{11},\quad \Kps{21}=r \Kps{21},\quad \Kps{12}=r^2 \Kps{12},\quad \Kps{22}=r^2 \Kps{22},
\end{align*}
which involve a scale factor of $r^n$ for the moment $k_{mn}^\prime$.

\item Compute pre-collision central moments
\newline
Transform the pre-collision raw moments $\Kps{mn}$ in $\mathbf{m}$ into pre-collision central moments $k_{mn}$ in $\mathbf{m}^c$ via
\begin{equation*}
\mathbf{m}^c=\tensor{F}\mathbf{m},
\end{equation*}
where $\tensor{F}$ is provided in Eq.~\eqref{eq:Fmatrix}. Thus, we get
\begin{eqnarray*}\label{eq:64}
  k_{00} &=& k_{00}^\prime, \nonumber\\
  k_{10} &=& k_{10}^\prime - u_x  k_{00}^\prime , \nonumber\\
  k_{01} &=& k_{01}^\prime - u_y  k_{00}^\prime, \nonumber\\
  k_{20} &=& k_{20}^\prime - 2 u_x  k_{10}^\prime + u_x^2 k_{00}^\prime , \nonumber\\
  k_{02} &=& k_{02}^\prime - 2 u_y k_{01}^\prime + u_y^2 k_{00}^\prime, \nonumber\\
  k_{11} &=& k_{11}^\prime - u_y  k_{10}^\prime - u_x k_{01}^\prime+ u_x u_y k_{00}^\prime , \nonumber\\
  k_{21} &=& k_{21}^\prime - 2 u_x k_{11}^\prime + u_x^2 k_{01}^\prime- u_y k_{20}^\prime + 2 u_x u_y k_{10}^\prime - u_x^2 u_y k_{00}^\prime , \nonumber\\
  k_{12} &=& k_{12}^\prime - 2 u_y k_{11}^\prime + u_y^2 k_{10}^\prime- u_x k_{02}^\prime + 2 u_x u_y k_{01}^\prime - u_x u_y^2 k_{00}^\prime , \nonumber\\
  k_{22} &=& k_{22}^\prime - 2 u_x k_{12}^\prime + u_x^2 k_{02}^\prime- 2 u_y k_{21}^\prime + 4 u_x u_y k_{11}^\prime - 2 u_x^2 u_y k_{01}^\prime + u_y^2 k_{20}^\prime - 2 u_x u_y^2 k_{10}^\prime + u_x^2 u_y^2 k_{00}^\prime. \nonumber\\
\end{eqnarray*}

\item Compute post-collision central moments: Relaxation under collision using preconditioned extended equilibria and source terms, and corrections
\newline
Perform the relaxations of central moments to their preconditioned equilibria, including the source terms for the body force, using the different relaxation rates given in $\tensor{\Lambda}$, and with corrections to eliminate the grid-anisotropy and non-GI truncation errors parameterized by the grid aspect ratio $r$ and the preconditioning parameter $\gamma$ reflecting the sub-step
\begin{equation*}
\tilde{\mathbf{m}}^c=\mathbf{m}^c + \tensor{B}^{-1}\left\{\tensor{\Lambda}\;\left(\; \tensor{B}\mathbf{m}^{c,eq}-\tensor{B}\mathbf{m}^c \;\right) + \left(\tensor{I} - \frac{\tensor{\Lambda}}{2}\right)  \tensor{B}\mathbf{\Phi}^c\Delta t\right\},
\end{equation*}
where combining certain moments for their independent evolutions under collision and their subsequent segregation following collision are formally shown via the operators $\tensor{B}$ and $\tensor{B}^{-1}$, respectively.

Thus, for the D2Q9 lattice, applying the operator $\tensor{B}$ implies combining the second order diagonal components of moments as
\begin{equation*}\label{eq:combinedmoments}
k_{2s} = \left(k_{20} + k_{02}\right) , \quad k_{2d} = \left(k_{20} - k_{02}\right),
\end{equation*}
which will relax to their equilibria at their own relaxation rates in what follows rather than with using individual components in this regard. From Sec.~\eqref{sec:2}, the preconditioned central moment equilibria for the rectangular D2Q9 lattice, including the necessary correction terms, can be written as
\begin{eqnarray*}\label{eq:centralmomentequilibriarectangularlattice}
k_{00}^{eq} &=& \rho, \quad\quad\quad\quad k_{10}^{eq}=0,\quad\quad\quad\quad k_{01}^{eq}=0,\\
k_{2s}^{eq} &=& k_{20}^{eq}+ k_{02}^{eq} = 2q^2 c_{s*}^2\rho +\left( \frac{1}{\gamma}-1 \right) \rho \left( u_x^2 + u_y^2 \right)+ \left( {\theta}_{bx}\partial_x u_x+ {\theta}_{by} \partial_y u_y + {\lambda}_{bx} \partial_x \rho  + {\lambda}_{by} \partial_y \rho \right) \Delta t,\\
k_{2d}^{eq} &=& k_{20}^{eq}- k_{02}^{eq} = \;\;\qquad\quad\quad\left( \frac{1}{\gamma}-1 \right) \rho \left( u_x^2 - u_y^2 \right)+ \left( \theta_{sx}\partial_x u_x - \theta_{sy} \partial_y u_y + \lambda_{sx} \partial_x \rho  + \lambda_{sy} \partial_y \rho \right) \Delta t,\\
k_{11}^{eq}&=&\left( \frac{1}{\gamma}-1 \right) \rho u_x u_y + \left( \tilde{{\theta}}_{sx} \partial_x u_x + \tilde{\theta}_{sy} \partial_y u_y + \tilde{\lambda}_{sx} \partial_x \rho  + \tilde{\lambda}_{sy} \partial_y \rho \right) \Delta t,\\
k_{21}^{eq}&=&\left( \dfrac{1}{\gamma^2}-\dfrac{3}{\gamma}+2 \right) \rho u_x^2 u_y,\qquad k_{12}^{eq}=\left( \dfrac{1}{\gamma^2}-\dfrac{3}{\gamma}+2 \right) \rho u_x u_y^2,\qquad k_{22}^{eq}= q^4 c_{s*}^4\rho.
\end{eqnarray*}
Here, and in the following, for better clarity, we use subscript `b' for the coefficients associated with the corrections for the trace of the diagonal components of the second order moments (for bulk viscosity $\xi$), and the subscript `s' for the corrections for those other moments related to the shear viscosity $\nu$, rather than the numerical subscripts used in Sec.~\ref{sec:2}. Then, rewriting Eqs.~\eqref{eq:38},~\eqref{eq:39}, and~\eqref{eq:40}, respectively, the coefficients appearing the above extended moment equilibria related to the corrections can be written as
\begin{align*}
&\theta_{bx}= -\Big[ P_\gamma u_x^2 +Q_\gamma u_y^2 +  \left(3 q^2 c_{s*}^2- 1\right) \Big]\rho \left( \frac{1}{\omega_{\xi}}- \frac{1}{2}\right),\\
&\theta_{by}= -\Big[ P_\gamma u_y^2 +Q_\gamma u_x^2 + \left(3 q^2 c_{s*}^2- r^2\right) \Big]\rho \left(\frac{1}{\omega_{\xi}}- \frac{1}{2}\right) ,\\
&\lambda_{bx}= - \left[U_\gamma q^2 c_{s*}^2 -1 \right] \left(\frac{1}{\omega_{\xi}}- \frac{1}{2}\right) u_x,\\
&\lambda_{3y}= - \left[U_\gamma q^2 c_{s*}^2 -r^2 \right] \left(\frac{1}{\omega_{\xi}}- \frac{1}{2}\right) u_y,
\end{align*}
\begin{align*}
&\theta_{sx}= -\Big[ P_\gamma u_x^2 - Q_\gamma u_y^2 +  \left(3 q^2 c_{s*}^2- 1\right) \Big]\rho \left( \frac{1}{\omega_{\nu}}- \frac{1}{2}\right),\\
&\theta_{sy}= +\Big[- P_\gamma u_y^2 + Q_\gamma u_x^2 - \left(3 q^2 c_{s*}^2- r^2\right) \Big]\rho \left(\frac{1}{\omega_{\nu}}- \frac{1}{2}\right) ,\\
&\lambda_{sx}= - \left[U_\gamma q^2 c_{s*}^2 -1 \right]  \left(\frac{1}{\omega_{\nu}}- \frac{1}{2}\right) u_x,\\
&\lambda_{3y}= + \left[U_\gamma q^2 c_{s*}^2 -r^2 \right] \left(\frac{1}{\omega_{\nu}}- \frac{1}{2}\right) u_y,
\end{align*}
and
\begin{align*}
&\tilde{\theta}_{sx}= -\left(\frac{1}{\gamma^2}-\frac{1}{\gamma} \right) \left( \frac{1}{\omega_{\nu}}- \frac{1}{2}\right)\rho u_x u_y ,\\
&\tilde{\theta}_{sy}= -\left(\frac{1}{\gamma^2}-\frac{1}{\gamma} \right) \left( \frac{1}{\omega_{\nu}}- \frac{1}{2}\right)\rho u_x u_y ,\\
&\tilde{\lambda}_{sx}= - \left(\frac{1}{\gamma}-1 \right) \left(\frac{1}{\omega_{\nu}}- \frac{1}{2}\right) q^2 c_{s*}^2 u_y,\\
&\tilde{\lambda}_{sy}= - \left(\frac{1}{\gamma}-1 \right) \left(\frac{1}{\omega_{\nu}}- \frac{1}{2}\right) q^2 c_{s*}^2 u_x,
\end{align*}
where
\begin{equation*}
P_\gamma =\frac{4}{\gamma^2}-\frac{1}{\gamma}, \qquad Q_\gamma =\frac{1}{\gamma^2}-\frac{1}{\gamma}, \qquad U_\gamma = \frac{2}{\gamma}+1,
\end{equation*}
and the relaxation parameters $\omega_\xi$ and $\omega_\nu$ determine the bulk and shear viscosities, respectively, which are shown at the end of this sub-step. The spatial derivatives of the density $\partial_x \rho$ and $\partial_y \rho$ appearing in the above moment correction terms are obtained from an isotropic second order finite difference scheme, while the spatial derivatives of the velocity field $\partial_x u_x$ and $\partial_y u_y$ in such terms are computed locally using non-equilibrium moments, by rewriting Eqs.~\eqref{eq:43},~\eqref{eq:45},~\eqref{eq:46}, and~\eqref{eq:48}, as follows. First, writing
\begin{eqnarray*}
 &&A= \cfrac{1}{2} \left(U_\gamma q^2 c_{s*}^2 - 1 \right) u_x,\qquad B= \cfrac{1}{2}\left(U_\gamma q^2 c_{s*}^2 - r^2 \right) u_y \\
 &&e_{b\rho}= -A \partial_x \rho - B \partial_y \rho, \qquad e_{s\rho}= -A \partial_x \rho + B \partial_y \rho,
\end{eqnarray*}
and subsequently defining the following intermediate quantities
\begin{equation*}
R_{2s} = \left({k_{20}} + {k_{02}}\right)- 2 q^2 c_{s*}^2 \rho +e_{b\rho}, \qquad R_{2d} = \left({k_{20}} - {k_{02}}\right)- e_{s\rho},
\end{equation*}
and
\begin{eqnarray*}
  C_{bx} &=& \left[-\frac{2 q^2 c_{s*}^2}{\omega_{\xi}} + \frac{P_\gamma u_x^2 + Q_\gamma u_y^2}{2} + \frac{(3 q^2 c_{s*}^2-1)}{2} \right]\rho,\quad C_{by} = \left[-\frac{2 q^2 c_{s*}^2}{\omega_{\xi}} + \frac{P_\gamma u_y^2 + Q_\gamma u_x^2}{2} + \frac{(3 q^2 c_{s*}^2-r^2)}{2} \right]\rho,\\
  C_{sx} &=& \left[-\frac{2q^2 c_{s*}^2}{\omega_{\nu}} + \frac{P_\gamma u_x^2 - Q_\gamma u_y^2}{2} + \frac{(3 q^2c_{s*}^2-1)}{2} \right]\rho,\quad C_{sy} = \left[\frac{2q^2 c_{s*}^2}{\omega_{\nu}} + \frac{-P_\gamma u_y^2 + Q_\gamma u_x^2}{2}- \frac{(3 q^2 c_{s*}^2-r^2)}{2}  \right]\rho,
\end{eqnarray*}
the required local expressions for the derivatives of the velocity field are then given by
\begin{equation*}
\partial_x u_x = \cfrac{\left( C_{sy} R_{2s} -C_{by} R_{2d}\right)}{\left( C_{bx} C_{sy} -C_{sx} C_{by}\right)}, \qquad \partial_y u_y = \cfrac{1}{C_{by}}\left(R_{2s} -C_{bx} \partial_x u_x \right).
\end{equation*}

With the above specifications, the post-collision central moments resulting from the relaxations of the various central moments to their central moment equilibria under collision and augmented by the effect of the source terms can now be written as follows:
\begin{eqnarray*}
\tilde{k}_{00} & = & k_{00},\\
\tilde{k}_{10} & = & k_{10}+\omega_1(k_{10}^{eq}-k_{10})+(1-\omega_1/2)\sigma_{10}\Delta t,\\
\tilde{k}_{01} & = & k_{01}+\omega_1(k_{01}^{eq}-k_{01})+(1-\omega_1/2)\sigma_{01}\Delta t,\\
\tilde{k}_{2s} & = & k_{2s}+\omega_\xi(k_{2s}^{eq}-k_{2s})+(1-\omega_\xi/2)\sigma_{2s}\Delta t,\\
\tilde{k}_{2d} & = & k_{2d}+\omega_\nu(k_{2d}^{eq}-k_{2d})+(1-\omega_\nu/2)\sigma_{2d}\Delta t,\\
\tilde{k}_{11} & = & k_{11}+\omega_\nu(k_{11}^{eq}-k_{11})+(1-\omega_\nu/2)\sigma_{11}\Delta t,\\
\tilde{k}_{21} & = & k_{21}+\omega_{21}(k_{21}^{eq}-k_{21}),\\
\tilde{k}_{12} & = & k_{12}+\omega_{12}(k_{12}^{eq}-k_{12}),\\
\tilde{k}_{22} & = & k_{22}+\omega_{22}(k_{22}^{eq}-k_{22}),
\end{eqnarray*}
where $\sigma_{2s}=\sigma_{20}+\sigma_{02}$ and $\sigma_{2d}=\sigma_{20}-\sigma_{02}$, and the central moments of the source terms relevant for recovering the preconditioned NS equations with a body force, $\sigma_{10}$, $\sigma_{01}$, $\sigma_{20}$, $\sigma_{02}$, and $\sigma_{11}$ are given in Eq.~\eqref{eq:13-central}. Here, the relaxation parameters $\omega_\nu$ and $\omega_\xi$ are related to the shear viscosity $\nu$ and bulk viscosity $\xi$, respectively, through the following expressions:
\begin{equation*}
\nu = \gamma \; q^2 c_{s*}^2\left(\frac{1}{\omega_{\nu}}-\frac{1}{2}\right)\Delta t, \qquad \xi = \gamma q^2 c_{s*}^2\left(\frac{1}{\omega_{\xi}}- \frac{1}{2}\right)\Delta t.
\end{equation*}
Notice that these transport coefficients are functions of the preconditioning parameter $\gamma$ and the grid aspect ratio $r$ (via $q$). In this work, we set the relaxation parameter associated with the bulk viscosity as well as those for the other, especially the higher order, moments to unity, i.e., $\omega_\xi=\omega_{1}=\omega_{21}=\omega_{12}=\omega_{22}=1.0$. To complete this sub-step for collision, we now segregate the post-collision combined central moments $\tilde{k}_{2s}$ and $\tilde{k}_{2d}$ into the bare central moments $\tilde{k}_{20}$ and $\tilde{k}_{02}$ (reflecting the application of the inverse operator $\tensor{B}^{-1}$) via
\begin{equation*}
\tilde {k}_{20} = \frac{1}{2}(\tilde{k}_{2s} + \tilde{k}_{2d}), \qquad \tilde {k}_{02} = \frac{1}{2}(\tilde{k}_{20} - \tilde{k}_{02}).
\end{equation*}
As a result, all the post-collision bare central moments supported by the rectangular D2Q9 lattice is now computed.

\item Compute post-collision raw moments
\newline
Evaluating
\begin{equation*}
\tilde{\mathbf{m}}=\tensor{F}^{-1}\tilde{\mathbf{m}}^c,
\end{equation*}
the post-collision raw moments $\tilde{k}_{mn}^\prime$ can be obtained from the corresponding central moments $\tilde{k}_{mn}$ computed in the previous sub-step, where $\tensor{F}^{-1}=\tensor{F}(-u_x,-u_y)$, with (see Eq.~\eqref{eq:Fmatrix} for $\tensor{F}$). Thus, we have
\begin{align}
  &\tilde{k}_{00} = \tilde{k}_{00}^\prime, \nonumber\\
  &\tilde{k}_{10} = \tilde{k}_{10}^\prime + u_x \tilde {k}_{00}^\prime , \nonumber\\
  &\tilde{k}_{01} = \tilde{k}_{01}^\prime + u_y \tilde{k}_{00}^\prime, \nonumber\\
  &\tilde{k}_{20} = \tilde{k}_{20}^\prime + 2 u_x  \tilde{k}_{10}^\prime + u_x^2 \tilde{k}_{00}^\prime , \nonumber\\
  &\tilde{k}_{02} = \tilde{k}_{02}^\prime + 2 u_y  \tilde{k}_{01}^\prime + u_y^2 \tilde{k}_{00}^\prime, \nonumber\\
  &\tilde{k}_{11} = \tilde{k}_{11}^\prime + u_y  \tilde{k}_{10}^\prime + u_x  \tilde{k}_{01}^\prime+ u_x u_y \tilde{k}_{00}^\prime , \nonumber\\
  &\tilde{k}_{21} = \tilde{k}_{21}^\prime + 2 u_x  \tilde{k}_{11}^\prime + u_x^2 \tilde{k}_{01}^\prime + u_y  \tilde{k}_{20}^\prime + 2 u_x u_y \tilde{k}_{10}^\prime + u_x^2 u_y \tilde{k}_{00}^\prime , \nonumber\\
  &\tilde{k}_{12} = \tilde{k}_{12}^\prime + 2 u_y  \tilde{k}_{11}^\prime + u_y^2 \tilde{k}_{10}^\prime + u_x  \tilde{k}_{02}^\prime + 2 u_x u_y \tilde{k}_{01}^\prime + u_x u_y^2 \tilde{k}_{00}^\prime , \nonumber\\
  &\tilde{k}_{22} = \tilde{k}_{22}^\prime + 2 u_x  \tilde{k}_{12}^\prime + u_x^2 \tilde{k}_{02}^\prime + 2 u_y  \tilde{k}_{21}^\prime + 4 u_x u_y \tilde{k}_{11}^\prime +2 u_x^2 u_y \tilde{k}_{01}^\prime + u_y^2 \tilde{k}_{20}^\prime+ 2 u_x u_y^2 \tilde{k}_{10}^\prime + u_x^2 u_y^2 \tilde{k}_{00}^\prime. \nonumber
\end{align}

\item Apply inverse scaling of post-collision raw moments for rectangular lattice
\newline
Perform
\begin{equation*}
\tilde{\mathbf{m}}\leftarrow\tensor{S}^{-1}\tilde{\mathbf{m}}
\end{equation*}
using Eq.~\eqref{eq:S-matrix-inverse} for the inverse scaling $\tensor{S}^{-1}$, which involves applying an inverse scale factor $r^{-n}$ for the raw moment $\tilde{k}_{mn}^\prime$ computed in the previous sub-step so that we have
\begin{align*}
&\tilde{k}_{00}^\prime=\tilde {k}_{00}^\prime, \quad \tilde{k}_{10}^\prime =\tilde{k}_{10}^\prime ,\quad \tilde{k}_{01}^\prime= \frac{1}{r}  \tilde{k}_{01}^\prime, \nonumber\\
&\tilde{k}_{20}^\prime= \tilde{k}_{20}^\prime,\quad \tilde{k}_{02}^\prime= \frac{1}{r^2} \tilde{k}_{02}^\prime,\quad
\tilde{k}_{11}^\prime=\frac{1}{r} \tilde{k}_{11}^\prime,\quad \tilde{k}_{21}^\prime=\frac{1}{r} \tilde{k}_{21}^\prime,\quad \tilde{k}_{12}^\prime=\frac{1}{r^2} \tilde{k}_{12}^\prime, \quad \tilde{k}_{22}^\prime=\frac{1}{r^2} \tilde{k}_{22}^\prime.
\end{align*}
This enables a more efficient transformation of the raw moments to distribution functions involving the inverse of simpler moment basis $\tensor{P}$ (see Eq.~\ref{eq:P-momentbasis}) for the square lattice, i.e., $\tensor{P}^{-1}$ in the next sub-step.

\item Compute post-collision distribution functions
\newline
Invoking the inverse mapping
\begin{equation*}
\tilde{\mathbf{f}}=\tensor{P}^{-1}\tilde{\mathbf{m}},
\end{equation*}
we then obtain the post-collision distribution functions on the rectangular D2Q9 lattice as follows:
\begin{eqnarray*}\label{eq:72}
  \tilde f_0 &=& \left(\tilde{k}_{00}^\prime - \tilde{k}_{20}^\prime - \tilde{k}_{02}^\prime + \tilde{k}_{22}^\prime\right), \\
  \tilde f_1 &=& \frac{1}{2} \left(\tilde{k}_{10}^\prime + \tilde{k}_{20}^\prime - \tilde{k}_{12}^\prime - \tilde{k}_{22}^\prime \right), \\
  \tilde f_2 &=& \frac{1}{2} \left(\tilde{k}_{01}^\prime + \tilde{k}_{02}^\prime - \tilde{k}_{21}^\prime - \tilde{k}_{22}^\prime \right), \\
  \tilde f_3 &=& \frac{1}{2} \left(-\tilde{k}_{10}^\prime + \tilde{k}_{20}^\prime + \tilde{k}_{12}^\prime - \tilde{k}_{22}^\prime \right), \\
  \tilde f_4 &=& \frac{1}{2} \left(-\tilde{k}_{01}^\prime + \tilde{k}_{02}^\prime + \tilde{k}_{21}^\prime - \tilde{k}_{22}^\prime \right), \\
  \tilde f_5 &=& \frac{1}{4} \left(\tilde{k}_{11}^\prime + \tilde{k}_{21}^\prime + \tilde{k}_{12}^\prime + \tilde{k}_{22}^\prime \right), \\
  \tilde f_6 &=& \frac{1}{4} \left(-\tilde{k}_{11}^\prime + \tilde{k}_{21}^\prime - \tilde{k}_{12}^\prime + \tilde{k}_{22}^\prime \right), \\
  \tilde f_7 &=& \frac{1}{4} \left(\tilde{k}_{11}^\prime - \tilde{k}_{21}^\prime - \tilde{k}_{12}^\prime + \tilde{k}_{22}^\prime \right), \\
  \tilde f_8 &=& \frac{1}{4} \left(-\tilde{k}_{11}^\prime - \tilde{k}_{21}^\prime + \tilde{k}_{12}^\prime + \tilde{k}_{22}^\prime \right).
\end{eqnarray*}

\item Stream distribution functions along particle characteristics
\newline
Performing perfect shift advection on the neighboring lattice nodes via
\begin{equation*}
f_\alpha(\bm{x},t+\Delta t)=\widetilde{f}_\alpha(\bm{x}-\bm{e}_\alpha\Delta t,t),
\end{equation*}
we obtain the updated distribution functions $f_\alpha$ at the end of time step $t+\Delta t$.

\item Update hydrodynamic fields
\newline
From the distribution functions obtained in the previous sub-step, update the fluid density and velocities, as well as the pressure field via
\begin{equation*}
\rho =\sum_{\alpha=0}^{8} f_\alpha, \quad     \rho \bm{u} =\sum_{\alpha=0}^{8} f_\alpha \bm{e}_\alpha + \frac{1}{2\gamma} \bm{F}\Delta t, \quad p= \gamma q^2 c_{s*}^2\rho.
\end{equation*}

\end{itemize}
It may be noted that for implementing moving wall no-slip boundary conditions for simulating shear flows on rectangular grids, the extension of the momentum-augmented half-way bounce back scheme with parametrization by the grid aspect ratio $r$ presented in our previous work~\cite{yahia2021central} can be used in the present LB formulation by noting that the speed of sound $c_s$ should be specified consistently using $c_s=q c_{s*}$, where $q=\mbox{min}(r,1)$.


\begin{thebibliography}{10}
\expandafter\ifx\csname url\endcsname\relax
  \def\url#1{\texttt{#1}}\fi
\expandafter\ifx\csname urlprefix\endcsname\relax\def\urlprefix{URL }\fi
\expandafter\ifx\csname href\endcsname\relax
  \def\href#1#2{#2} \def\path#1{#1}\fi

\bibitem{mcnamara1988use}
G.~R. McNamara, G.~Zanetti, Use of the {B}oltzmann equation to simulate
  lattice-gas automata, Physical review letters 61~(20) (1988) 2332.

\bibitem{benzi1992lattice}
R.~Benzi, S.~Succi, M.~Vergassola, The lattice {B}oltzmann equation: theory and
  applications, Physics Reports 222~(3) (1992) 145--197.

\bibitem{lallemand2021lattice}
P.~Lallemand, L.-S. Luo, M.~Krafczyk, W.-A. Yong, The lattice {B}oltzmann
  method for nearly incompressible flows, Journal of Computational Physics 431
  (2021) 109713.

\bibitem{qian1992lattice}
Y.-H. Qian, D.~d'Humi{\`e}res, P.~Lallemand, Lattice {BGK} models for
  navier-stokes equation, EPL (Europhysics Letters) 17~(6) (1992) 479.

\bibitem{d2002multiple}
D.~d'Humieres, Multiple--relaxation--time lattice {B}oltzmann models in three
  dimensions, Philosophical Transactions of the Royal Society of London. Series
  A: Mathematical, Physical and Engineering Sciences 360~(1792) (2002)
  437--451.

\bibitem{geier2006cascaded}
M.~Geier, A.~Greiner, J.~G. Korvink, Cascaded digital lattice {B}oltzmann
  automata for high {R}eynolds number flow, Physical Review E 73~(6) (2006)
  066705.

\bibitem{geier2015cumulant}
M.~Geier, M.~Sch{\"o}nherr, A.~Pasquali, M.~Krafczyk, The cumulant lattice
  {B}oltzmann equation in three dimensions: Theory and validation, Computers \&
  Mathematics with Applications 70~(4) (2015) 507--547.

\bibitem{karlin1999perfect}
I.~V. Karlin, A.~Ferrante, H.~C. {\"O}ttinger, Perfect entropy functions of the
  lattice {B}oltzmann method, Europhys. Lett. 47~(2) (1999) 182.

\bibitem{kruger2017lattice}
T.~Kr{\"u}ger, H.~Kusumaatmaja, A.~Kuzmin, O.~Shardt, G.~Silva, E.~M. Viggen,
  The lattice {B}oltzmann method, Springer International Publishing 10 (2017)
  978--3.

\bibitem{lallemand2000theory}
P.~Lallemand, L.-S. Luo, Theory of the lattice {B}oltzmann method: Dispersion,
  dissipation, isotropy, galilean invariance, and stability, Physical Review E
  61~(6) (2000) 6546.

\bibitem{koelman1991simple}
J.~Koelman, A simple lattice {B}oltzmann scheme for navier-stokes fluid flow,
  EPL (Europhysics Letters) 15~(6) (1991) 603.

\bibitem{hegele2013rectangular}
L.~A. Hegele~Jr, K.~Mattila, P.~C. Philippi, Rectangular lattice-boltzmann
  schemes with bgk-collision operator, Journal of Scientific Computing 56~(2)
  (2013) 230--242.

\bibitem{peng2016lattice}
C.~Peng, Z.~Guo, L.-P. Wang, A lattice-{BGK} model for the {N}avier-{S}tokes
  equations based on a rectangular grid, Computers \& Mathematics with
  Applications (2016).

\bibitem{wang2019simulating}
Z.~Wang, J.~Zhang, Simulating anisotropic flows with isotropic lattice models
  via coordinate and velocity transformation, International Journal of Modern
  Physics C 30~(10) (2019) 1941001.

\bibitem{bouzidi2001lattice}
M.~Bouzidi, D.~d'Humi{\`e}res, P.~Lallemand, L.-S. Luo, Lattice {B}oltzmann
  equation on a two-dimensional rectangular grid, Journal of Computational
  Physics 172~(2) (2001) 704--717.

\bibitem{zhou2012mrt}
J.~G. Zhou, {MRT} rectangular lattice {B}oltzmann method, International Journal
  of Modern Physics C 23~(05) (2012) 1250040.

\bibitem{peng2016hydrodynamically}
C.~Peng, H.~Min, Z.~Guo, L.-P. Wang, A hydrodynamically-consistent {MRT}
  lattice {B}oltzmann model on a 2d rectangular grid, Journal of Computational
  Physics 326 (2016) 893--912.

\bibitem{asinari2008generalized}
P.~Asinari, Generalized local equilibrium in the cascaded lattice {B}oltzmann
  method, Physical Review E 78~(1) (2008) 016701.

\bibitem{premnath2009incorporating}
K.~N. Premnath, S.~Banerjee, Incorporating forcing terms in cascaded lattice
  boltzmann approach by method of central moments, Physical Review E 80~(3)
  (2009) 036702.

\bibitem{premnath2011three}
K.~N. Premnath, S.~Banerjee, On the three-dimensional central moment lattice
  {B}oltzmann method, Journal of Statistical Physics 143~(4) (2011) 747--794.

\bibitem{ning2016numerical}
Y.~Ning, K.~N. Premnath, D.~V. Patil, Numerical study of the properties of the
  central moment lattice {B}oltzmann method, International Journal for
  Numerical Methods in Fluids 82~(2) (2016) 59--90.

\bibitem{de2017non}
A.~De~Rosis, Non-orthogonal central moments relaxing to a discrete equilibrium:
  A d2q9 lattice {B}oltzmann model, EPL (Europhysics Letters) 116~(4) (2017)
  44003.

\bibitem{de2017nonorthogonal}
A.~De~Rosis, Nonorthogonal central-moments-based lattice {B}oltzmann scheme in
  three dimensions, Physical Review E 95~(1) (2017) 013310.

\bibitem{fei2017consistent}
L.~Fei, K.~H. Luo, Consistent forcing scheme in the cascaded lattice
  {B}oltzmann method, Physical Review E 96~(5) (2017) 053307.

\bibitem{fei2018three}
L.~Fei, K.~H. Luo, Q.~Li, Three-dimensional cascaded lattice {B}oltzmann
  method: Improved implementation and consistent forcing scheme, Physical
  Review E 97~(5) (2018) 053309.

\bibitem{Hajabdollahi201897}
F.~Hajabdollahi, K.~N. Premnath, {G}alilean-invariant preconditioned
  central-moment lattice {B}oltzmann method without cubic velocity errors for
  efficient steady flow simulations, Phys. Rev. E 97 (2018) 053303.

\bibitem{HAJABDOLLAHI2018838}
F.~Hajabdollahi, K.~N. Premnath, Central moments-based cascaded lattice
  {B}oltzmann method for thermal convective flows in three-dimensions, Int. J.
  Heat Mass Transf. 120 (2018) 838 -- 850.

\bibitem{chavez2018improving}
M.~Ch{\'a}vez-Modena, E.~Ferrer, G.~Rubio, Improving the stability of
  multiple-relaxation lattice {B}oltzmann methods with central moments, Comput.
  Fluids 172 (2018) 397--409.

\bibitem{hajabdollahi2019cascaded}
F.~Hajabdollahi, K.~N. Premnath, S.~W. Welch, Cascaded lattice {B}oltzmann
  method based on central moments for axisymmetric thermal flows including
  swirling effects, International Journal of Heat and Mass Transfer 128 (2019)
  999--1016.

\bibitem{fei2020mesoscopic}
L.~Fei, J.~Yang, Y.~Chen, H.~Mo, K.~H. Luo, Mesoscopic simulation of
  three-dimensional pool boiling based on a phase-change cascaded lattice
  {B}oltzmann method, Physics of Fluids 32~(10) (2020) 103312.

\bibitem{hajabdollahi2021central}
F.~Hajabdollahi, K.~N. Premnath, S.~W. Welch, Central moment lattice
  {B}oltzmann method using a pressure-based formulation for multiphase flows at
  high density ratios and including effects of surface tension and {M}arangoni
  stresses, Journal of Computational Physics 425 (2021) 109893.

\bibitem{adam2019numerical}
S.~Adam, K.~N. Premnath, Numerical investigation of the cascaded central moment
  lattice {B}oltzmann method for non-newtonian fluid flows, Journal of
  Non-Newtonian Fluid Mechanics 274 (2019) 104188.

\bibitem{adam2021cascaded}
S.~Adam, F.~Hajabdollahi, K.~N. Premnath, Cascaded lattice {B}oltzmann modeling
  and simulations of three-dimensional non-newtonian fluid flows, Computer
  Physics Communications (2021) 107858.

\bibitem{yahia2021central}
E.~Yahia, K.~N. Premnath, Central moment lattice {B}oltzmann method on a
  rectangular lattice, Physics of Fluids 33~(5) (2021) 057110.

\bibitem{yahia2021three}
E.~Yahia, W.~Schupbach, K.~N. Premnath, Three-dimensional central moment
  lattice boltzmann method on a cuboid lattice for anisotropic and
  inhomogeneous flows, Fluids 6~(9) (2021) 326.

\bibitem{dubois2015stability}
F.~Dubois, T.~F{\'e}vrier, B.~Graille, On the stability of a relative velocity
  lattice {B}oltzmann scheme for compressible navier--stokes equations, Comptes
  Rendus M{\'e}canique 343~(10-11) (2015) 599--610.

\bibitem{turkel1987preconditioned}
E.~Turkel, Preconditioned methods for solving the incompressible and low speed
  compressible equations, Journal of computational physics 72~(2) (1987)
  277--298.

\bibitem{turkel1999preconditioning}
E.~Turkel, Preconditioning techniques in computational fluid dynamics, Annual
  Review of Fluid Mechanics 31~(1) (1999) 385--416.

\bibitem{guo2004preconditioned}
Z.~Guo, T.~Zhao, Y.~Shi, Preconditioned lattice-{B}oltzmann method for steady
  flows, Physical Review E 70~(6) (2004) 066706.

\bibitem{premnath2009steady}
K.~N. Premnath, M.~J. Pattison, S.~Banerjee, Steady state convergence
  acceleration of the generalized lattice {B}oltzmann equation with forcing
  term through preconditioning, Journal of Computational Physics 228~(3) (2009)
  746--769.

\bibitem{izquierdo2009optimal}
S.~Izquierdo, N.~Fueyo, Optimal preconditioning of lattice {B}oltzmann methods,
  Journal of Computational Physics 228~(17) (2009) 6479--6495.

\bibitem{meng2018preconditioned}
X.~Meng, L.~Wang, X.~Yang, Z.~Guo, Preconditioned multiple-relaxation-time
  lattice {B}oltzmann equation model for incompressible flow in porous media,
  Physical Review E 98~(5) (2018) 053309.

\bibitem{hajabdollahi2019improving}
F.~Hajabdollahi, K.~N. Premnath, Improving the low mach number steady state
  convergence of the cascaded lattice {B}oltzmann method by preconditioning,
  Computers \& Mathematics with Applications 78~(4) (2019) 1115--1130.

\bibitem{walsh2020preconditioned}
B.~Walsh, F.~J. Boyle, A preconditioned lattice {B}oltzmann flux solver for
  steady flows on unstructured hexahedral grids, Computers \& Fluids 210 (2020)
  104634.

\bibitem{hajabdollahi2018symmetrized}
F.~Hajabdollahi, K.~N. Premnath, Symmetrized operator split schemes for force
  and source modeling in cascaded lattice {B}oltzmann methods for flow and
  scalar transport, Phys. Rev. E 97~(6) (2018) 063303.

\bibitem{chapman1990mathematical}
S.~Chapman, T.~G. Cowling, D.~Burnett, The mathematical theory of non-uniform
  gases: an account of the kinetic theory of viscosity, thermal conduction and
  diffusion in gases, Cambridge university press, 1990.

\bibitem{zong2016designing}
Y.~Zong, C.~Peng, Z.~Guo, L.-P. Wang, Designing correct fluid hydrodynamics on
  a rectangular grid using {MRT} lattice {B}oltzmann approach, Computers \&
  Mathematics with Applications 72~(2) (2016) 288--310.

\bibitem{fei2018modeling}
L.~Fei, K.~H. Luo, C.~Lin, Q.~Li, Modeling incompressible thermal flows using a
  central-moments-based lattice {B}oltzmann method, Int. J. Heat Mass Transf.
  120 (2018) 624--634.

\bibitem{fei2018cascaded}
L.~Fei, K.~H. Luo, Cascaded lattice {B}oltzmann method for incompressible
  thermal flows with heat sources and general thermal boundary conditions,
  Computers \& Fluids 165 (2018) 89--95.

\bibitem{hajabdollahi2020local}
F.~Hajabdollahi, K.~N. Premnath, Local vorticity computation approach in double
  distribution functions based lattice {B}oltzmann methods for flow and scalar
  transport, International Journal of Heat and Fluid Flow 83 (2020) 108577.

\bibitem{ghia1982high}
U.~Ghia, K.~N. Ghia, C.~Shin, High-{R}e solutions for incompressible flow using
  the navier-stokes equations and a multigrid method, Journal of computational
  physics 48~(3) (1982) 387--411.

\end{thebibliography}

\end{document}